%% file: capacity_analysis_bbm_channel_model.tex
\documentclass[journal, final, twoside]{IEEEtran}
\usepackage{cite}

\ifCLASSINFOpdf
\else
\fi
\usepackage[cmex10]{amsmath}
\usepackage{centernot}
\usepackage{amssymb}
\usepackage{graphics}
\usepackage{tikz}
\usepackage{pgfplots}

\usetikzlibrary{pgfplots.groupplots, patterns, decorations.pathreplacing}

\newtheorem{IEEEtheorem}{Theorem}
\newtheorem{IEEElemma}{Lemma}
\newtheorem{IEEEprop}{Proposition}
\newtheorem{IEEEdefinition}{Definition}
\newtheorem{IEEEpropcccap}{(\cite{Wolfowitz_1959},~\cite{Blackwell_1959}) Proposition}

\usepackage{bm}

\DeclareMathOperator*{\argmin}{arg\,min}

\DeclareMathOperator{\E}{E}
\DeclareMathOperator{\Var}{Var}

\usepackage{algorithm}
\usepackage[noend]{algorithmic}

\usepackage{multirow}
\usepackage{hhline}

\usepackage{subfiles}

\newcommand{\ca}{{\sim}}
\newcommand{\kcount}[2]{\ensuremath{K_{#1}^{(#2)}}}

\newcommand{\ZOerror} {\mbox{\ensuremath{0 \rightarrow 1}}}
\newcommand{\OZerror} {\mbox{\ensuremath{1 \rightarrow 0}}}
\newcommand{\indep}{\mathrel\bot\joinrel\mspace{-8mu}\mathrel\bot}

\usepackage{fancyhdr}

\fancypagestyle{plain}{
  \fancyhf{}
  \fancyhead[C]{Conference on \LaTeX}     
  \fancyfoot[L]{This is a notice}

}
\usepackage{eso-pic}

\pgfmathdeclarefunction{gauss}{2}{%
	\pgfmathparse{1/(#2*sqrt(2*pi))*exp(-((x-#1)^2)/(2*#2^2))}%
}

\begin{document}
\bstctlcite{IEEEexample:BSTcontrol}


\title{On the Capacity of the Beta-Binomial Channel Model for Multi-Level Cell Flash Memories}
%
%
%


\author{Veeresh Taranalli,~\IEEEmembership{Student Member, IEEE,}
        Hironori Uchikawa,~\IEEEmembership{Member, IEEE,}
        Paul H. Siegel,~\IEEEmembership{Fellow, IEEE}
\thanks{
This work was supported by National Science Foundation (NSF) Grants CCF-1116739, CCF-1405119 and the Center for Memory and Recording Research, UC San Diego.}
\thanks{V. Taranalli, and P. H. Siegel are with the University of California, San Diego, La Jolla, CA 92093-0401, USA (e-mail: {vtaranalli, psiegel}@ucsd.edu).}
\thanks{H. Uchikawa is with Toshiba Corporation, Tokyo 247-8585, Japan (e-mail: hironori.uchikawa@toshiba.co.jp).}}

\maketitle
%
\begin{abstract}
The beta-binomial (BBM) channel model was recently proposed to model the overdispersed statistics of empirically observed bit errors in multi-level cell (MLC) flash memories. In this paper, we study the capacity of the BBM channel model for MLC flash memories. Using the compound channel approach, we first show that the BBM channel model capacity is zero. However, through empirical observation, this appears to be a very pessimistic estimate of the flash memory channel capacity. We propose a refined channel model called the truncated-support beta-binomial (\mbox{TS-BBM}) channel model and derive its capacity. Using empirical error statistics from \mbox{1X-nm} and \mbox{2Y-nm} MLC flash memories, we numerically estimate the \mbox{TS-BBM} channel model capacity as a function of the program/erase (P/E) cycling stress. The capacity of the \mbox{2-TS-BBM} channel model provides an upper bound on the coding rates for the flash memory chip assuming a single binary error correction code is used.
\end{abstract}


%
\IEEEpeerreviewmaketitle

\subfile{sections/introduction}

%

\subfile{sections/capacity_bbm_channel_model.tex}

%
\subfile{sections/truncated_support_bbm_channel_model.tex}

%

\subfile{sections/capacity_ts_bbm_channel_model.tex}
%

\subfile{sections/conclusion}
%


\subfile{sections/appendix}





%





\ifCLASSOPTIONcaptionsoff
  \newpage
\fi



\bibliographystyle{IEEEtran}
\bibliography{IEEEabrv,references}
%
%



%

\begin{IEEEbiography}[{\includegraphics[width=1.0in,height=1.25in,clip]{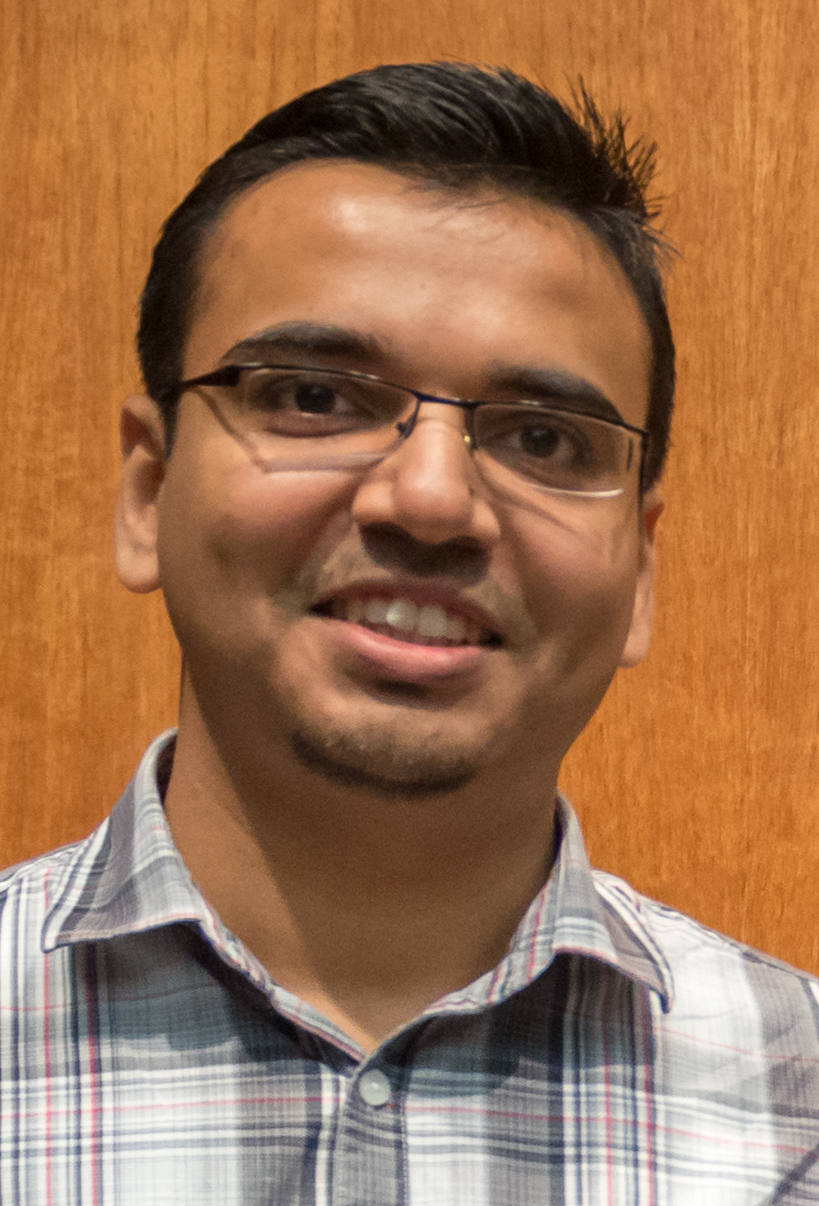}}]{Veeresh Taranalli}
(S'06) received the B.Tech. degree in electronics and communication engineering from the National Institute of Technology Karnataka, Surathkal, India, in 2009, and the M.S. degree in electrical engineering from the University of California, San Diego, CA, USA, in 2013. He is currently pursuing the Ph.D. degree in electrical engineering with the University of California, San Diego. His research interests include error characterization, channel modeling and study of modern coding techniques for NAND flash memory applications. He received the 2015 Shannon Memorial Fellowship awarded by the University of California, San Diego.
\end{IEEEbiography}


\begin{IEEEbiography}[{\includegraphics[width=1.0in,height=1.25in,clip]{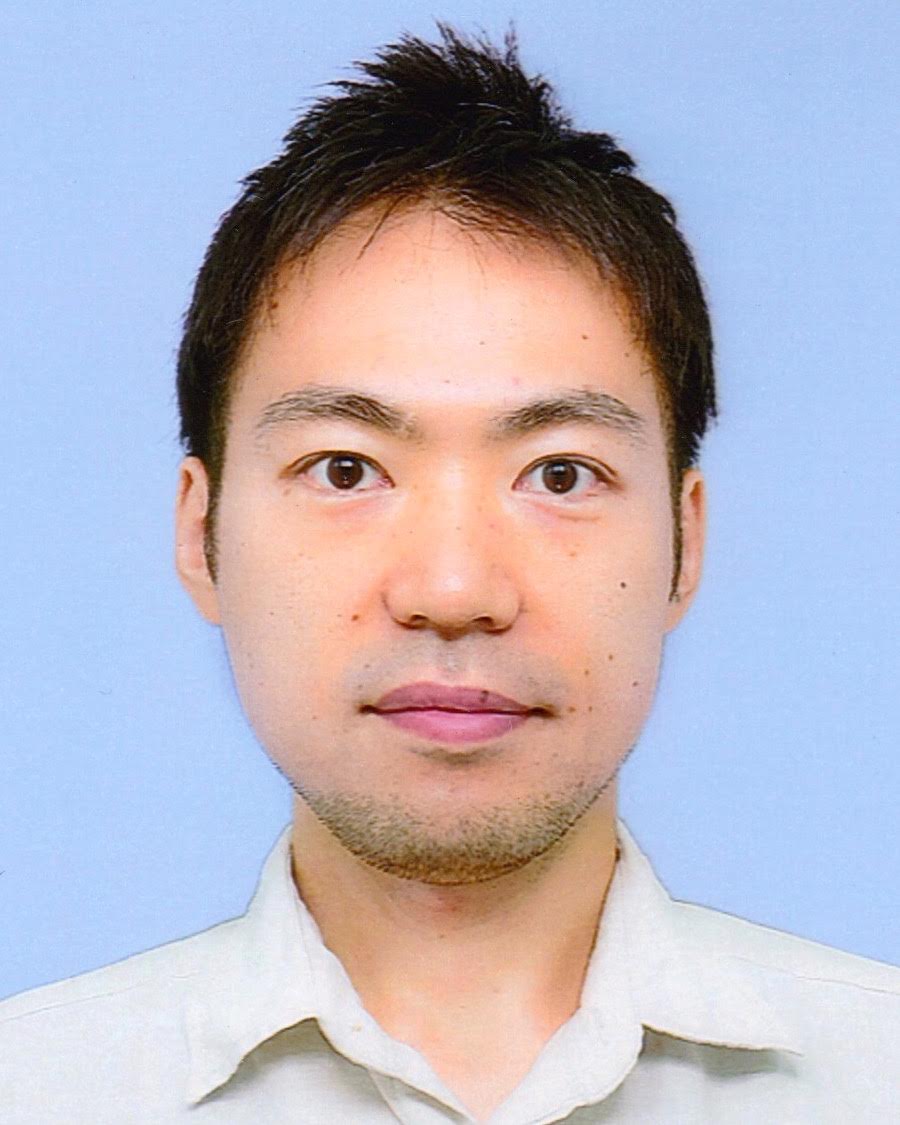}}]{Hironori Uchikawa}
received the B.E. and M.E. degrees in electrical engineering from Yokohama National University in 2001 and 2003, respectively, and the Ph.D. degree in electrical engineering from the Tokyo Institute of Technology in 2012. Since 2003, he has been with Toshiba Corporation. From 2013 to 2014, he was a Visiting Scholar with the Center for Memory Recording Research, University of California, San Diego. His research interests include coding theory, information theory, communication theory, and their application to storage systems. He received the 2008 Young Researcher’s Award of the Institute of Electronics, Information and Communication Engineers of Japan.
\end{IEEEbiography}

\vfill

\begin{IEEEbiography}[{\includegraphics[width=1.0in,height=1.25in,clip]{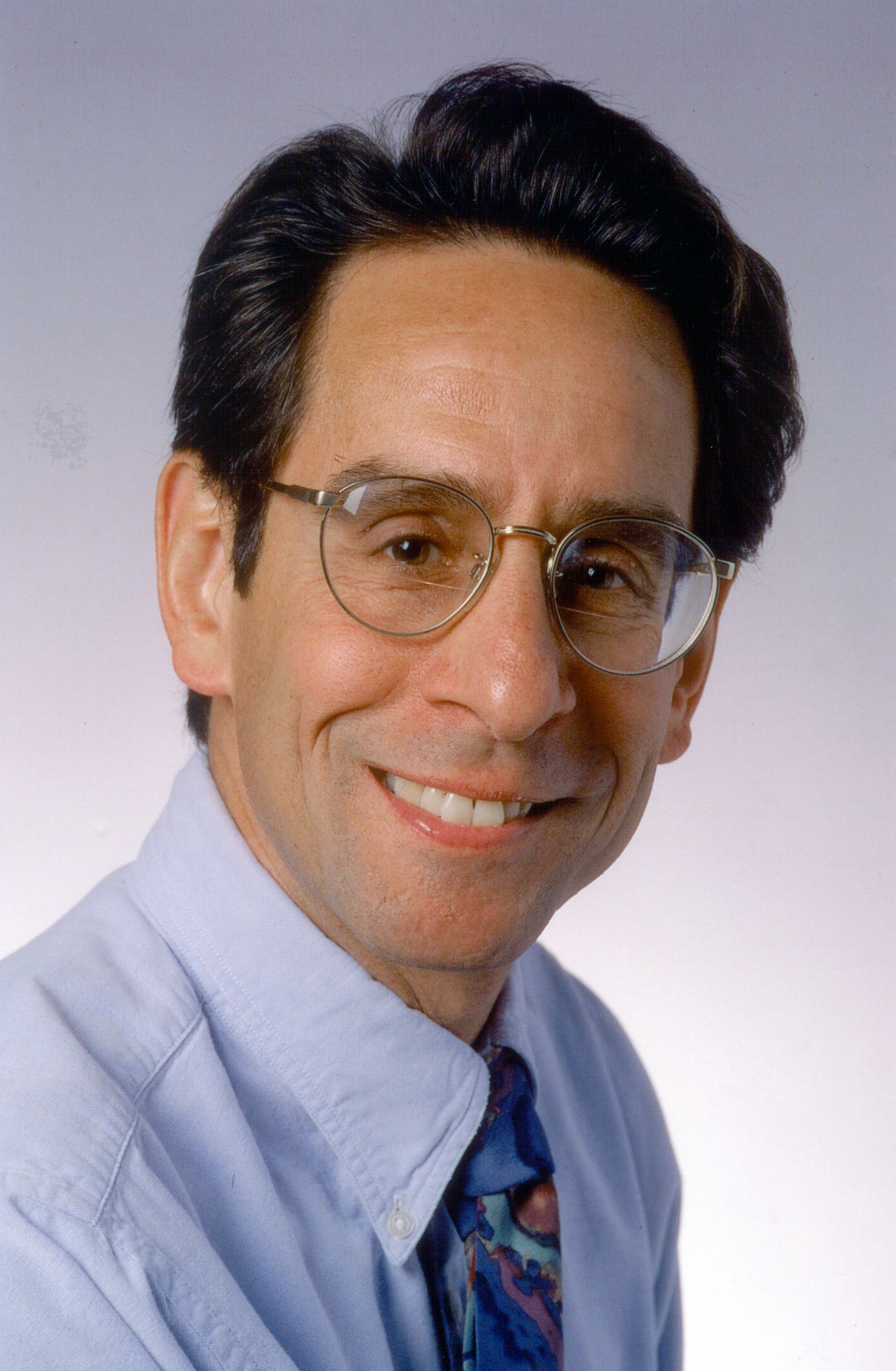}}]{Paul H. Siegel}
(M'82-SM'90-F'97) received the S.B. and Ph.D. degrees in mathematics from the Massachusetts Institute of Technology, Cambridge, MA, USA, in 1975 and 1979, respectively. He held a Chaim Weizmann Postdoctoral Fellowship with the Courant Institute, New York University, New York, NY, USA. He was with the IBM Research Division, San Jose, CA, USA, from 1980 to 1995. He joined the faculty at the University of California, San Diego, CA, USA, in 1995, where he is currently a Professor of electrical and computer engineering with the Jacobs School of Engineering. He is affiliated with the Center for Memory and Recording Research where he holds an Endowed Chair and served as a Director from 2000 to 2011. His research interests include information theory and communications, particularly coding and modulation techniques, with applications to digital data storage and transmission. He is a Member of the National Academy of Engineering. He was a Member of the Board of Governors of the IEEE Information Theory Society from 1991 to 1996 and from 2009 to 2014. He was the 2015 Padovani Lecturer of the IEEE Information Theory Society. He was a recipient of the 2007 Best Paper Award in Signal Processing and Coding for Data Storage from the Data Storage Technical Committee of the IEEE Communications Society. He was the co-recipient of the 1992 IEEE Information Theory Society Paper Award and the 1993 IEEE Communications Society Leonard G. Abraham Prize Paper Award. He served as a Co-Guest Editor of the 1991 Special Issue on Coding for Storage Devices of the IEEE TRANSACTIONS ON INFORMATION THEORY. He served as an Associate Editor of Coding Techniques of the IEEE TRANSACTIONS ON INFORMATION THEORY from 1992 to 1995, and as an Editor-in-Chief from 2001 to 2004. He was also a Co-Guest Editor of the 2001 two-part issue on The Turbo Principle: From Theory to Practice and the 2016 issue on Recent Advances in Capacity Approaching Codes of the IEEE JOURNAL ON SELECTED AREAS IN COMMUNICATIONS.
\end{IEEEbiography}

\vfill






\end{document}

%% file: sections/introduction.tex
\section{Introduction}
\label{sec:introduction}
\IEEEPARstart{T}{he} development of channel models for NAND flash memories is essential for improved design, decoding and performance evaluation of error-correcting codes (ECC) and error-mitigating codes.
Channel models for flash memories are also important as mathematical tools to approximate/estimate the capacity of the underlying flash memory channel. In this paper, we focus on the information-theoretic aspect of flash memory channel models and study the capacity of the recently proposed 2-Beta-Binomial (2-BBM) channel model for multi-level cell (MLC) flash memories~\cite{Taranalli_2016}.
%
%
\subsection{Overview of the Problem}
The beta-binomial (BBM) channel model for MLC flash memories was first proposed in~\cite{Taranalli_2016} to accurately model the overdispersed statistics of the number of bit errors per frame observed empirically during program/erase (P/E) cycling. It was also shown in~\cite{Taranalli_2016} that the 2-BBM channel model for MLC flash memories provides more accurate ECC frame error rate (FER) performance estimation than a \mbox{2-Binary Asymmetric Channel} (\mbox{2-BAC}) model. Therefore, the 2-BBM channel model presents itself as a natural candidate for estimating the actual MLC flash memory channel capacity. However, as we will show in this paper, the capacity of the BBM channel model is zero. The main goal of this paper is to derive a non-zero capacity channel model for MLC flash memories which is also representative of the empirically observed overdispersed error statistics and provides accurate ECC FER performance estimation. We note that the BBM channel model can be shown to be equivalent to an urn based channel model, similar to the zero-capacity urn based contagion channel model proposed in~\cite{Alajaji_1994}. In~\cite{Alajaji_1994}, an alternative finite memory version of the zero-capacity urn based contagion channel model was proposed and this alternative channel model was shown to have a non-zero capacity. Another approach to studying zero-capacity real world channels is provided by the example of slow fading channels in wireless communications. The Shannon capacity of a slow fading wireless communication channel is known to be zero and hence, in practice the $\epsilon-$outage capacity is used as an alternative performance measure for slow fading channels~\cite{Tse_2005}.
%

Previously, information-theoretic studies of NAND flash memory channel models have been proposed in~\cite{Dong_2012, Huang_2013, Li_2016}. In~\cite{Dong_2012}, an approximate channel model that incorporates P/E cycling, inter-cell interference (ICI), and data retention effects for MLC flash memories is proposed, and bounds on its capacity are established. In~\cite{Huang_2013}, the problem of estimating flash memory capacity with an underlying m-AM (amplitude modulation) channel model with input dependent Gaussian noise variance is studied. An information-theoretic study of a one dimensional causal channel model for ICI in flash memories is presented in~\cite{Li_2016}. It is important to note that all these previous information-theoretic studies used cell level based channel models. However in practice, the ECC is applied at the bit (page) level and hence we consider a binary-input binary-output channel model such as the BBM channel model for the study of MLC flash memory capacity. The problem of how well the channel models used in~\cite{Dong_2012, Huang_2013} represent the empirical error characteristics has also not been addressed. In this context, the BBM channel proposed in~\cite{Taranalli_2016} has been shown to provide accurate bit error rate and ECC FER performance prediction under P/E cycling. This makes the BBM channel model an ideal candidate for an information-theoretic study of the flash memory channel.

\subsection{Summary of Contributions}
We observe that the BBM channel model for MLC flash memories proposed in~\cite{Taranalli_2016} is a compound channel model. Using well known results for compound channel capacity~\cite{Wolfowitz_1959, Blackwell_1959}, we show that the capacity of the BBM channel model is zero.

Using the BBM channel model parameters derived from empirical error characterization results in~\cite{Taranalli_2016}, we observe that the BBM channel model is a very pessimistic channel model for MLC flash memories with respect to the problem of capacity estimation. This is because the probability mass of the beta-distributed $\ZOerror$ and $\OZerror$ bit error probabilities in the BBM channel model is concentrated in a small interval of the support $[0, 1]$. This observation allows us to define a
truncated-support beta random variable and correspondingly a truncated-support beta-binomial (TS-BBM) channel model. We derive analytically the relationships between the statistics of the number of bit errors per frame resulting from a TS-BBM channel model and those corresponding to the BBM channel model. These results are then used to propose an approximate search algorithm to identify the truncation intervals necessary to obtain a TS-BBM channel model from a BBM channel model.

Using Monte-Carlo simulations, we show LDPC and polar code FER performance results using the TS-BBM channel model. Comparing with the results obtained using the BBM channel model~\cite{Taranalli_2016}, we observe that the proposed TS-BBM channel model is also able to provide accurate ECC FER performance estimation by modeling the overdispersed statistics of the number of bit errors per frame.

We derive the capacity of the proposed TS-BBM channel model using the compound channel approach. We then present non-zero capacity estimates corresponding to the TS-BBM channel models derived from empirical error characterization results under P/E cycling stress~\cite{Taranalli_2016}. The capacity estimates are obtained for \mbox{1X-nm} and \mbox{2Y-nm} feature size MLC flash memory chips from two different vendors referred to as vendor-A and vendor-B, respectively. To the best of our knowledge, this is the first study that presents capacity estimates for a binary-input binary-output channel model that has been shown to accurately fit the empirical error characteristics in MLC flash memories~\cite{Taranalli_2016}. However, note that the BBM and the TS-BBM channel models do not explicitly model the data dependence of cell/bit errors due to ICI~\cite{Taranalli_2015} and assume independent random bit errors in lower and upper pages of a MLC flash memory. Therefore these constraints of the TS-BBM channel model have to be considered when using the corresponding capacity estimates in practical applications.

\subsection{Organization of the Paper}
The rest of the paper is organized as follows. In Section~\ref{sec:capacity_bbm_channel_model}, we briefly review the BBM channel model proposed in~\cite{Taranalli_2016} and show that the capacity of the BBM channel model is zero. In Section~\ref{sec:ts_bbm_channel_model}, the \mbox{TS-BBM} channel model is defined and statistical properties of the model are derived. Results of FER performance simulation of LDPC and polar codes using the \mbox{TS-BBM} channel model are also presented. In Section~\ref{sec:capacity_ts_bbm_channel_model}, we derive and numerically compute the capacity of a \mbox{TS-BBM} channel model. We also provide a brief discussion of open problems related to coding for the \mbox{TS-BBM} channel model. Section~\ref{sec:conclusion} provides concluding remarks.

%% file: sections/capacity_bbm_channel_model.tex
\section{Capacity of the Beta-Binomial Channel Model}
\label{sec:capacity_bbm_channel_model}
In this section, we briefly review the BBM channel model presented in~\cite{Taranalli_2016} and derive the capacity of a BBM channel model. The BBM channel model is based on the beta-binomial probability distribution which is defined as the probability distribution of counts resulting from a binomial distribution when the probability of success varies according to the beta distribution between sets of trials\cite{Skellam_1948}.
\begin{figure}
    \centering
    \begin{tikzpicture}[yscale=0.3, xscale=1.2, node distance=0.2cm, auto, thick]
        \draw[-, thick] (-1.0, 0.0) -- (1.0, 0.0);
        \draw[-, thick] (-1.0, -5.0) -- (1.0, -5.0);
        \draw[-, thick] (-1.0, 0.0) -- (1.0, -5.0);
        \draw[-, thick] (-1.0, -5.0) -- (1.0, 0.0);

        \node at (0.0, 1.0) {\scriptsize{$1 - p$}};
        \node at (0.0, -6.0) {\scriptsize{$1 - q$}};
        \node at (-0.1, -1.0) {\scriptsize{$p$}};
        \node at (-0.2, -4.0) {\scriptsize{$q$}};

        \node at (-1.1, 0.0) {\scriptsize{0}};
        \node at (-1.1, -5.0) {\scriptsize{1}};
        \node at (1.1, 0.0) {\scriptsize{0}};
        \node at (1.1, -5.0) {\scriptsize{1}};

        \node at (-1.5, -2.5) {$X$};
        \node at (1.5, -2.5) {$Y$};

    \end{tikzpicture}
    \vspace*{-0.9em}
    \caption{Binary Asymmetric Channel}
    \label{fig:bac}
\end{figure}
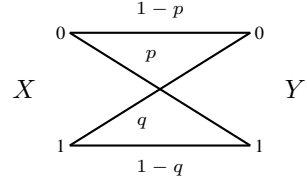

Let $\textrm{BAC}(\mathcal{X}, \mathcal{Y}, s)$ denote a binary asymmetric channel with state $s = (p, q)$ where $p = \Pr(y = 1|x = 0)$ and $q = \Pr(y = 0|x = 1)$, as shown in Fig.~\ref{fig:bac}. Let $\mathcal{X}$ and $\mathcal{Y}$ denote the input and output alphabets, respectively, both of which are binary. Using the notation from~\cite{Taranalli_2016}, we denote the number of $\ZOerror$ and $\OZerror$ bit errors in a frame with $m$ zeros (and $N-m$ ones) by $\kcount{m}{0}$ and $\kcount{N-m}{1}$, respectively. In the BBM channel model, $\kcount{m}{0}$ and $\kcount{N-m}{1}$ are modeled as independent random variables distributed according to the beta-binomial distribution i.e.,
{\allowdisplaybreaks
\begin{IEEEeqnarray}{rCl}
    p & \sim & \textrm{Beta}(a, b) \nonumber \\
    \kcount{m}{0}~|~p & \sim & \textrm{Binomial}(m, p) \nonumber \\
    \kcount{m}{0} & \sim & \textrm{Beta-Binomial}(m, a, b) ;\\
    q & \sim & \textrm{Beta}(c, d)  \nonumber \\
    \kcount{N-m}{1}~|~q & \sim & \textrm{Binomial}(N-m, q)  \nonumber \\
    \kcount{N-m}{1} & \sim & \textrm{Beta-Binomial}(N-m, c, d) ;\\
    \kcount{m}{0} & \indep & \kcount{N-m}{1} \label{eqn:bbm_xy_indep}
\end{IEEEeqnarray}}where $(a, b)$ (and similarly ($c$, $d$)) correspond to the parameters of a beta distribution whose probability density function is defined as
\begin{IEEEeqnarray}{rCl}
    f(\theta; \alpha, \beta) & = & \frac{\theta^{\alpha-1} (1 - \theta)^{\beta-1}}{B(\alpha, \beta)}~~0 \leq \theta \leq 1\\
    B(\alpha, \beta) & = & \int_{0}^{1} \lambda^{\alpha-1} (1 - \lambda)^{\beta-1} \textrm{d}\lambda
\end{IEEEeqnarray}
where $B(\alpha, \beta)$ represents the beta function.
Table~\ref{table:bbm_parameter_table} shows the estimated upper page BBM channel model parameters for both \mbox{vendor-A} and \mbox{vendor-B} MLC flash memory chips (cf.~Figures~6, 7 in \cite{Taranalli_2016}) from empirical data. Note that a frame length of $N = 8192$ was used to estimate the parameters of the underlying beta distributions of the BBM channel model. As Fig.~\ref{fig:bbm_param_vendor_a} shows, the parameter values of the beta distributions are independent of the frame length used to estimate them.
\begin{table}[!ht]
    \caption{Parameters corresponding to the upper page BBM channel models for vendor-A and vendor-B chips. $N = 8192$.}
	\label{table:bbm_parameter_table}
	\centering
	\bgroup
	\ifCLASSOPTIONonecolumn
    	\begin{tabular}{|c|c c c|c c c|}
    		\hline
    		\textbf{P/E} & \multicolumn{6}{|c|}{\textbf{Upper Page}} \\
    		\hhline{~------}
    		\textbf{Cycles} & \multicolumn{6}{|c|}{\textbf{Vendor-A}} \\
    		\hhline{~------}
    		& a & b & $\zeta_{a, b}$ & c & d & $\zeta_{c, d}$ \\
    		\hline
    		6000 & 22.67 & 7596.71 & $1.22 \times 10^{-3}$ & 18.16 & 11890.14 &  $0.69 \times 10^{-3}$ \\
    		\hline
    		8000 & 20.72 & 4143.52 & $2.13 \times 10^{-3}$ & 22.28 & 7821.13 &  $1.18 \times 10^{-3}$   \\
    		\hline
    		10000 & 21.36 & 2819.03 & $3.17 \times 10^{-3}$ & 26.12 & 5890.35 &  $1.69 \times 10^{-3}$ \\
    		\hline
             & \multicolumn{6}{|c|}{\textbf{Vendor-B}} \\
            \hline
            6000 & 15.58 & 20535.47 & $0.37 \times 10^{-3}$ & 7.16 & 7193.92 &  $0.69 \times 10^{-3}$\\
    		\hline
    		8000 & 15.28 & 9068.43 & $0.83 \times 10^{-3}$ & 7.58 & 4092.87 &  $1.25 \times 10^{-3}$\\
    		\hline
    		10000 & 13.36 & 4142.23 & $1.69 \times 10^{-3}$ & 9.28 & 2938.88 & $1.95 \times 10^{-3}$\\
    		\hline
    	\end{tabular}
	\else
		\def\arraystretch{1.5}
        \tabcolsep=0.11cm
    	\resizebox{0.98\columnwidth}{!}{
    	\begin{tabular}{|c|c c c|c c c|}
    		\hline
    		\textbf{P/E} & \multicolumn{6}{|c|}{\textbf{Upper Page}} \\
    		\hhline{~------}
    		\textbf{Cycles} & \multicolumn{6}{|c|}{\textbf{Vendor-A}} \\
    		\hhline{~------}
    		& a & b & $\zeta_{a, b}$ & c & d & $\zeta_{c, d}$ \\
    		\hline
    		6000 & 22.67 & 7596.71 & $1.22 \times 10^{-3}$ & 18.16 & 11890.14 &  $0.69 \times 10^{-3}$ \\
    		\hline
    		8000 & 20.72 & 4143.52 & $2.13 \times 10^{-3}$ & 22.28 & 7821.13 &  $1.18 \times 10^{-3}$   \\
    		\hline
    		10000 & 21.36 & 2819.03 & $3.17 \times 10^{-3}$ & 26.12 & 5890.35 &  $1.69 \times 10^{-3}$ \\
    		\hline
             & \multicolumn{6}{|c|}{\textbf{Vendor-B}} \\
            \hline
            6000 & 15.58 & 20535.47 & $0.37 \times 10^{-3}$ & 7.16 & 7193.92 &  $0.69 \times 10^{-3}$\\
    		\hline
    		8000 & 15.28 & 9068.43 & $0.83 \times 10^{-3}$ & 7.58 & 4092.87 &  $1.25 \times 10^{-3}$\\
    		\hline
    		10000 & 13.36 & 4142.23 & $1.69 \times 10^{-3}$ & 9.28 & 2938.88 & $1.95 \times 10^{-3}$\\
    		\hline
    	\end{tabular}
    	}
	\fi
	\egroup
\end{table}
\ifCLASSOPTIONonecolumn
    \begin{figure}[h]
        \centering
        \includegraphics[width=0.7\textwidth]{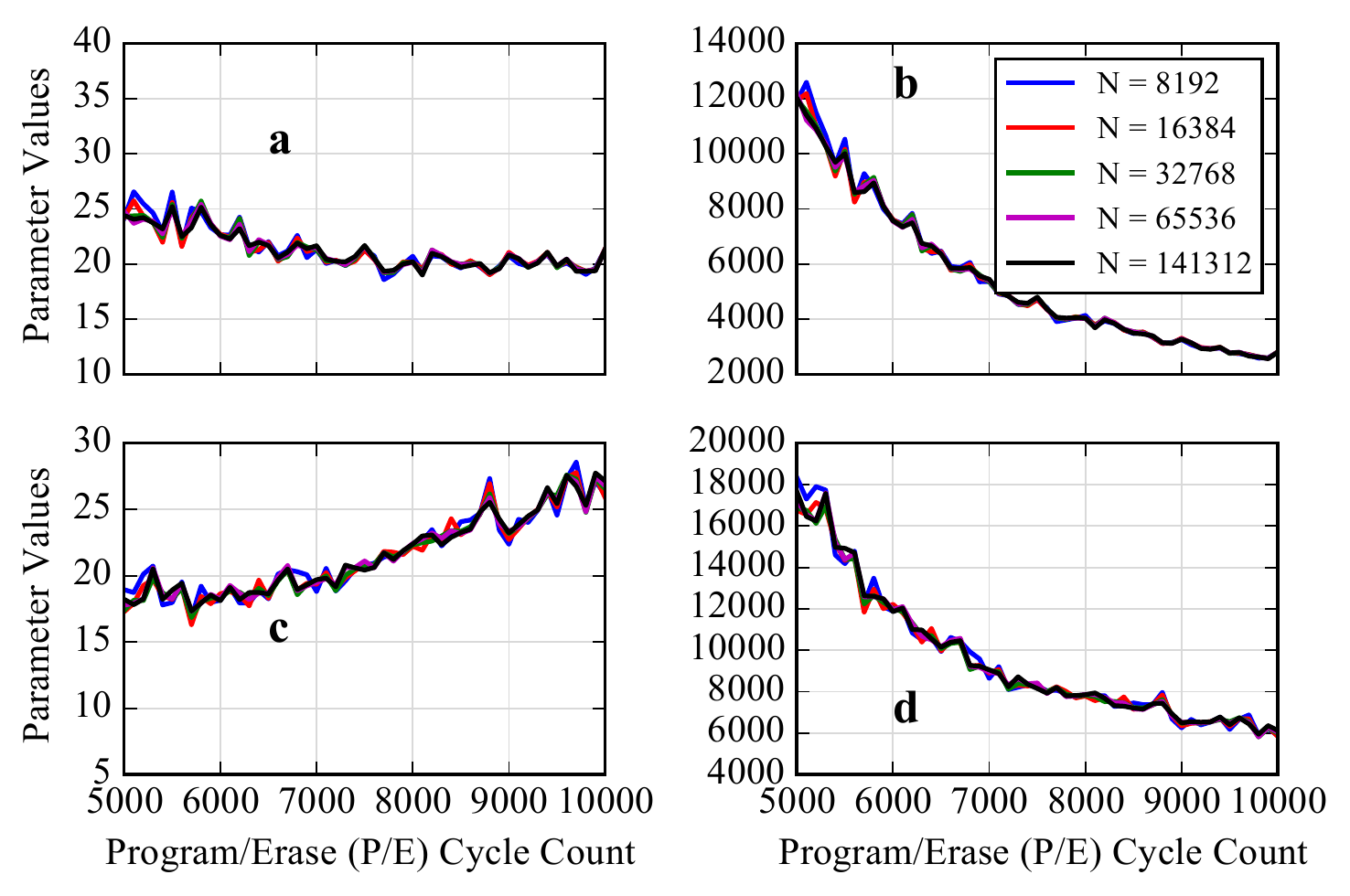}
        \caption{Parameter estimates for the upper page BBM channel model (($a$, $b$) for $\ZOerror$ error, ($c$, $d$) for $\OZerror$ error) using different frame lengths for \mbox{vendor-A} chip.}
        \label{fig:bbm_param_vendor_a}
    \end{figure}
\else
    \begin{figure}[h]
    	\centering
    	\includegraphics[width=0.47\textwidth]{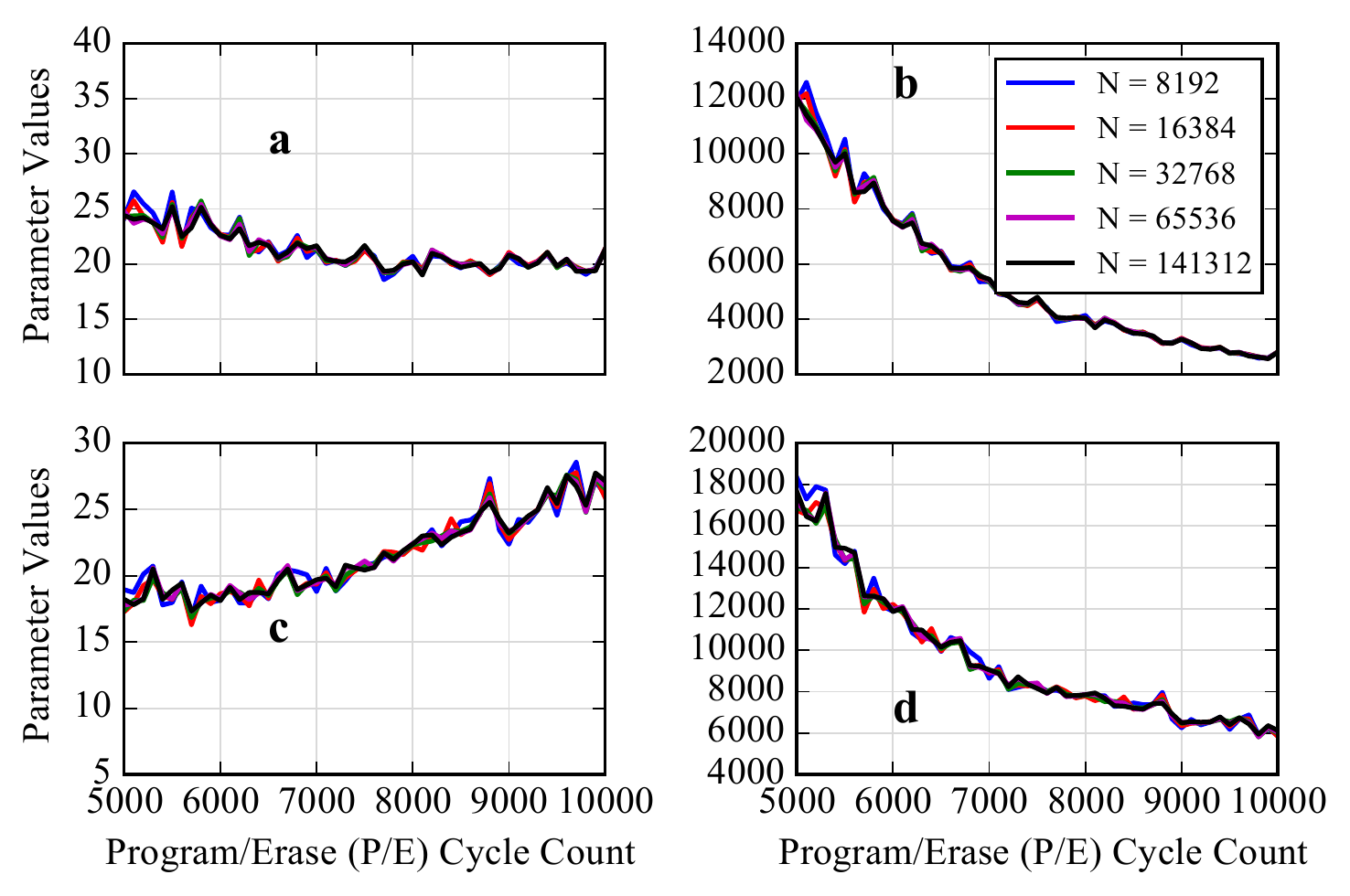}
    	\caption{Parameter estimates for the upper page BBM channel model (($a$, $b$) for $\ZOerror$ error, ($c$, $d$) for $\OZerror$ error) using different frame lengths for \mbox{vendor-A} chip.}
    	\label{fig:bbm_param_vendor_a}
    \end{figure}
\fi
Next, we observe that the BBM channel model is a compound channel model and derive its capacity.
\begin{IEEEdefinition}
A compound channel is defined as a set of discrete memoryless channels with state, $\textrm{DMC}(\mathcal{X}, \mathcal{Y}, s)$, with input alphabet $\mathcal{X}$, output alphabet $\mathcal{Y}$ and state $s \in \mathcal{S}$, where the channel state is chosen at random and fixed throughout the entire transmission block/frame i.e., $\Pr(y^N | x^N, s) = \Pi_{i=1}^{N} \Pr(y_i | x_i, s)$.
\end{IEEEdefinition}
\begin{IEEEdefinition}
    The BBM channel is a compound channel consisting of a set of BACs with state, BAC($\mathcal{X}$, $\mathcal{Y}$, $s$), indexed by the state variable,
    $s \in \mathcal{S} = \{(p, q)~|~p\sim\textrm{Beta}(a, b),~q\sim\textrm{Beta}(c, d) \}$.
    \label{defn:BBM_compound_channel}
\end{IEEEdefinition}
%
%
%
\begin{IEEEpropcccap}
The capacity of a compound channel assuming no state information is available at the encoder or the decoder is given by 
\begin{IEEEeqnarray}{rCl}
    C_{\textrm{CC}} & = & \sup_{\bar{\pi}} \inf_{s \in \mathcal{S}} I_{\bar{\pi}, s}(X;Y)
    \label{eqn:cap_cc}
\end{IEEEeqnarray}
where $I_{\bar{\pi}, s}(X;Y)$ denotes the mutual information corresponding to a DMC with state $s$ and input probability distribution, $\bar{\pi}$.
\label{prop:compound_channel_capacity}
\end{IEEEpropcccap}
%
\begin{IEEEtheorem}
    The capacity of a BBM channel is $0$.
\end{IEEEtheorem}
\begin{IEEEproof}
    From Definition~\ref{defn:BBM_compound_channel} and Proposition~\ref{prop:compound_channel_capacity},
    \begin{IEEEeqnarray}{rCl}
        C_{\textrm{BBM}} & = & \sup_{\bar{\pi}} \inf_{s \in \mathcal{S}} I_{\bar{\pi}, s}(X;Y)
    \end{IEEEeqnarray}
    where $I_{\bar{\pi}, s}(X;Y) = h(\pi_0(1-p)+\pi_1q) - \pi_0h(p) - \pi_1h(q)$ is the mutual information for a BAC with state $s = (p, q)$ and input probability distribution $\bar{\pi} = (\pi_0, \pi_1)$, i.e., $\pi_0 = \Pr(x=0), \pi_1 = \Pr(x=1)$. Here $h(\cdot)$ denotes the binary entropy function. Using the max-min inequality~\cite{Boyd_2004},
    \begin{IEEEeqnarray}{rCl}
        C_{\textrm{BBM}} & \leq & \inf_{s \in \mathcal{S}} \sup_{\bar{\pi}} I_{\bar{\pi}, s}(X;Y) = \inf_{s \in \mathcal{S}} C_s
    \end{IEEEeqnarray}
    where $C_s$ is the capacity of a BAC with state $s$. As $p, q \in [0, 1]$, $C_s = 0$ when the state $s = (\frac{1}{2}, \frac{1}{2})$. Therefore,
    \begin{IEEEeqnarray}{rCl}
        C_{\textrm{BBM}} \leq 0 \implies C_{\textrm{BBM}} = 0.
    \end{IEEEeqnarray}
\end{IEEEproof}

%% file: sections/truncated_support_bbm_channel_model.tex
\section{Truncated-Support Beta-Binomial Channel Model for MLC Flash Memories}
\label{sec:ts_bbm_channel_model}
%
\subsection{Motivation}
In the previous section, we showed that the capacity of a BBM channel model for MLC flash memories is zero. However empirical observations suggest that the BBM channel model is a very pessimistic model in this respect. To elaborate, from Table~\ref{table:bbm_parameter_table},
we observe that the parameters satisfy $a > 2$, $b > 2$, $c > 2$ and $d > 2$. It is known~\cite{Gupta_2004} that the beta probability density function (pdf) with parameters $\alpha$ and $\beta$ is unimodal and bell-shaped with two inflection points in $[0, 1]$ if $\alpha > 2$ and $\beta > 2$. The inflection points about the mode are given by
\begin{IEEEeqnarray}{rCl}
    \frac{\alpha-1}{\alpha+\beta-2} \pm \frac{\sqrt{\frac{(\alpha-1)(\beta-1)}{\alpha+\beta-3}}}{\alpha+\beta-2}.
\end{IEEEeqnarray}
The difference between the two inflection points, $\zeta_{\alpha, \beta}$, can be used as a measure of spread of the beta density function, where
\begin{IEEEeqnarray}{rCl}
    \zeta_{\alpha, \beta} & = & 2\frac{\sqrt{\frac{(\alpha-1)(\beta-1)}{\alpha+\beta-3}}}{\alpha+\beta-2}.
\end{IEEEeqnarray}
Table~\ref{table:bbm_parameter_table} also shows the values of $\zeta_{a, b}$ and $\zeta_{c, d}$ for the upper page BBM channel model for vendor-A and vendor-B chips. Both $\zeta_{a, b}$ and $\zeta_{c, d}$ are small indicating that the probability mass corresponding to the $\textrm{Beta}(a,b)$ and $\textrm{Beta}(c,d)$ pdf's of the BBM channel model is concentrated in a small interval. For example, from a visual inspection of a typical beta pdf corresponding to the BBM channel model, as shown in Fig.~\ref{fig:beta_pdf}, it is clear that the pdf value is negligible outside the bit error probability range of [$0.002$, $0.009$].
These observations motivate us to propose and define a truncated-support beta-binomial (\mbox{TS-BBM}) channel model for MLC flash memories. The primary goal of the TS-BBM channel model is to provide a non-zero capacity estimate which can be reasonably interpreted in the context of MLC flash memories, while modeling the empirically observed distributions of the number of bit errors per frame as accurately as possible. This latter constraint is essential for accurate ECC FER performance estimation as shown in~\cite{Taranalli_2016}.
\subsection{Definition and Statistics of the TS-BBM Channel Model}
Before defining a TS-BBM channel model, we define truncated-support beta (TS-Beta) and truncated-support beta-binomial (TS-BBM) random variables as follows.
\begin{IEEEdefinition}
A TS-Beta random variable is defined by the probability density function
\begin{IEEEeqnarray}{rCl}
    g(\theta; \alpha, \beta) & = & \frac{\theta^{\alpha-1} (1 - \theta)^{\beta-1}}{B_{\theta_u}(\alpha, \beta) - B_{\theta_l}(\alpha, \beta)}.
\end{IEEEeqnarray}
Here the support of the beta density function is truncated, i.e., $\theta \in [\theta_l, \theta_u]$, $\theta_l < \theta_u$ and $B_{\theta_l}(\alpha, \beta)$ and $B_{\theta_u}(\alpha, \beta)$ are incomplete beta functions defined as
\begin{IEEEeqnarray}{rCl}
    B_{\theta}(\alpha, \beta) & = & \int_{0}^{\theta} \lambda^{\alpha-1} (1 - \lambda)^{\beta-1} \textrm{d}\lambda.
\end{IEEEeqnarray}
Note that the TS-Beta probability density function is obtained by normalizing the beta probability density function as follows
\begin{IEEEeqnarray}{rCl}
    g(\theta; \alpha, \beta) & = &  \left\{ \,
                            \begin{IEEEeqnarraybox}[][c]{l?s}
                                \IEEEstrut
                                    0 & $,~\theta \notin [\theta_l, \theta_u]$ \\
                                    \frac{f(\theta; \alpha, \beta)}{\eta_\theta} & $,~\theta \in [\theta_l, \theta_u]$
                                \IEEEstrut
                            \end{IEEEeqnarraybox}
                            \right.
\end{IEEEeqnarray}
where
\begin{IEEEeqnarray}{rCl}
    \eta_{\theta} = \Pr(\theta_l \leq \theta \leq \theta_u) & = & \frac{B_{\theta_u}(\alpha, \beta) - B_{\theta_l}(\alpha, \beta)}{B(\alpha, \beta)}.
    \label{eqn:etatheta_def}
\end{IEEEeqnarray}
\end{IEEEdefinition}
\begin{IEEEdefinition}
A TS-BBM random variable $Z$ is defined by the probability distribution
\ifCLASSOPTIONonecolumn
    \begin{IEEEeqnarray}{rCl}
        \Pr(Z = z) & = & {n \choose z} \Bigg(\frac{B_{\theta_u}(\alpha+z, \beta+n-z) - B_{\theta_l}(\alpha+z, \beta+n-z)}{B_{\theta_u}(\alpha, \beta) - B_{\theta_l}(\alpha, \beta)} \Bigg).
    \end{IEEEeqnarray}
\else
    \begin{IEEEeqnarray}{rCl}
        \Pr(Z = z) & = & {n \choose z} \Bigg(\frac{B_{\theta_u}(\alpha+z, \beta+n-z)}{B_{\theta_u}(\alpha, \beta) - B_{\theta_l}(\alpha, \beta)} - \nonumber \\
        & & \frac{B_{\theta_l}(\alpha+z, \beta+n-z)}{B_{\theta_u}(\alpha, \beta) - B_{\theta_l}(\alpha, \beta)} \Bigg).
    \end{IEEEeqnarray}
\fi
Here $z \in \{0, 1, \ldots, n\}$ where $n$ is the number of trials and $\theta_u$, $\theta_l$ are the upper and lower limits of the success probability distributed as a TS-Beta random variable.
\label{defn:ts_bbm_dist}
\end{IEEEdefinition}
Before we compute the mean and variance of a TS-BBM random variable, we define the functions $\delta_\theta$, $\phi_\theta$ in terms of the beta density function, which will be useful in simplifying the notation and interpretation of the results:
\ifCLASSOPTIONonecolumn
    \begin{IEEEeqnarray}{rCl}
        \delta_\theta & = & \rho(f(\theta_u; \alpha+1, \beta+1) - f(\theta_l; \alpha+1, \beta+1)) \label{eqn:deltatheta_def}\\
        \phi_\theta & = & \rho(\theta_uf(\theta_u; \alpha+1, \beta+1) - \theta_lf(\theta_l; \alpha+1, \beta+1)) \\
        \rho & = & \frac{\alpha\beta}{(\alpha+\beta)(\alpha+\beta+1)}. \label{eqn:phitheta_def}
    \end{IEEEeqnarray}
\else
    \begin{IEEEeqnarray}{C}
        \delta_\theta = \rho(f(\theta_u; \alpha+1, \beta+1) - f(\theta_l; \alpha+1, \beta+1)) \label{eqn:deltatheta_def}\\
        \hspace*{-0.4em}\phi_\theta = \rho(\theta_uf(\theta_u; \alpha+1, \beta+1) - \theta_lf(\theta_l; \alpha+1, \beta+1)) \\
        \rho = \frac{\alpha\beta}{(\alpha+\beta)(\alpha+\beta+1)}. \label{eqn:phitheta_def}
    \end{IEEEeqnarray}
\fi
\begin{IEEEprop}
    The mean and variance of a TS-BBM random variable $Z$ are given by
    \ifCLASSOPTIONonecolumn
        \begin{IEEEeqnarray}{rCl}
        \E[Z] & = &  n\Bigg( \frac{B_{\theta_u}(\alpha+1, \beta) - B_{\theta_l}(\alpha+1, \beta)}{B_{\theta_u}(\alpha, \beta) - B_{\theta_l}(\alpha, \beta)}  \Bigg) \\
        & = & \E[\tilde{Z}] - \left(\frac{n}{\alpha+\beta}\right)\frac{\delta_\theta}{\eta_\theta} \\
        \Var[Z] & = & n \Bigg( \frac{B_{\theta_u}(\alpha+1, \beta) - B_{\theta_l}(\alpha+1, \beta)}{B_{\theta_u}(\alpha, \beta) - B_{\theta_l}(\alpha, \beta)} \Bigg) \Bigg(1 - n \Bigg( \frac{B_{\theta_u}(\alpha+1, \beta) - B_{\theta_l}(\alpha+1, \beta)}{B_{\theta_u}( \alpha, \beta) - B_{\theta_l}(\alpha, \beta)}\Bigg)\Bigg) + \nonumber \\
        & & n(n-1)\Bigg( \frac{B_{\theta_u}(\alpha+2, \beta) - B_{\theta_l}(\alpha+2, \beta)}{B_{\theta_u}(\alpha, \beta) - B_{\theta_l}(\alpha, \beta)} \Bigg) \\
        & = & \Var[\tilde{Z}] - \left(\frac{n\beta(\alpha + \beta + n)}{(\alpha + \beta)^2(\alpha + \beta + 1)}\right)\frac{\delta_\theta}{\eta_\theta} - \frac{n^2}{(\alpha+\beta)^2}\left(\frac{\delta_\theta}{\eta_\theta}\right)^2 - \left(\frac{n(n-1)}{\alpha+\beta+1}\right) \frac{\phi_\theta}{\eta_\theta}
        \end{IEEEeqnarray}
    \else
        \begin{IEEEeqnarray}{rCl}
        \E[Z] & = &  n\Bigg( \frac{B_{\theta_u}(\alpha+1, \beta) - B_{\theta_l}(\alpha+1, \beta)}{B_{\theta_u}(\alpha, \beta) - B_{\theta_l}(\alpha, \beta)}  \Bigg) \\
        & = & \E[Z'] - \left(\frac{n}{\alpha+\beta}\right)\frac{\delta_\theta}{\eta_\theta} \\
        \Var[Z] & = & n \Bigg( \frac{B_{\theta_u}(\alpha+1, \beta) - B_{\theta_l}(\alpha+1, \beta)}{B_{\theta_u}(\alpha, \beta) - B_{\theta_l}(\alpha, \beta)} \Bigg) \nonumber \\
        & & \Bigg(1 - n \Bigg( \frac{B_{\theta_u}(\alpha+1, \beta) - B_{\theta_l}(\alpha+1, \beta)}{B_{\theta_u}( \alpha, \beta) - B_{\theta_l}(\alpha, \beta)}\Bigg)\Bigg) + \nonumber \\
        & & n(n-1)\Bigg( \frac{B_{\theta_u}(\alpha+2, \beta) - B_{\theta_l}(\alpha+2, \beta)}{B_{\theta_u}(\alpha, \beta) - B_{\theta_l}(\alpha, \beta)} \Bigg) \\
        & = & \Var[Z'] - \left(\frac{n\beta(\alpha + \beta + n)}{(\alpha + \beta)^2(\alpha + \beta + 1)}\right)\frac{\delta_\theta}{\eta_\theta} \nonumber \\
        & & -  \frac{n^2}{(\alpha+\beta)^2}\left(\frac{\delta_\theta}{\eta_\theta}\right)^2 - \left(\frac{n(n-1)}{\alpha+\beta+1}\right) \frac{\phi_\theta}{\eta_\theta}
        \end{IEEEeqnarray}
    \fi
    where $\tilde{Z}$ is a beta-binomial random variable with parameters $(n, \alpha, \beta)$.
    \label{prop:ts_bbm_mean_var}
\end{IEEEprop}
\begin{IEEEproof}
    See Appendix~\ref{app:proof_ts_bbm_mean_var}.
\end{IEEEproof}
To define the \mbox{2-TS-BBM} channel model for MLC flash memories, we adopt the definitions and notation from \cite{Taranalli_2016} which were reviewed in Section~\ref{sec:capacity_bbm_channel_model}. The $\ZOerror$ and $\OZerror$ bit error counts for a given input frame containing $m$ zeros are denoted by $\kcount{m}{0}$ and $\kcount{N-m}{1}$, respectively. Similarly, $K^{(0)}$ and $K^{(1)}$ denote the total number of $\ZOerror$ and $\OZerror$
bit errors, respectively. The variables $\kcount{m}{0}$ and $\kcount{N-m}{1}$ are modeled as being distributed according to the TS-BBM distribution i.e.,
\begin{IEEEeqnarray}{rCl}
    p & \sim & \textrm{TS-Beta}(p_l, p_u; a, b) \nonumber \\
    \kcount{m}{0}~|~p & \sim & \textrm{Binomial}(m, p) \nonumber \\
    \kcount{m}{0} & \sim & \textrm{TS-BBM}(p_l, p_u; m, a, b); \\
    q & \sim & \textrm{TS-Beta}(q_l, q_u; c, d)  \nonumber \\
    \kcount{N-m}{1}~|~q & \sim & \textrm{Binomial}(N-m, q)  \nonumber \\
    \kcount{N-m}{1} & \sim & \textrm{TS-BBM}(q_l, q_u; N-m, c, d); \\
    \kcount{m}{0} & \indep & \kcount{N-m}{1} \label{eqn:ts_bbm_xy_indep}.
\end{IEEEeqnarray}
Before we derive expressions for the mean and variance of the number of bit errors per frame resulting from a \mbox{TS-BBM} channel model, we introduce some simplifying notation as follows. Let
\begin{IEEEeqnarray}{rCl}
    U_{p_j}^{(i)} & = & B_{p_j}(a+i, b)  \\
    V_{q_j}^{(i)} & = & B_{q_j}(c+i, d)
\end{IEEEeqnarray}
where $j \in \{l, u\}$ and $i \in \{0, 1, 2\}$.
\begin{IEEEprop}
    The mean and variance of $K^{(0)}$ and $K^{(1)}$ for a \mbox{TS-BBM} channel model are given by
    \ifCLASSOPTIONonecolumn
        \begin{IEEEeqnarray}{rCl}
            \E[K^{(0)}] & = & \frac{N}{2} \Bigg(\frac{U_{p_u}^{(1)} - U_{p_l}^{(1)}}{U_{p_u}^{(0)} - U_{p_l}^{(0)}}\Bigg) \\
            \Var[K^{(0)}] & = & \frac{N}{2}\Bigg(\frac{U_{p_u}^{(1)} - U_{p_l}^{(1)}}{U_{p_u}^{(0)} - U_{p_l}^{(0)}}\Bigg) \Bigg(1 - \frac{N}{2}\Bigg(\frac{U_{p_u}^{(1)} - U_{p_l}^{(1)}}{U_{p_u}^{(0)} - U_{p_l}^{(0)}}\Bigg)\Bigg) + \frac{N(N-1)}{4} \Bigg(\frac{U_{p_u}^{(2)} - U_{p_l}^{(2)}}{U_{p_u}^{(0)} - U_{p_l}^{(0)}}\Bigg); \\
            \E[K^{(1)}] & = & \frac{N}{2} \Bigg(\frac{V_{q_u}^{(1)} - V_{q_l}^{(1)}}{V_{q_u}^{(0)} - V_{q_l}^{(0)}}\Bigg)\\
            \Var[K^{(1)}] & = & \frac{N}{2}\Bigg(\frac{V_{q_u}^{(1)} - V_{q_l}^{(1)}}{V_{q_u}^{(0)} - V_{q_l}^{(0)}}\Bigg) \Bigg(1 - \frac{N}{2}\Bigg(\frac{V_{q_u}^{(1)} - V_{q_l}^{(1)}}{V_{q_u}^{(0)} - V_{q_l}^{(0)}}\Bigg)\Bigg) + \frac{N(N-1)}{4} \Bigg(\frac{V_{q_u}^{(2)} - V_{q_l}^{(2)}}{V_{q_u}^{(0)} - V_{q_l}^{(0)}}\Bigg).
        \end{IEEEeqnarray}
    \else
        \begin{IEEEeqnarray}{rCl}
            \E[K^{(0)}] & = & \frac{N}{2} \Bigg(\frac{U_{p_u}^{(1)} - U_{p_l}^{(1)}}{U_{p_u}^{(0)} - U_{p_l}^{(0)}}\Bigg) \\
            \Var[K^{(0)}] & = & \frac{N}{2}\Bigg(\frac{U_{p_u}^{(1)} - U_{p_l}^{(1)}}{U_{p_u}^{(0)} - U_{p_l}^{(0)}}\Bigg) \Bigg(1 - \frac{N}{2}\Bigg(\frac{U_{p_u}^{(1)} - U_{p_l}^{(1)}}{U_{p_u}^{(0)} - U_{p_l}^{(0)}}\Bigg)\Bigg) \nonumber \\ & & + \frac{N(N-1)}{4} \Bigg(\frac{U_{p_u}^{(2)} - U_{p_l}^{(2)}}{U_{p_u}^{(0)} - U_{p_l}^{(0)}}\Bigg); \\
            \E[K^{(1)}] & = & \frac{N}{2} \Bigg(\frac{V_{q_u}^{(1)} - V_{q_l}^{(1)}}{V_{q_u}^{(0)} - V_{q_l}^{(0)}}\Bigg)\\
            \Var[K^{(1)}] & = & \frac{N}{2}\Bigg(\frac{V_{q_u}^{(1)} - V_{q_l}^{(1)}}{V_{q_u}^{(0)} - V_{q_l}^{(0)}}\Bigg) \Bigg(1 - \frac{N}{2}\Bigg(\frac{V_{q_u}^{(1)} - V_{q_l}^{(1)}}{V_{q_u}^{(0)} - V_{q_l}^{(0)}}\Bigg)\Bigg) \nonumber \\ & & + \frac{N(N-1)}{4} \Bigg(\frac{V_{q_u}^{(2)} - V_{q_l}^{(2)}}{V_{q_u}^{(0)} - V_{q_l}^{(0)}}\Bigg).
        \end{IEEEeqnarray}
    \fi
    \label{prop:mean_var_k0_k1_ts_bbm_model}
\end{IEEEprop}
\begin{IEEEproof}
    See Appendix~\ref{app:proof_mean_var_k0_k1_ts_bbm_model}.
\end{IEEEproof}
\begin{IEEEprop}
	The mean and the variance of $K$ for a \mbox{TS-BBM} channel model are given by
    \ifCLASSOPTIONonecolumn
        \begin{IEEEeqnarray}{rCl}
            \E[K] & = & \frac{N}{2}\Bigg(\frac{U_{p_u}^{(1)} - U_{p_l}^{(1)}}{U_{p_u}^{(0)} - U_{p_l}^{(0)}} + \frac{V_{q_u}^{(1)} - V_{q_l}^{(1)}}{V_{q_u}^{(0)} - V_{q_l}^{(0)}} \Bigg) \\
            \Var[K] & = & \frac{N}{2}\Bigg(\frac{U_{p_u}^{(1)} - U_{p_l}^{(1)}}{U_{p_u}^{(0)} - U_{p_l}^{(0)}}\Bigg) \Bigg(1 - \frac{N}{2}\Bigg(\frac{U_{p_u}^{(1)} - U_{p_l}^{(1)}}{U_{p_u}^{(0)} - U_{p_l}^{(0)}}\Bigg)\Bigg) \nonumber \\
            & & + \frac{N}{2}\Bigg(\frac{V_{q_u}^{(1)} - V_{q_l}^{(1)}}{V_{q_u}^{(0)} - V_{q_l}^{(0)}}\Bigg) \Bigg(1 - \frac{N}{2}\Bigg(\frac{V_{q_u}^{(1)} - V_{q_l}^{(1)}}{V_{q_u}^{(0)} - V_{q_l}^{(0)}}\Bigg)\Bigg) \nonumber \\
            & & + \frac{N(N-1)}{4} \Bigg(\frac{U_{p_u}^{(2)} - U_{p_l}^{(2)}}{U_{p_u}^{(0)} - U_{p_l}^{(0)}} + \frac{V_{q_u}^{(2)} - V_{q_l}^{(2)}}{V_{q_u}^{(0)} - V_{q_l}^{(0)}} \Bigg) \nonumber \\
            & & - \frac{N}{2} \Bigg(\frac{U_{p_u}^{(1)} - U_{p_l}^{(1)}}{U_{p_u}^{(0)} - U_{p_l}^{(0)}}\Bigg)\Bigg(\frac{V_{q_u}^{(1)} - V_{q_l}^{(1)}}{V_{q_u}^{(0)} - V_{q_l}^{(0)}} \Bigg).
        \end{IEEEeqnarray}
    \else
        \begin{IEEEeqnarray}{rCl}
            \E[K] & = & \frac{N}{2}\Bigg(\frac{U_{p_u}^{(1)} - U_{p_l}^{(1)}}{U_{p_u}^{(0)} - U_{p_l}^{(0)}} + \frac{V_{q_u}^{(1)} - V_{q_l}^{(1)}}{V_{q_u}^{(0)} - V_{q_l}^{(0)}} \Bigg) \\
            \Var[K] & = & \frac{N}{2}\Bigg(\frac{U_{p_u}^{(1)} - U_{p_l}^{(1)}}{U_{p_u}^{(0)} - U_{p_l}^{(0)}}\Bigg) \Bigg(1 - \frac{N}{2}\Bigg(\frac{U_{p_u}^{(1)} - U_{p_l}^{(1)}}{U_{p_u}^{(0)} - U_{p_l}^{(0)}}\Bigg)\Bigg) \nonumber \\
            & & + \frac{N}{2}\Bigg(\frac{V_{q_u}^{(1)} - V_{q_l}^{(1)}}{V_{q_u}^{(0)} - V_{q_l}^{(0)}}\Bigg) \Bigg(1 - \frac{N}{2}\Bigg(\frac{V_{q_u}^{(1)} - V_{q_l}^{(1)}}{V_{q_u}^{(0)} - V_{q_l}^{(0)}}\Bigg)\Bigg) \nonumber \\
            & & + \frac{N(N-1)}{4} \Bigg(\frac{U_{p_u}^{(2)} - U_{p_l}^{(2)}}{U_{p_u}^{(0)} - U_{p_l}^{(0)}} + \frac{V_{q_u}^{(2)} - V_{q_l}^{(2)}}{V_{q_u}^{(0)} - V_{q_l}^{(0)}} \Bigg) \nonumber \\
            & & - \frac{N}{2} \Bigg(\frac{U_{p_u}^{(1)} - U_{p_l}^{(1)}}{U_{p_u}^{(0)} - U_{p_l}^{(0)}}\Bigg)\Bigg(\frac{V_{q_u}^{(1)} - V_{q_l}^{(1)}}{V_{q_u}^{(0)} - V_{q_l}^{(0)}} \Bigg).
        \end{IEEEeqnarray}
    \fi
    \label{prop:mean_var_ts_bbm_model}
\end{IEEEprop}
\begin{IEEEproof}
    See Appendix~\ref{app:proof_mean_var_ts_bbm_model}.
\end{IEEEproof}
\begin{IEEEprop}
    The mean and variance of $K^{(0)}$, $K^{(1)}$, $K$ corresponding to a \mbox{TS-BBM} channel model can be expressed in terms of the mean and variance of $\tilde{K}^{(0)}$, $\tilde{K}^{(1)}$, $\tilde{K}$ corresponding to a \mbox{BBM} channel model with the same parameters, as
    \begin{IEEEeqnarray}{rCl}
        \E[K^{(0)}] & = & \E[\tilde{K}^{(0)}] - \Delta^{(0)}_{mean} \\
        \Var[K^{(0)}] & = & \Var[\tilde{K}^{(0)}] - \Delta^{(0)}_{var}; \\
        \E[K^{(1)}] & = & \E[\tilde{K}^{(1)}] - \Delta^{(1)}_{mean} \\
        \Var[K^{(1)}] & = & \Var[\tilde{K}^{(1)}] - \Delta^{(1)}_{var}; \\
        \E[K] & = & \E[\tilde{K}] - \Delta_{mean} \\
        \Var[K] & = & \Var[\tilde{K}] - \Delta_{var}
    \end{IEEEeqnarray}
    where
    \ifCLASSOPTIONonecolumn
        \begin{IEEEeqnarray}{rCl}
            \Delta^{(0)}_{mean} & = & \frac{N}{2} \frac{1}{(a+b)} \frac{\delta_p}{\eta_p} \\
            \Delta^{(0)}_{var} & = & \frac{N}{4} \Bigg(w_1\frac{\delta_p}{\eta_p} + w_2\frac{\phi_p}{\eta_p} + w_3 \left(\frac{\delta_p}{\eta_p}\right)^2 \Bigg); \\
            \Delta^{(1)}_{mean} & = & \frac{N}{2} \frac{1}{(c+d)} \frac{\delta_q}{\eta_q} \\
            \Delta^{(1)}_{var} & = & \frac{N}{4} \Bigg(w_4\frac{\delta_q}{\eta_q} + w_5\frac{\phi_q}{\eta_q} + w_6 \left(\frac{\delta_q}{\eta_q}\right)^2 \Bigg); \\
            \Delta_{mean} & = &  \frac{N}{2}\left(\frac{1}{(a+b)} \frac{\delta_p}{\eta_p} + \frac{1}{(c+d)}  \frac{\delta_q}{\eta_q} \right) \\
            \Delta_{var} & = & \frac{N}{4} \Bigg(w_7 \frac{\delta_p}{\eta_p} + w_8 \frac{\delta_q}{\eta_q} + w_2 \frac{\phi_p}{\eta_p} + w_5 \frac{\phi_q}{\eta_q} + w_3 \left(\frac{\delta_p}{\eta_p}\right)^2 + w_6 \left(\frac{\delta_q}{\eta_q}\right)^2 + w_9\frac{\delta_p}{\eta_p}\frac{\delta_q}{\eta_q}\Bigg),
        \end{IEEEeqnarray}
        \begin{IEEEeqnarray}{CCC}
            w_1 = \frac{(a+b)(a+2b+1) + N(b - a(a+b+1))}{(a+b)^2 (a+b+1)} \\
            w_2 = \frac{N-1}{a+b+1} ~~~~~ w_3 = \frac{N}{(a+b)^2} \\
            w_4 = \frac{(c+d)(c+2d+1) + N(d - c(c+d+1))}{(c+d)^2 (c+d+1)} \\
            w_5 = \frac{N-1}{c+d+1} ~~~~~ w_6 = \frac{N}{(c+d)^2} \\
            w_7 = \frac{N(b - a(a+b+1))}{(a+b)^2(a+b+1)} + \frac{2d}{(a+b)(c+d)} \\
            w_8 = \frac{N(d - c(c+d+1))}{(c+d)^2(c+d+1)} + \frac{2b}{(a+b)(c+d)} \\
            w_9 = \frac{2}{(a+b)(c+d)}.
        \end{IEEEeqnarray}
    \else
        \begin{IEEEeqnarray}{rCl}
            \Delta^{(0)}_{mean} & = & \frac{N}{2} \frac{1}{(a+b)} \frac{\delta_p}{\eta_p} \\
            \Delta^{(0)}_{var} & = & \frac{N}{4} \Bigg(w_1\frac{\delta_p}{\eta_p} + w_2\frac{\phi_p}{\eta_p} + w_3 \left(\frac{\delta_p}{\eta_p}\right)^2 \Bigg); \\
            \Delta^{(1)}_{mean} & = & \frac{N}{2} \frac{1}{(c+d)} \frac{\delta_q}{\eta_q} \\
            \Delta^{(1)}_{var} & = & \frac{N}{4} \Bigg(w_4\frac{\delta_q}{\eta_q} + w_5\frac{\phi_q}{\eta_q} + w_6 \left(\frac{\delta_q}{\eta_q}\right)^2 \Bigg); \\
            \Delta_{mean} & = &  \frac{N}{2}\left(\frac{1}{(a+b)} \frac{\delta_p}{\eta_p} + \frac{1}{(c+d)}  \frac{\delta_q}{\eta_q} \right) \\
            \Delta_{var} & = & \frac{N}{4} \Bigg(w_7 \frac{\delta_p}{\eta_p} + w_8 \frac{\delta_q}{\eta_q} + w_2 \frac{\phi_p}{\eta_p} + w_5 \frac{\phi_q}{\eta_q} \nonumber \\
            & & + w_3 \left(\frac{\delta_p}{\eta_p}\right)^2 + w_6 \left(\frac{\delta_q}{\eta_q}\right)^2 + w_9\frac{\delta_p}{\eta_p}\frac{\delta_q}{\eta_q}\Bigg),
        \end{IEEEeqnarray}
        \begin{IEEEeqnarray}{CCC}
            w_1 = \frac{(a+b)(a+2b+1) + N(b - a(a+b+1))}{(a+b)^2 (a+b+1)} \\
            w_2 = \frac{N-1}{a+b+1} ~~~~~ w_3 = \frac{N}{(a+b)^2} \\
            w_4 = \frac{(c+d)(c+2d+1) + N(d - c(c+d+1))}{(c+d)^2 (c+d+1)} \\
            w_5 = \frac{N-1}{c+d+1} ~~~~~ w_6 = \frac{N}{(c+d)^2} \\
            w_7 = \frac{N(b - a(a+b+1))}{(a+b)^2(a+b+1)} + \frac{2d}{(a+b)(c+d)} \\
            w_8 = \frac{N(d - c(c+d+1))}{(c+d)^2(c+d+1)} + \frac{2b}{(a+b)(c+d)} \\
            w_9 = \frac{2}{(a+b)(c+d)}.
        \end{IEEEeqnarray}
    \fi
    \label{prop:mean_var_k0_k1_k_ts_bbm_model_bbm_model}
\end{IEEEprop}
\begin{IEEEproof}
    See Appendix~\ref{app:proof_mean_var_k0_k1_k_ts_bbm_model_bbm_model}.
\end{IEEEproof}
\ifCLASSOPTIONonecolumn
    \begin{figure}
        \centering
        \includegraphics[width=0.7\textwidth]{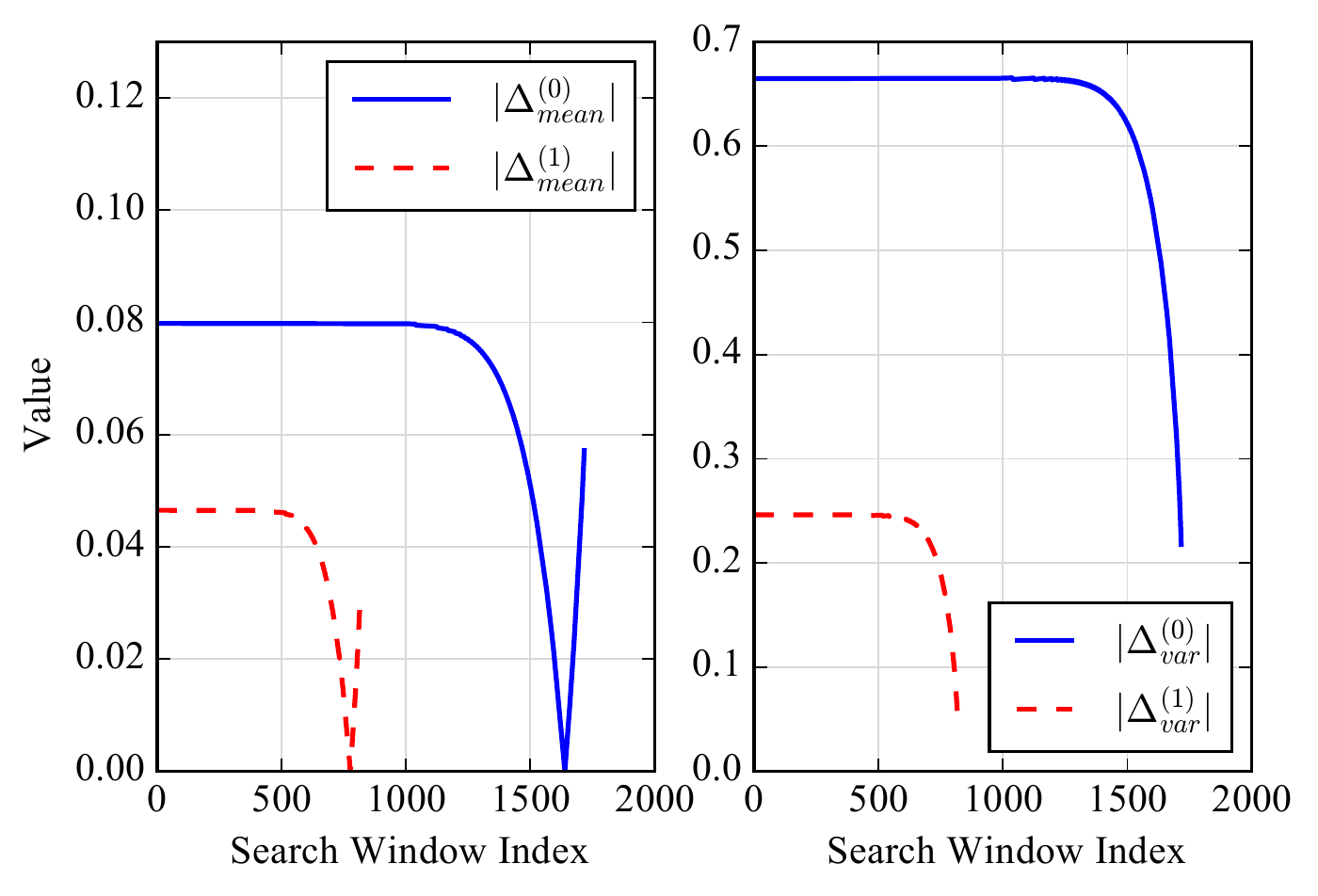}
        \caption{Plot showing $|\Delta^{(x)}_{mean}|$ and $|\Delta^{(x)}_{var}|$ obtained using Algorithm~\ref{algo:opt_truncation_interval}, corresponding to the beta distributions at 6,000 P/E cycles for vendor-A chip. $N = 8192$, $\mu = 10^{-6}$, $\epsilon = 0.01$.}
        \label{fig:delta_mean_var}
    \end{figure}
\else
    \begin{figure}
    	\centering
    	\includegraphics[width=0.47\textwidth]{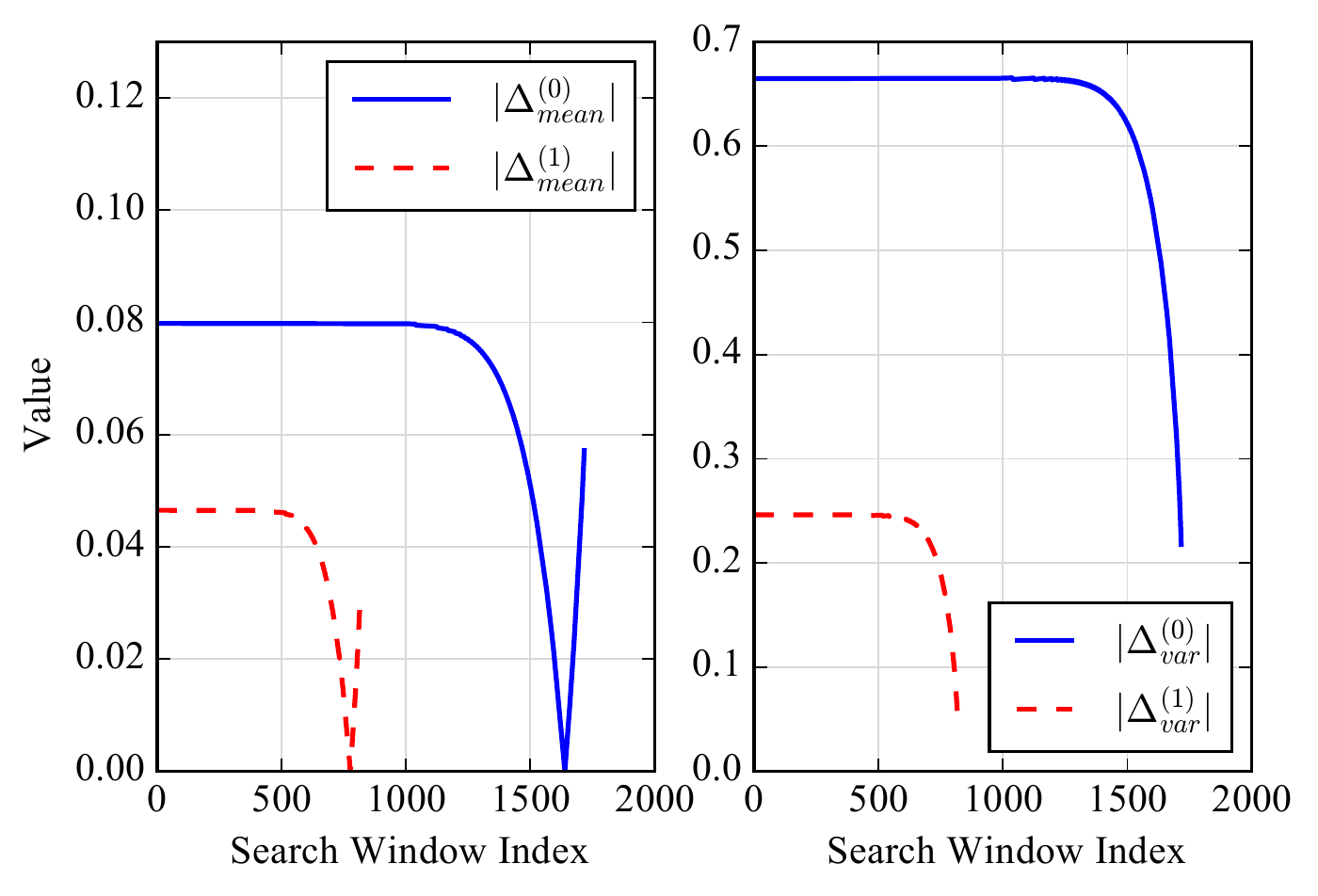}
    	\caption{Plot showing $|\Delta^{(x)}_{mean}|$ and $|\Delta^{(x)}_{var}|$ obtained using Algorithm~\ref{algo:opt_truncation_interval}, corresponding to the beta distributions at 6,000 P/E cycles for vendor-A chip. $N = 8192$, $\mu = 10^{-6}$, $\epsilon = 0.01$.}
    	\label{fig:delta_mean_var}
    \end{figure}
\fi
\subsection{Choosing the Truncation Intervals}
As shown in Proposition~\ref{prop:mean_var_k0_k1_k_ts_bbm_model_bbm_model}, the mean and variance statistics of the number of bit errors per frame
for the \mbox{TS-BBM} channel model differ from those of the BBM channel model and this difference depends on the functions $\eta_p$, $\eta_q$, $\delta_p$, $\delta_q$,
$\phi_p$, $\phi_q$ of the truncation points $(p_l, p_u)$, $(q_l, q_u)$ of the \mbox{TS-BBM} channel model. Recall that $\eta_p$, $\eta_q$ represent the area under the beta pdf curve (probability) between the chosen truncation points. For choosing the truncation intervals of a \mbox{TS-BBM} channel model, we let $\eta_p$, $\eta_q$ be equal to some large probability value, e.g., $\eta_p = 1-\epsilon$ and $\eta_q = 1-\epsilon$ where $\epsilon \ll 1$. Subject to the constraints on $\eta_p$, $\eta_q$, the choice of the truncation intervals should be such that the respective differences between the mean and the variance of the number of bit errors per frame for the TS-BBM and the BBM channel models are minimized; i.e., $|\Delta^{(0)}_{mean}|$, $|\Delta^{(1)}_{mean}|$ and $|\Delta^{(0)}_{var}|$, $|\Delta^{(1)}_{var}|$ are minimized. Ideally, we would want to solve the following optimization problems to determine the optimal truncation intervals for the TS-BBM channel model:
\ifCLASSOPTIONonecolumn
    \begin{IEEEeqnarray}{rC}
    \textrm{(P1)} & ~\textrm{Choose points~} p^{\ast}_l, p^{\ast}_u \textrm{~s.t.}~\eta_p = 1-\epsilon~\textrm{and}~|\Delta^{(0)}_{mean}|~\textrm{and}~|\Delta^{(0)}_{var}|~\textrm{are minimized.}
    \end{IEEEeqnarray}
\else
    \begin{IEEEeqnarray}{rC}
    \textrm{(P1)} & ~\textrm{Choose points~} p^{\ast}_l, p^{\ast}_u \textrm{~s.t.}~\eta_p = 1-\epsilon~\textrm{and} \nonumber \\
    & |\Delta^{(0)}_{mean}|~\textrm{and}~|\Delta^{(0)}_{var}|~\textrm{are minimized.}
    \end{IEEEeqnarray}
\fi
\ifCLASSOPTIONonecolumn
    \begin{IEEEeqnarray}{rC}
    \textrm{(P2)} & ~\textrm{Choose points~} q^{\ast}_l, q^{\ast}_u \textrm{~s.t.}~\eta_q = 1-\epsilon~\textrm{and}~|\Delta^{(1)}_{mean}|~\textrm{and}~|\Delta^{(1)}_{var}|~\textrm{are minimized.}
    \end{IEEEeqnarray}
\else
    \begin{IEEEeqnarray}{rC}
    \textrm{(P2)} & ~\textrm{Choose points~} q^{\ast}_l, q^{\ast}_u \textrm{~s.t.}~\eta_q = 1-\epsilon~\textrm{and} \nonumber \\
     & |\Delta^{(1)}_{mean}|~\textrm{and}~|\Delta^{(1)}_{var}|~\textrm{are minimized.}
    \end{IEEEeqnarray}
\fi
However there are some major issues we encounter when trying to solve (P1) and (P2) directly. The first issue is that the constraints in the above optimization problems involve incomplete beta functions (expressions for $\eta_p$ and $\eta_q$) which do not have a closed form and the constraints are nonlinear. The second issue is that it is not known if $|\Delta^{(0)}_{mean}|$ and $|\Delta^{(0)}_{var}|$ (and similarly $|\Delta^{(1)}_{mean}|$ and $|\Delta^{(1)}_{var}|$) can be minimized simultaneously subject to the given constraints; i.e., it is not known if feasible solutions exist for problems (P1) and (P2).

Hence we propose a numerical discrete search algorithm to find the truncation intervals that minimize $|\Delta^{(0)}_{mean}|$ (resp., $|\Delta^{(1)}_{mean}|$) and $|\Delta^{(0)}_{var}|$ (resp., $|\Delta^{(1)}_{var}|$) separately, subject to the relaxed constraints $1 - \epsilon \leq \eta_p < 1$ and $1 - \epsilon \leq \eta_q < 1$. The justification for the relaxed constraints is that, in the context of deriving a \mbox{TS-BBM} channel model from a BBM channel model, the specific values of $\eta_p$ and $\eta_q$
are not very important in practice as long as they are close to 1, which ensures that the \mbox{TS-BBM} channel model preserves the variability (randomness) of the original BBM channel model as much as possible.
The numerical discrete search algorithm is described by Algorithms~\ref{algo:opt_truncation_interval} and \ref{algo:search_candidates}. The search for the truncation interval is done by dividing the continuous interval $[0, 1]$ into a set of discrete points with resolution $\mu$. Using pairs of these points as endpoints, windows that satisfy the relaxed constraints are identified. The algorithm then computes $|\Delta^{(x)}_{mean}|$ or $|\Delta^{(x)}_{var}|$, and picks the window with a minimum value as the truncation interval. Fig.~\ref{fig:delta_mean_var} shows a plot of the $|\Delta^{(x)}_{mean}|$ and $|\Delta^{(x)}_{var}|$ obtained using the numerical discrete search algorithm. We note that it appears that an absolute minimum exists for $|\Delta^{(x)}_{mean}|$ functions whereas for $|\Delta^{(x)}_{var}|$ functions, it does not appear to be so. However choosing a small $\mu$ ensures that the algorithm provides accurate results for practical purposes, as shown in Table~\ref{table:mean_var_table_vendor_a}.
%
The complexity of Algorithm~\ref{algo:opt_truncation_interval} is $O(l\log{}l)$ where $l = 1/\mu$.
\begin{algorithm}[!ht]
	\caption{Get Truncation Interval}
	\label{algo:opt_truncation_interval}
	\algsetup{
		linenosize=\small
	}
	\begin{algorithmic}[1]
        \REQUIRE $N$, $\alpha$, $\beta$, $\epsilon$, $\mu$, optParam
        \ENSURE Truncation interval points $\hat{\theta}_l$ and $\hat{\theta}_u$
        \STATE cIntervals = \textbf{searchCandidates}($\alpha$, $\beta$, $\epsilon$, $\mu$)
        \STATE Initialize $\bar{\Delta}$ as an empty vector.
        \STATE Initialize $\textrm{intervalCount} = 0$.
        \FOR{$\theta_l, \theta_u \in \textrm{cIntervals}$}
            \STATE Compute $\eta_\theta$ using (\ref{eqn:etatheta_def}).
            \STATE Compute $\delta_\theta$ using (\ref{eqn:deltatheta_def}) and $\phi_\theta$ using (\ref{eqn:phitheta_def}).
            \IF{optParam == ``mean''}
                \STATE $\bar{\Delta}[\textrm{intervalCount}] = |\Delta^{(x)}_{mean}|$
            \ELSE
                \STATE $\bar{\Delta}[\textrm{intervalCount}] = |\Delta^{(x)}_{var}|$
            \ENDIF
            \STATE $\textrm{intervalCount} = \textrm{intervalCount} + 1$
        \ENDFOR
        \STATE $\textrm{minimumIndex} = \argmin \bar{\Delta}$
        \RETURN $\hat{\theta}_l$, $\hat{\theta}_u = \textrm{cIntervals[minimumIndex]}$
    \end{algorithmic}
\end{algorithm}
\begin{algorithm}[!ht]
	\caption{searchCandidates($\alpha$, $\beta$, $\epsilon$, $\mu$)}
	\label{algo:search_candidates}
	\algsetup{
		linenosize=\small
	}
	\begin{algorithmic}[1]
        \REQUIRE $\alpha$, $\beta$, $\epsilon$, $\mu$
        \ENSURE List of candidate truncation interval pairs $[\theta_l, \theta_u]$.
        \STATE Divide the unit interval $[0, 1]$ into $l$ equal intervals where $l~=~1/\mu$ and store the points in a list candidatePoints.
        \STATE Compute cumulative distribution function $F$ of $\textrm{Beta}(\alpha, \beta)$ at every point in candidatePoints.
        \STATE Initialize candidateList as an empty list.
        \FOR{$\textrm{startPoint} \in \textrm{candidatePoints}$}
            \STATE Using binary search, find the smallest interval \mbox{[startPoint, endPoint]} such that \\ $F(\textrm{endPoint}) - F(\textrm{startPoint}) \geq 1 - \epsilon$.
            \STATE Add [startPoint, endPoint] to candidateList.
        \ENDFOR
        \RETURN candidateList
    \end{algorithmic}
\end{algorithm}
\subsection{Results}
\begin{table}[!ht]
    \caption{Truncation intervals for the \mbox{TS-BBM} channel models obtained using Algorithm~\ref{algo:opt_truncation_interval} using $N = 8192$, $\mu = 10^{-6}$, $\epsilon = 0.01$ for vendor-A chip. All truncation interval points should be multiplied by~$10^{-3}$.}
	\label{table:truncation_points_vendor_a}
	\centering
	\bgroup
	\ifCLASSOPTIONonecolumn
    	\begin{tabular}{|c|c c |c c|c c|c c|}
    		\hline
            \textbf{P/E} & \multicolumn{8}{|c|}{\textbf{Vendor-A, Upper Page}} \\
    		\hhline{~--------}
            \textbf{Cycles} & \multicolumn{4}{c|}{min $|\Delta^{(x)}_{mean}|$} & \multicolumn{4}{c|}{min $|\Delta^{(x)}_{var}|$} \\
            \hline
            & \multicolumn{2}{c|}{$x = 0$} & \multicolumn{2}{c|}{$x = 1$} & \multicolumn{2}{c|}{$x = 0$} & \multicolumn{2}{c|}{$x = 1$} \\
            \hline
    		& $p_l$ & $p_u$ & $q_l$ & $q_u$ & $p_l$ & $p_u$ & $q_l$ & $q_u$ \\
    		\hline
    		6000 & 1.64 & 4.89 & 0.78 & 2.64 & 1.71 & 6.16 & 0.82 & 3.36 \\
    		\hline
    		8000 & 2.66 & 8.35 & 1.56 & 4.69 & 2.79 & 11.02 & 1.63 & 6.01 \\
    		\hline
    		10000 & 4.06 & 12.51 & 2.54 & 7.03 & 4.26 & 16.18 & 2.66 & 8.81 \\
    		\hline
             & \multicolumn{8}{|c|}{\textbf{Vendor-A, Lower Page}} \\
            \hline
            6000 & 0.000 & 0.064 & 2.76 & 9.14 & 0.000 & 0.064 & 2.90 & 12.34 \\
    		\hline
    		8000 & 0.001 & 0.083 & 3.94 & 12.56 & 0.001 & 0.083 & 4.13 & 16.79 \\
    		\hline
    		10000 & 0.001 & 0.111 & 5.64 & 16.34 & 0.001 & 0.111 & 5.91 & 21.87\\
    		\hline
    	\end{tabular}
	\else
		\def\arraystretch{1.5}
        \tabcolsep=0.11cm
    	\resizebox{0.9\columnwidth}{!}{
    	\begin{tabular}{|c|c c |c c|c c|c c|}
    		\hline
            \textbf{P/E} & \multicolumn{8}{|c|}{\textbf{Vendor-A, Upper Page}} \\
    		\hhline{~--------}
            \textbf{Cycles} & \multicolumn{4}{c|}{min $|\Delta^{(x)}_{mean}|$} & \multicolumn{4}{c|}{min $|\Delta^{(x)}_{var}|$} \\
            \hline
            & \multicolumn{2}{c|}{$x = 0$} & \multicolumn{2}{c|}{$x = 1$} & \multicolumn{2}{c|}{$x = 0$} & \multicolumn{2}{c|}{$x = 1$} \\
            \hline
    		& $p_l$ & $p_u$ & $q_l$ & $q_u$ & $p_l$ & $p_u$ & $q_l$ & $q_u$ \\
    		\hline
    		6000 & 1.64 & 4.89 & 0.78 & 2.64 & 1.71 & 6.16 & 0.82 & 3.36 \\
    		\hline
    		8000 & 2.66 & 8.35 & 1.56 & 4.69 & 2.79 & 11.02 & 1.63 & 6.01 \\
    		\hline
    		10000 & 4.06 & 12.51 & 2.54 & 7.03 & 4.26 & 16.18 & 2.66 & 8.81 \\
    		\hline
             & \multicolumn{8}{|c|}{\textbf{Vendor-A, Lower Page}} \\
            \hline
            6000 & 0.000 & 0.064 & 2.76 & 9.14 & 0.000 & 0.064 & 2.90 & 12.34 \\
    		\hline
    		8000 & 0.001 & 0.083 & 3.94 & 12.56 & 0.001 & 0.083 & 4.13 & 16.79 \\
    		\hline
    		10000 & 0.001 & 0.111 & 5.64 & 16.34 & 0.001 & 0.111 & 5.91 & 21.87\\
    		\hline
    	\end{tabular}
    	}
	\fi
	\egroup
\end{table}

\begin{table}[!ht]
    \caption{Truncation intervals for the \mbox{TS-BBM} channel models obtained using Algorithm~\ref{algo:opt_truncation_interval} using $N = 8192$, $\mu = 10^{-6}$, $\epsilon = 0.01$ for vendor-B chip. All truncation interval points should be multiplied by~$10^{-3}$.}
	\label{table:truncation_points_vendor_b}
	\centering
	\bgroup
	\ifCLASSOPTIONonecolumn
    	\begin{tabular}{|c|c c |c c|c c|c c|}
    		\hline
            \textbf{P/E} & \multicolumn{8}{|c|}{\textbf{Vendor-B, Upper Page}} \\
    		\hhline{~--------}
            \textbf{Cycles} & \multicolumn{4}{c|}{min $|\Delta^{(x)}_{mean}|$} & \multicolumn{4}{c|}{min $|\Delta^{(x)}_{var}|$} \\
            \hline
            & \multicolumn{2}{c|}{$x = 0$} & \multicolumn{2}{c|}{$x = 1$} & \multicolumn{2}{c|}{$x = 0$} & \multicolumn{2}{c|}{$x = 1$} \\
            \hline
    		& $p_l$ & $p_u$ & $q_l$ & $q_u$ & $p_l$ & $p_u$ & $q_l$ & $q_u$ \\
    		\hline
    		6000 & 0.36 & 1.37 & 0.31 & 2.29 & 0.38 & 1.64 & 0.34 & 3.41 \\
    		\hline
    		8000 & 0.80 & 3.05 & 0.60 & 4.18 & 0.84 & 3.78 & 0.65 & 5.56 \\
    		\hline
    		10000 & 1.44 & 6.05 & 1.17 & 6.63 & 1.53 & 7.83 & 1.25 & 10.84 \\
    		\hline
             & \multicolumn{8}{|c|}{\textbf{Vendor-B, Lower Page}} \\
            \hline
            6000 & 0.000 & 0.004 & 1.70 & 6.63 & 0.000 & 0.004 & 1.80 & 8.68  \\
     		\hline
     		8000 & 0.000 & 0.004 &  3.58 & 13.04 & 0.000 & 0.004 & 3.78 & 18.90 \\
     		\hline
     		10000 & 0.000 & 0.004 & 6.33 & 22.71 & 0.000 & 0.004 & 6.68 & 30.54\\
     		\hline
    	\end{tabular}
	\else
		\def\arraystretch{1.5}
        \tabcolsep=0.11cm
    	\resizebox{0.9\columnwidth}{!}{
        \begin{tabular}{|c|c c |c c|c c|c c|}
    		\hline
            \textbf{P/E} & \multicolumn{8}{|c|}{\textbf{Vendor-B, Upper Page}} \\
    		\hhline{~--------}
            \textbf{Cycles} & \multicolumn{4}{c|}{min $|\Delta^{(x)}_{mean}|$} & \multicolumn{4}{c|}{min $|\Delta^{(x)}_{var}|$} \\
            \hline
            & \multicolumn{2}{c|}{$x = 0$} & \multicolumn{2}{c|}{$x = 1$} & \multicolumn{2}{c|}{$x = 0$} & \multicolumn{2}{c|}{$x = 1$} \\
            \hline
    		& $p_l$ & $p_u$ & $q_l$ & $q_u$ & $p_l$ & $p_u$ & $q_l$ & $q_u$ \\
    		\hline
    		6000 & 0.36 & 1.37 & 0.31 & 2.29 & 0.38 & 1.64 & 0.34 & 3.41 \\
    		\hline
    		8000 & 0.80 & 3.05 & 0.60 & 4.18 & 0.84 & 3.78 & 0.65 & 5.56 \\
    		\hline
    		10000 & 1.44 & 6.05 & 1.17 & 6.63 & 1.53 & 7.83 & 1.25 & 10.84 \\
    		\hline
            & \multicolumn{8}{|c|}{\textbf{Vendor-B, Lower Page}} \\
            \hline
            6000 & 0.000 & 0.004 & 1.70 & 6.63 & 0.000 & 0.004 & 1.80 & 8.68  \\
      		\hline
      		8000 & 0.000 & 0.004 &  3.58 & 13.04 & 0.000 & 0.004 & 3.78 & 18.90 \\
      		\hline
      		10000 & 0.000 & 0.004 & 6.33 & 22.71 & 0.000 & 0.004 & 6.68 & 30.54\\
      		\hline
    	\end{tabular}
    	}
	\fi
	\egroup
\end{table}

\begin{table}[!ht]
    \caption{Comparison of mean and variance of the number of bit errors per frame obtained from experiment, BBM and TS-BBM channel models using $N = 8192$, $\epsilon = 0.01$ and $\mu = 10^{-6}$ for vendor-A chip.}
	\label{table:mean_var_table_vendor_a}
	\centering
	\bgroup
	\ifCLASSOPTIONonecolumn
    	\begin{tabular}{|c|c c |c c|c c|c c|}
    		\hline
    		\textbf{P/E} & \multicolumn{8}{|c|}{\textbf{Vendor-A, Upper Page}} \\
    		\hhline{~--------}
    		\textbf{Cycles} & \multicolumn{2}{c|}{\textbf{Experiment}} & \multicolumn{2}{c|}{\textbf{BBM}} & \multicolumn{2}{c|}{\textbf{TS-BBM}} & \multicolumn{2}{c|}{\textbf{TS-BBM}} \\
    		\hhline{~~~~~~~~~}
            & \multicolumn{2}{c|}{} & \multicolumn{2}{c|}{} & \multicolumn{2}{c|}{min $|\Delta^{(x)}_{mean}|$} & \multicolumn{2}{c|}{min $|\Delta^{(x)}_{var}|$} \\
            \hhline{~--------}
    		& \textbf{Mean} & \textbf{Variance} & \textbf{Mean} & \textbf{Variance} & \textbf{Mean} & \textbf{Variance} & \textbf{Mean} & \textbf{Variance} \\
    		\hline
    		6000 & 18.43 & 30.06 & 18.43 & 27.06 & 18.43 & 26.42 & 18.52 & 26.79 \\
    		\hline
    		8000 & 32.01 & 66.43 & 32.01 & 57.88 & 32.01 & 55.96 & 32.17 & 56.97 \\
    		\hline
    		10000 & 48.88 & 125.99 & 48.88 & 105.10 & 48.88 & 100.92 & 49.11 & 102.97 \\
    		\hline
             & \multicolumn{8}{|c|}{\textbf{Vendor-A, Lower Page}} \\
            \hline
            6000 & 21.90 & 46.71 & 21.90 & 46.90 & 21.90 & 45.04 & 22.01 & 46.01 \\
    		\hline
    		8000 & 30.55 & 75.89 & 30.55 & 76.25 & 30.55 & 72.86 & 30.71 & 74.54 \\
    		\hline
    		10000 & 41.37 & 111.35 & 41.37 & 111.92 & 41.37 & 106.68 & 41.56 & 109.13 \\
    		\hline
    	\end{tabular}
	\else
		\def\arraystretch{1.5}
        \tabcolsep=0.11cm
    	\resizebox{0.99\columnwidth}{!}{
    	\begin{tabular}{|c|c c |c c|c c|c c|}
    		\hline
    		\textbf{P/E} & \multicolumn{8}{|c|}{\textbf{Vendor-A, Upper Page}} \\
    		\hhline{~--------}
    		\textbf{Cycles} & \multicolumn{2}{c|}{\textbf{Experiment}} & \multicolumn{2}{c|}{\textbf{BBM}} & \multicolumn{2}{c|}{\textbf{TS-BBM}} & \multicolumn{2}{c|}{\textbf{TS-BBM}} \\
    		\hhline{~~~~~~~~~}
            & \multicolumn{2}{c|}{} & \multicolumn{2}{c|}{} & \multicolumn{2}{c|}{min $|\Delta^{(x)}_{mean}|$} & \multicolumn{2}{c|}{min $|\Delta^{(x)}_{var}|$} \\
            \hhline{~--------}
    		& \textbf{Mean} & \textbf{Variance} & \textbf{Mean} & \textbf{Variance} & \textbf{Mean} & \textbf{Variance} & \textbf{Mean} & \textbf{Variance} \\
    		\hline
    		6000 & 18.43 & 30.06 & 18.43 & 27.06 & 18.43 & 26.42 & 18.52 & 26.79 \\
    		\hline
    		8000 & 32.01 & 66.43 & 32.01 & 57.88 & 32.01 & 55.96 & 32.17 & 56.97 \\
    		\hline
    		10000 & 48.88 & 125.99 & 48.88 & 105.10 & 48.88 & 100.92 & 49.11 & 102.97 \\
    		\hline
             & \multicolumn{8}{|c|}{\textbf{Vendor-A, Lower Page}} \\
            \hline
            6000 & 21.90 & 46.71 & 21.90 & 46.90 & 21.90 & 45.04 & 22.01 & 46.01 \\
    		\hline
    		8000 & 30.55 & 75.89 & 30.55 & 76.25 & 30.55 & 72.86 & 30.71 & 74.54 \\
    		\hline
    		10000 & 41.37 & 111.35 & 41.37 & 111.92 & 41.37 & 106.68 & 41.56 & 109.13 \\
    		\hline
    	\end{tabular}
    	}
	\fi
	\egroup
\end{table}

\ifCLASSOPTIONonecolumn
    \begin{figure}[h]
        \centering
        \includegraphics[width=0.7\textwidth]{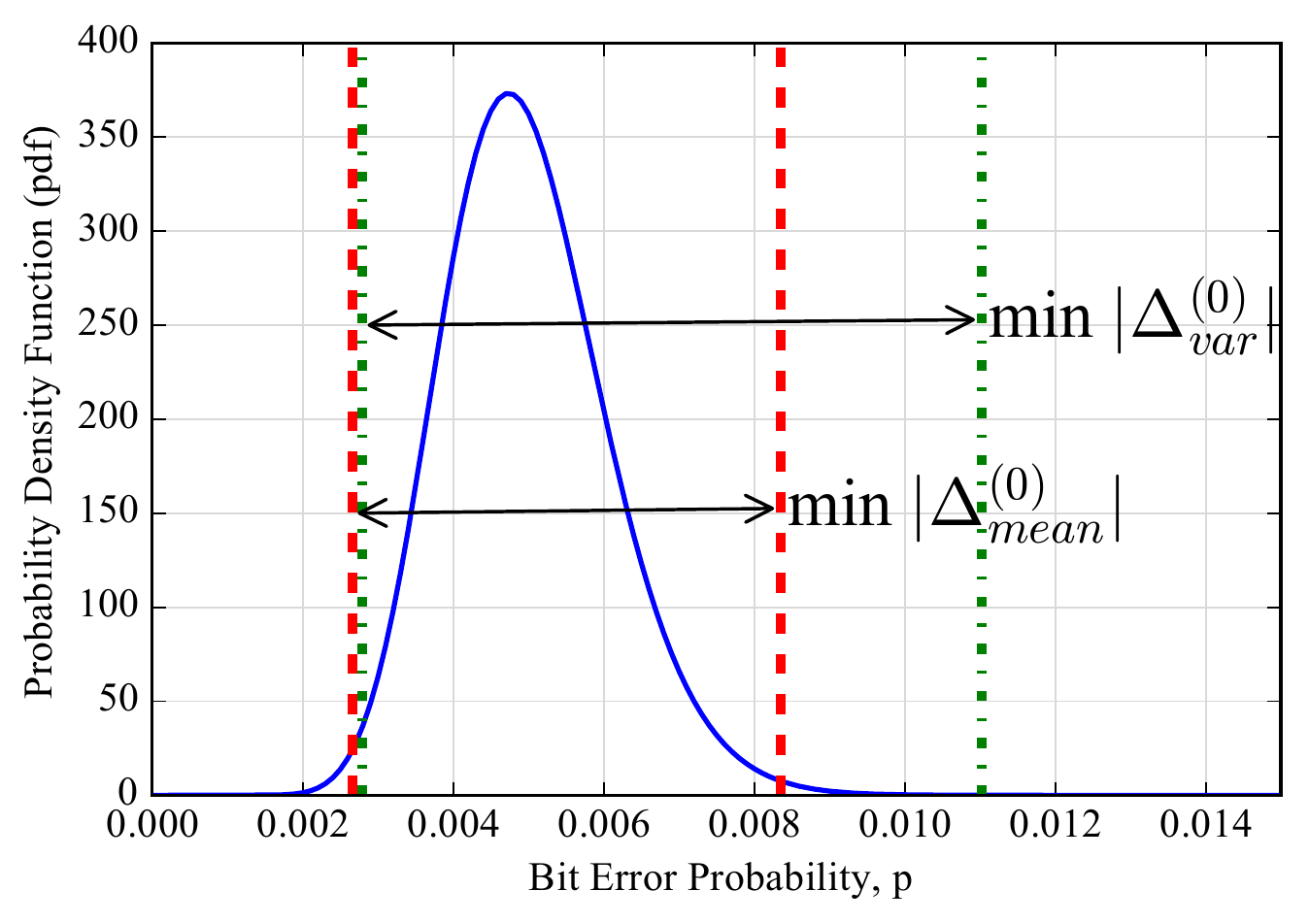}
        \caption{Plot showing the beta pdf corresponding to the upper page $0 \rightarrow 1$ errors BBM channel model and the truncation interval points, $p_l$ and $p_u$ corresponding to TS-BBM channel models in Table~\ref{table:truncation_points_vendor_a} at 8,000 P/E cycles for vendor-A chip.}
        \label{fig:beta_pdf}
    \end{figure}
\else
    \begin{figure}[h]
    	\centering
    	\includegraphics[width=0.47\textwidth]{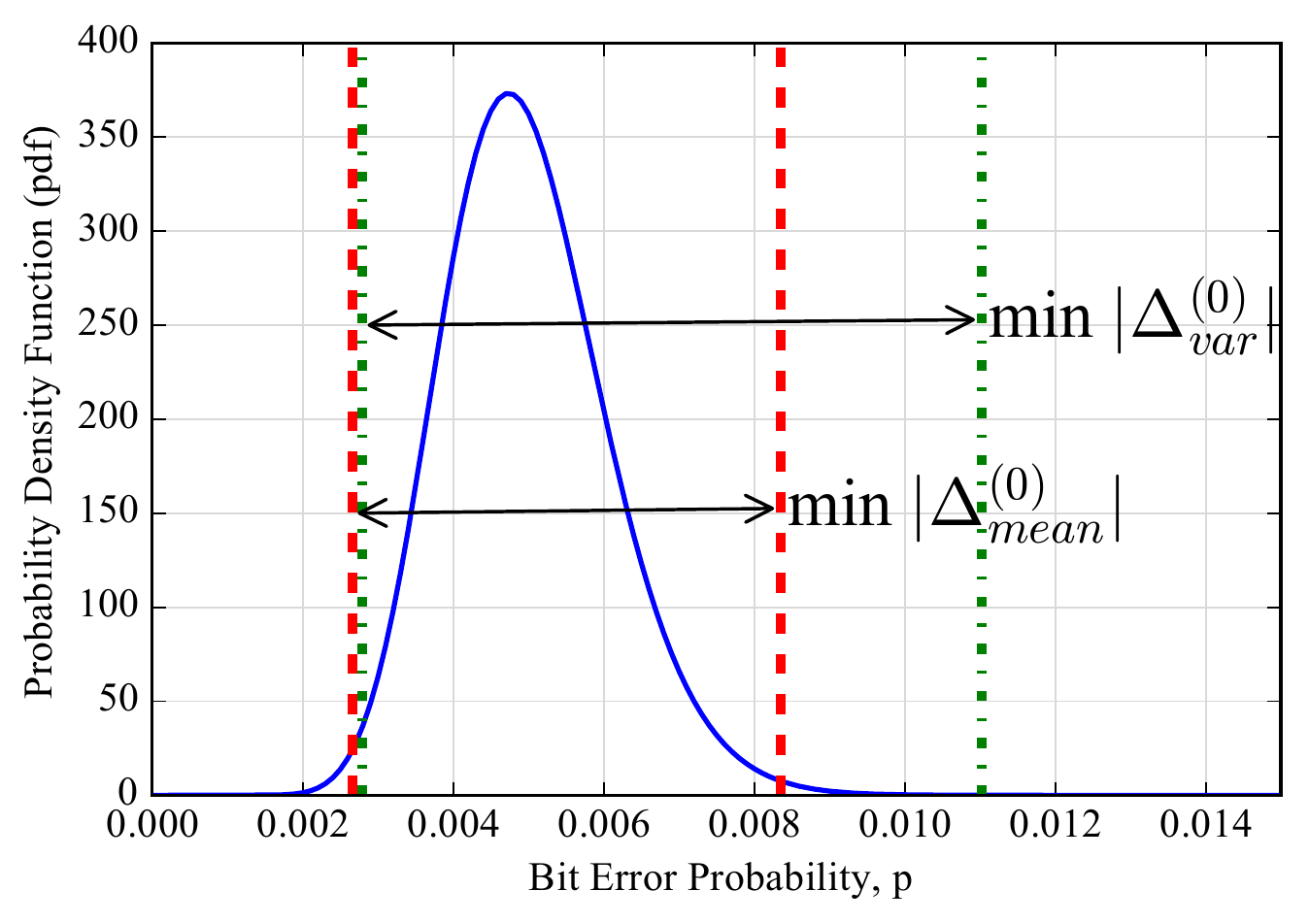}
    	\caption{Plot showing the beta pdf corresponding to the upper page $0 \rightarrow 1$ errors BBM channel model and the truncation interval points, $p_l$ and $p_u$ corresponding to TS-BBM channel models in Table~\ref{table:truncation_points_vendor_a} at 8,000 P/E cycles for vendor-A chip.}
    	\label{fig:beta_pdf}
    \end{figure}
\fi
Table~\ref{table:truncation_points_vendor_a} and Table~\ref{table:truncation_points_vendor_b} show the truncation intervals for the \mbox{TS-BBM} channel models obtained using Algorithm~\ref{algo:opt_truncation_interval} for \mbox{vendor-A} and \mbox{vendor-B} flash memory chips, respectively. We observe that the truncation intervals corresponding to the $\textrm{min}~|\Delta^{(x)}_{var}|$ constraint are wider than those corresponding to the $\textrm{min}~|\Delta^{(x)}_{mean}|$ constraint. This is expected because a wider support would lead to a larger variance for the number of bit errors per frame in the \mbox{TS-BBM} channel model. Fig.~\ref{fig:beta_pdf} shows the beta pdf and the truncation intervals at 8,000 P/E cycles. We also observe that for the TS-Beta distributions of the lower page $\ZOerror$ bit errors, the truncation intervals for both the $\textrm{min}~|\Delta^{(x)}_{mean}|$ and $\textrm{min}~|\Delta^{(x)}_{var}|$ constraints are exactly the same. This is because the $\ZOerror$ bit error probability in the lower page is extremely small during P/E cycling, as was observed in~\cite{Taranalli_2016}, which leads to a very narrow-width beta distribution. Because we have $\epsilon = 0.01$, in this case there is only one truncation interval that satisfies the probability mass condition given in step~5 of Algorithm~\ref{algo:search_candidates}.
The mean and variance of the total number of bit errors per frame for the \mbox{TS-BBM} channel models in Table~\ref{table:truncation_points_vendor_a} are shown in Table~\ref{table:mean_var_table_vendor_a} and are compared with the statistics obtained from empirical data and the BBM channel models. We observe that the \mbox{TS-BBM} channel models optimized for the $\textrm{min}~|\Delta^{(x)}_{mean}|$ constraint are able to match the mean corresponding to the BBM channel models, which yields a precise estimate of the average raw bit error rate (RBER).

Fig.~\ref{fig:ldpc_polar_vendor_b} shows the FER performance of a regular QC-LDPC code and a Polar code, respectively, using empirical data, as well as the 2-BAC, 2-BBM and 2-TS-BBM channel models for vendor-B chip. The empirical FER performance estimates are obtained from the error data collected from MLC flash memory chips during P/E cycling experiments.
To estimate the FER performance using the proposed channel models, Monte-Carlo simulations are used where pseudo-random codewords of the code are generated and transmitted through the appropriate channel model and the received codeword is decoded. At least 400 frame errors are recorded for FER estimation.
The simulation of the \mbox{TS-BBM} channel model is implemented using rejection sampling and the beta random variates are generated using the SciPy library~\cite{SciPy_2016}.

The regular quasi-cyclic LDPC (QC-LDPC) code parameters are $N=8192$, $k=7683$, with $d_c = 64$ and $d_v = 4$, where $d_c$ and $d_v$ refer to the check node and variable node degrees, respectively, in the parity check matrix.
A sum-product belief propagation decoder with a maximum of $50$ iterations and early termination is used to decode the QC-LDPC code. A detailed description of construction of the QC-LDPC code is given in~\cite{Taranalli_2016}.
The polar code parameters are $N=8192$, $k=7684$ and it is optimized for a binary symmetric channel (BSC) with bit error probability $p=0.001$ using the construction technique proposed in~\cite{Tal_Vardy_2013}. The successive cancellation list (SC-List) decoder proposed in~\cite{Tal_Vardy_2015} is used for decoding the polar code with list size 8.

We observe that the ECC FER performance estimates obtained using the $\textrm{min}~|\Delta^{(x)}_{var}|$ \mbox{2-TS-BBM} channel model are slightly more accurate (with respect to the empirical results) than those obtained using the $\textrm{min}~|\Delta^{(x)}_{mean}|$ \mbox{2-TS-BBM} channel model. We also observe that the $\textrm{min}~|\Delta^{(x)}_{var}|$ \mbox{2-TS-BBM} channel model is essentially the same as the BBM channel model with respect to ECC FER performance estimation.
\ifCLASSOPTIONonecolumn
    \begin{figure}[h]
    	\centering
    	\includegraphics[width=0.7\textwidth]{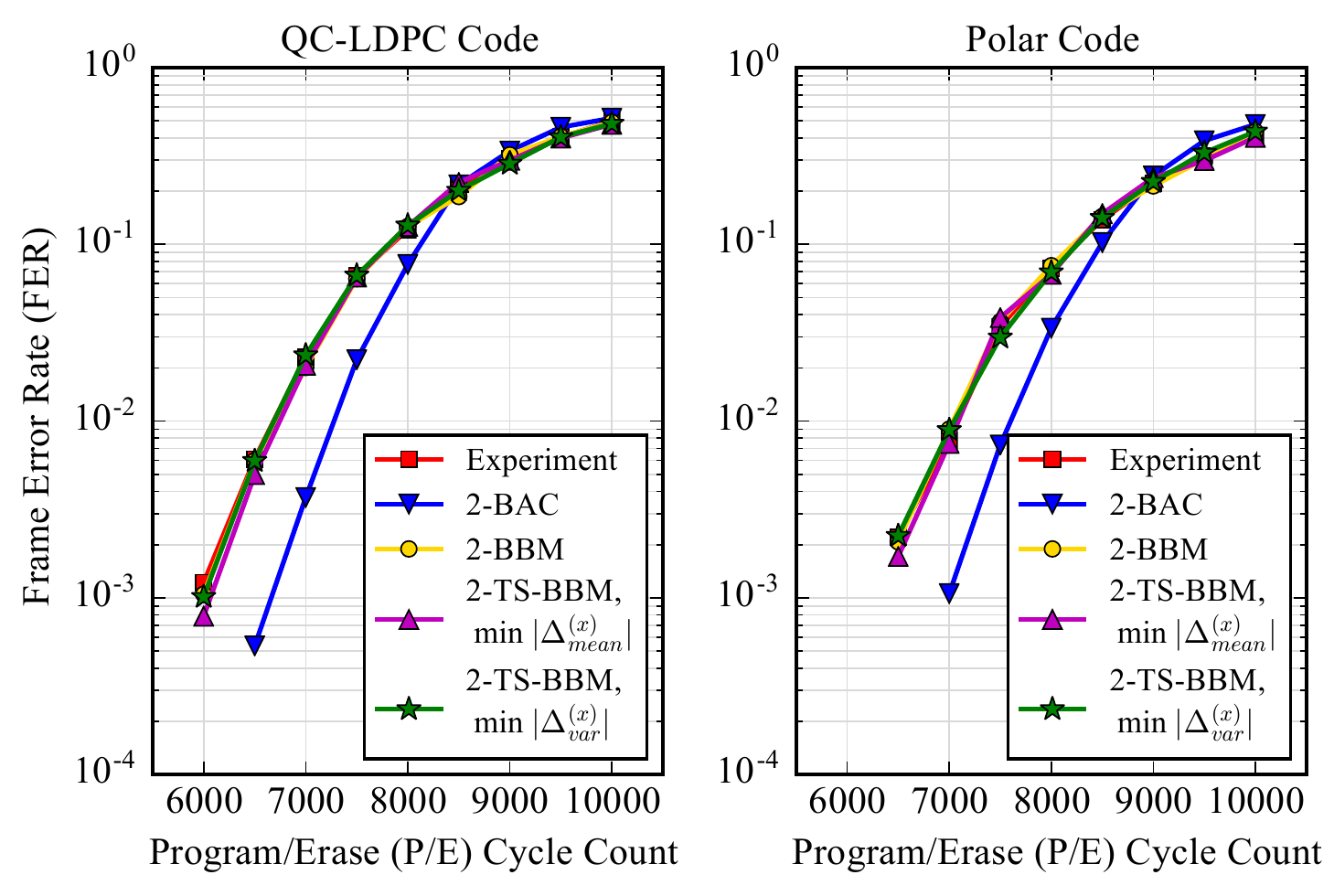}
    	\caption{Comparison of FER performance of a regular QC-LDPC code and a Polar code using empirical error data and simulated error data from the 2-BAC, 2-BBM and 2-TS-BBM channel models for \mbox{vendor-B} chip.}
    	\label{fig:ldpc_polar_vendor_b}
    \end{figure}
\else
    \begin{figure}[h]
        \centering
        \includegraphics[width=0.47\textwidth]{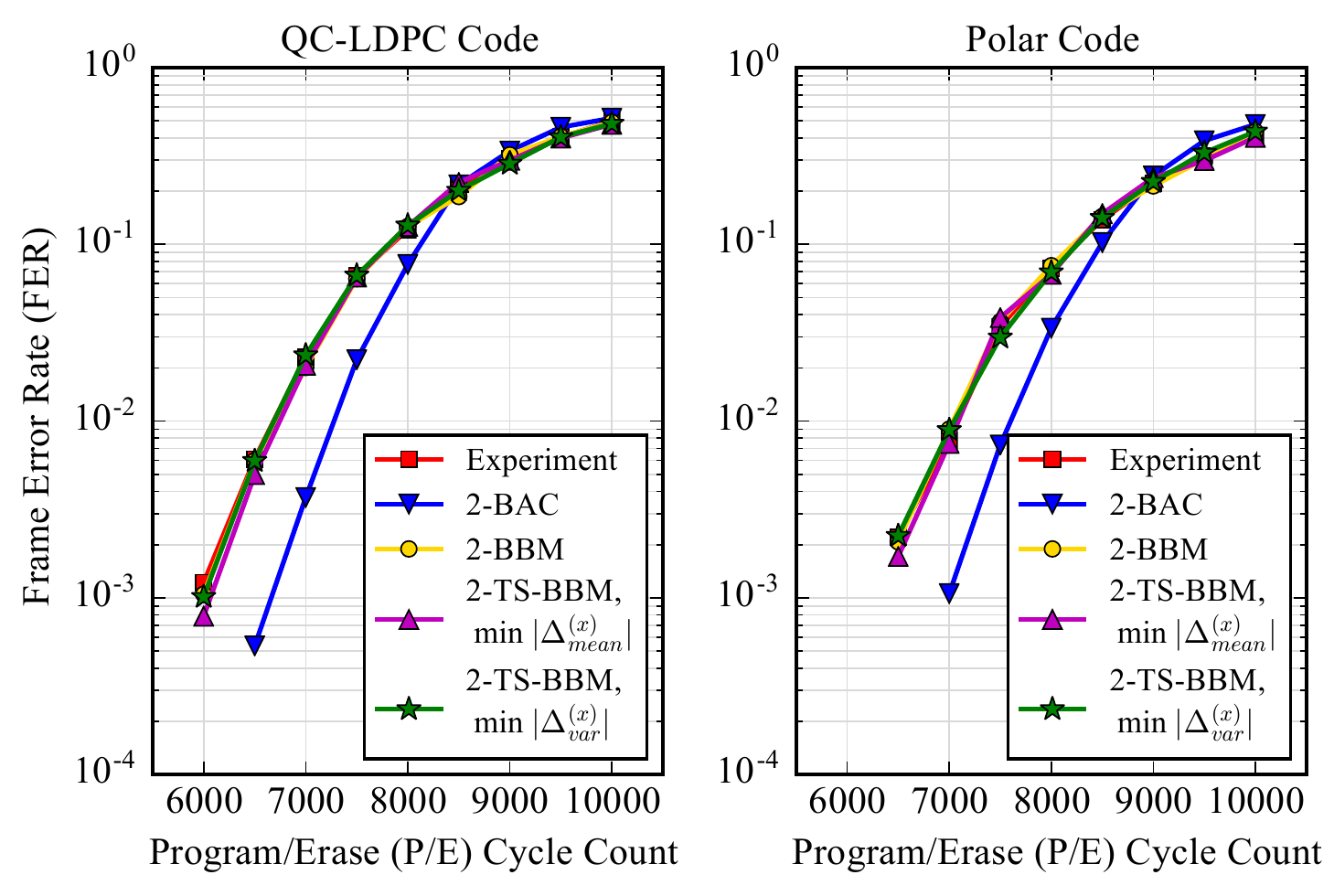}
        \caption{Comparison of FER performance of a regular QC-LDPC code and a Polar code using empirical error data and simulated error data from the 2-BAC, 2-BBM and 2-TS-BBM channel models for \mbox{vendor-B} chip.}
        \label{fig:ldpc_polar_vendor_b}
    \end{figure}
\fi

%% file: sections/capacity_ts_bbm_channel_model.tex
\section{Capacity of the Truncated-Support Beta-Binomial Channel Model}
\label{sec:capacity_ts_bbm_channel_model}

The capacity of the \mbox{TS-BBM} channel model can be derived using the compound channel approach presented in Section~\ref{sec:capacity_bbm_channel_model}, as shown in Lemma~\ref{lemma:degrading_bac} and Theorem~\ref{theorem:tsbbm_cap} below. Lemma~\ref{lemma:degrading_bac} shows that a pair of BACs can be ordered with respect to their mutual information irrespective of the input probability distribution, when their $\ZOerror$ and $\OZerror$ bit error probability pairs $(p, q)$, can be ordered componentwise. This result is key to showing, in Theorem~\ref{theorem:tsbbm_cap}, that the capacity of the \mbox{TS-BBM} channel, which is a compound channel consisting of a set of BACs, achieves the upper bound given by the max-min inequality~\cite{Boyd_2004} on the compound channel capacity.
\begin{IEEElemma}
    Given two BACs, $\textrm{BAC}(\mathcal{X}, \mathcal{Z}, (p_1, q_1))$, $\textrm{BAC}(\mathcal{X}, \mathcal{Y}, (p_2, q_2))$ with $0 \leq p_2 \leq p_1 < \frac{1}{2}$ and $0 \leq q_2 \leq q_1 < \frac{1}{2}$, there always exists a degrading channel $\textrm{BAC}(\mathcal{Y}, \mathcal{Z}, (p', q'))$ such that,
    \begin{IEEEeqnarray}{rCl}
        \Pr(z|x) & = & \sum_{y \in \mathcal{Y}} \Pr(y|x)\Pr(z|y)
        \label{eqn:degrading_channel}
    \end{IEEEeqnarray}
    and $I(X;Z) \leq I(X;Y)$.
    \label{lemma:degrading_bac}
\end{IEEElemma}
\begin{IEEEproof}
    For the first part, it is sufficient to show that $p'$ and $q'$ obtained by solving the set of equations resulting from (\ref{eqn:degrading_channel}) are always positive. From (\ref{eqn:degrading_channel}),
    \begin{IEEEeqnarray}{rCCCl}
        (1-p_2)p' & + & p_2(1-q') & = & p_1 \label{eqn:system1}\\
        q_2(1-p') & + & (1-q_2)q' & = & q_1 \label{eqn:system2}.
    \end{IEEEeqnarray}
    Solving (\ref{eqn:system1}) and (\ref{eqn:system2}),
    \begin{IEEEeqnarray}{rCl}
        q' & = & \frac{(q_1 - q_2)(1 - p_2) + (p_1 - p_2)q_2}{1 - p_2 - q_2} \\
        p' & = & \frac{(p_1 - p_2)(1 - q_2) + (q_1 - q_2)p_2}{1 - p_2 - q_2}.
    \end{IEEEeqnarray}
    Clearly, $q' \geq 0$ and $p' \geq 0$. \\
    Since $X \rightarrow Y \rightarrow Z$ form a Markov chain, the data processing inequality~\cite{Cover_2006} implies that we have $I(X;Z) \leq I(X;Y)$.
\end{IEEEproof}

Assume, without loss of generality,
\begin{IEEEeqnarray}{rCl}
    0 \leq p_l < p_u < \frac{1}{2} ~~~~ 0 \leq q_l < q_u < \frac{1}{2}.
\end{IEEEeqnarray}
\begin{IEEEtheorem}
    The capacity of a TS-BBM channel is equal to the capacity of a $\textrm{BAC}(\mathcal{X}, \mathcal{Y}, (p_u, q_u))$ given by
    \ifCLASSOPTIONonecolumn
        \begin{IEEEeqnarray}{rCl}
            C_{\textrm{TS-BBM}} & = & \Bigg(\frac{p_u}{1-p_u-q_u}\Bigg)h(q_u) - \Bigg(\frac{1-q_u}{1-p_u-q_u}\Bigg)h(p_u) + \textrm{log}_2\Bigg(1 + 2^{\frac{h(p_u) - h(q_u)}{1 - p_u - q_u}}\Bigg).
            \label{eqn:ts_bbm_cap_thm}
        \end{IEEEeqnarray}
    \else
        \begin{IEEEeqnarray}{rCl}
            C_{\textrm{TS-BBM}} & = & \Bigg(\frac{p_u}{1-p_u-q_u}\Bigg)h(q_u) - \Bigg(\frac{1-q_u}{1-p_u-q_u}\Bigg)h(p_u) + \nonumber \\ & & \textrm{log}_2\Bigg(1 + 2^{\frac{h(p_u) - h(q_u)}{1 - p_u - q_u}}\Bigg).
            \label{eqn:ts_bbm_cap_thm}
        \end{IEEEeqnarray}
    \fi
    \label{theorem:tsbbm_cap}
\end{IEEEtheorem}
\begin{IEEEproof}
    The TS-BBM channel is also a compound channel consisting of a set of BACs with varying states $s~\in~ \mathcal{S}$ where $\mathcal{S}=\{(p, q)|p\sim\textrm{TS-Beta}(p_l, p_u; a, b), q\sim\textrm{TS-Beta}(q_l, q_u; c, d)\} $.
    Using the compound channel approach, the capacity of the TS-BBM channel is given by
    \begin{IEEEeqnarray}{rCl}
        C_{\textrm{TS-BBM}} & = & \sup_{\bar{\pi}} \inf_{s \in \mathcal{S}} I_{\bar{\pi}, s}(X;Y)
    \end{IEEEeqnarray}
    Using Lemma~1, $\forall s \in \mathcal{S} \textrm{~and~} \bar{\pi}, I_{\bar{\pi}, s}(X;Y) \geq I_{\bar{\pi}, s'}(X;Y)$, where $s' = (p_u, q_u)$. Therefore,
    \begin{IEEEeqnarray}{rCl}
        \inf_{s \in \mathcal{S}} I_{\bar{\pi}, s}(X;Y) & = & I_{\bar{\pi}, s'}(X;Y)
    \end{IEEEeqnarray}
    so
    \begin{IEEEeqnarray}{rCl}
        C_{\textrm{TS-BBM}} & = & \sup_{\bar{\pi}} I_{\bar{\pi}, s'}(X;Y)
        \label{eqn:ts_bbm_cap_bac}
    \end{IEEEeqnarray}
    From (\ref{eqn:ts_bbm_cap_bac}), $C_{\textrm{TS-BBM}}$ is equal to the capacity of $\textrm{BAC}(\mathcal{X}, \mathcal{Y}, (p_u, q_u))$ and is given by (\ref{eqn:ts_bbm_cap_thm}).
\end{IEEEproof}
\ifCLASSOPTIONonecolumn
    \begin{figure}
        \centering
        \includegraphics[width=0.7\textwidth]{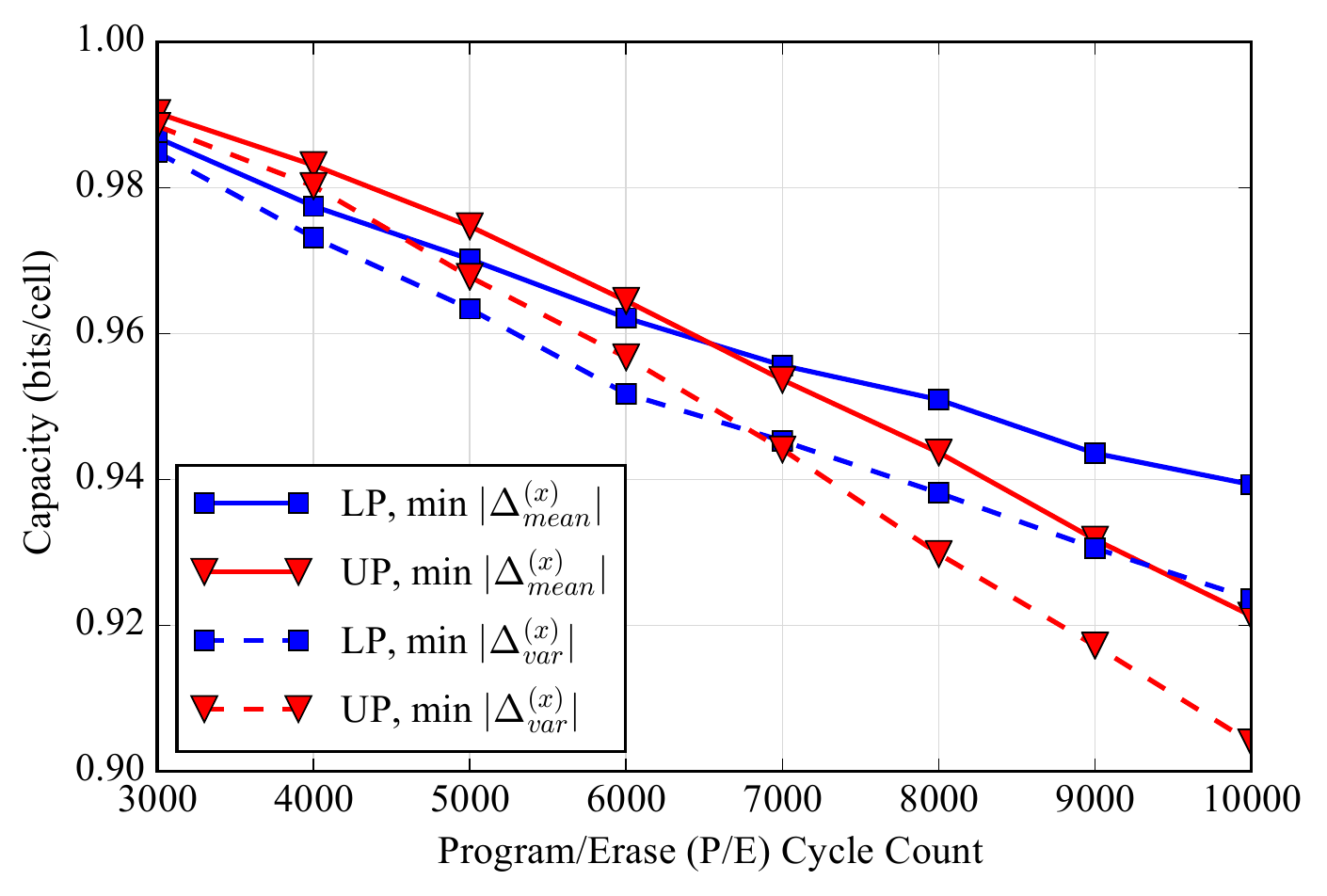}
        \caption{Plot showing the evolution of TS-BBM channel model capacities as a function of P/E cycle count for lower and upper pages of vendor-A flash memory chip. $\epsilon = 0.01$, $\mu = 10^{-6}$.}
        \label{fig:lp_up_tsbbm_capacity_vendorA}
    \end{figure}
\else
    \begin{figure}
        \centering
        \includegraphics[width=0.47\textwidth]{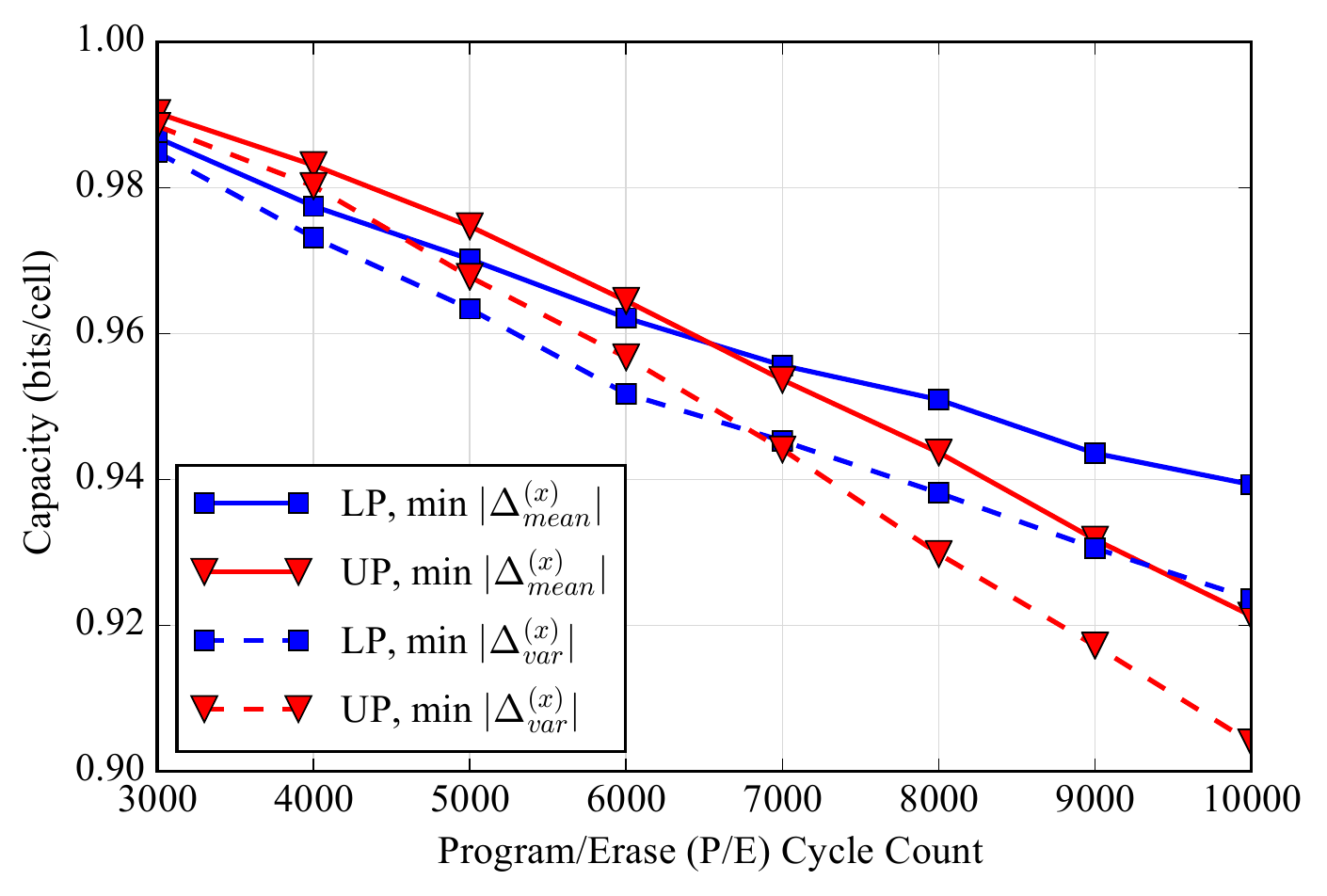}
        \caption{Plot showing the evolution of TS-BBM channel model capacities as a function of P/E cycle count for lower and upper pages of vendor-A flash memory chip. $\epsilon = 0.01$, $\mu = 10^{-6}$.}
        \label{fig:lp_up_tsbbm_capacity_vendorA}
    \end{figure}
\fi
\ifCLASSOPTIONonecolumn
    \begin{figure}
        \centering
        \includegraphics[width=0.7\textwidth]{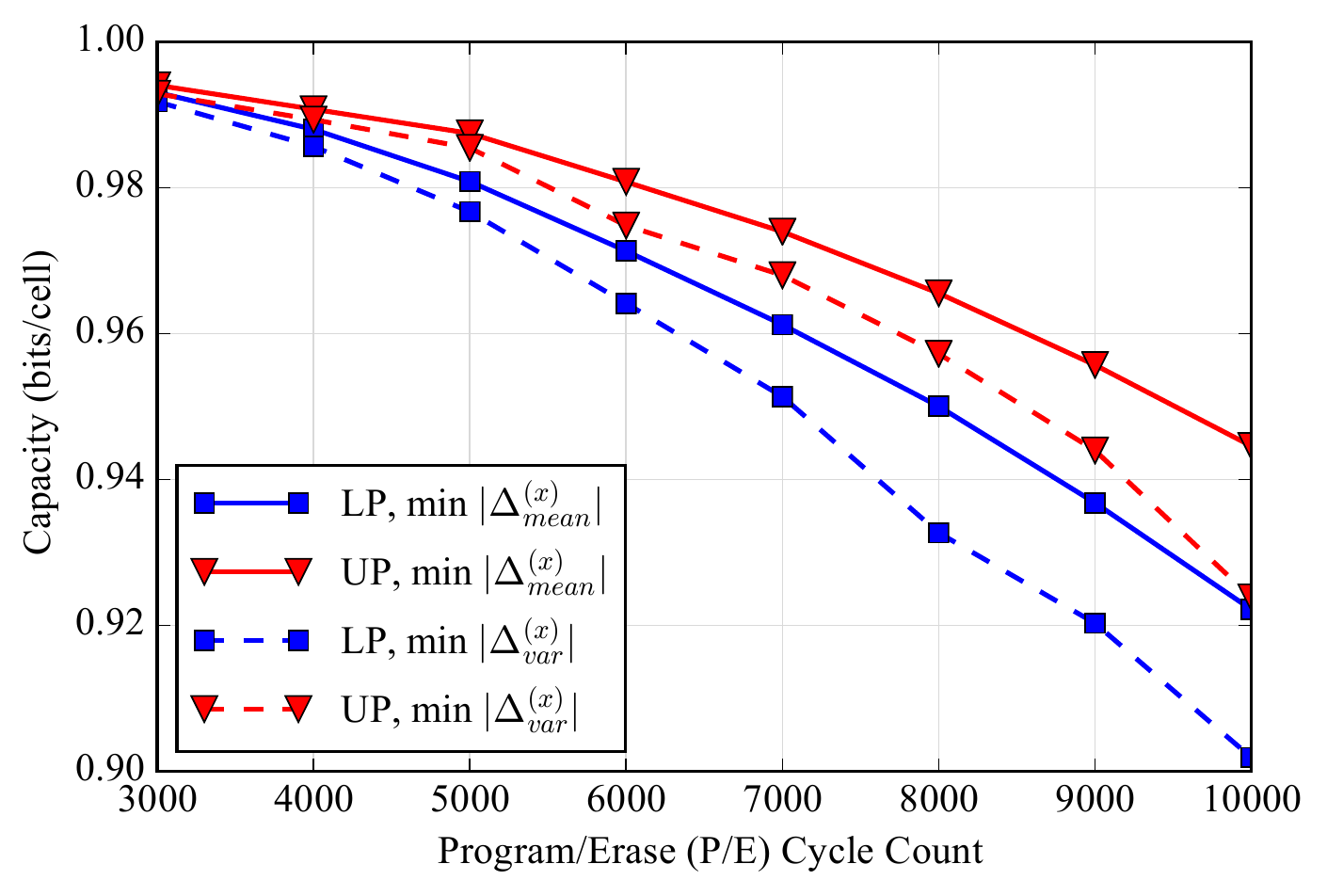}
        \caption{Plot showing the evolution of TS-BBM channel model capacities as a function of P/E cycle count for lower and upper pages of vendor-B flash memory chip. $\epsilon = 0.01$, $\mu = 10^{-6}$.}
        \label{fig:lp_up_tsbbm_capacity_vendorB}
    \end{figure}
\else
    \begin{figure}
        \centering
        \includegraphics[width=0.47\textwidth]{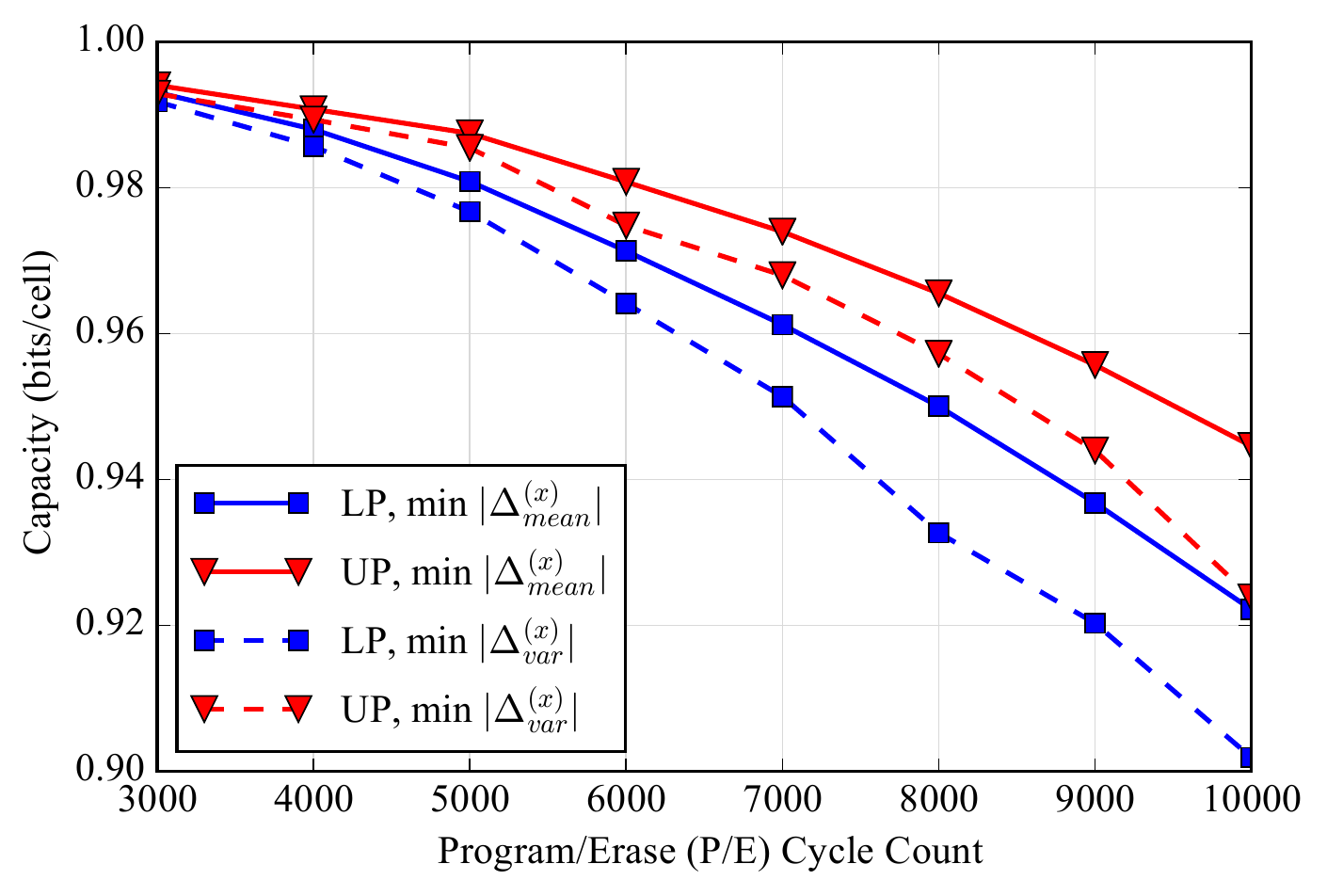}
        \caption{Plot showing the evolution of TS-BBM channel model capacities as a function of P/E cycle count for lower and upper pages of vendor-B flash memory chip. $\epsilon = 0.01$, $\mu = 10^{-6}$.}
        \label{fig:lp_up_tsbbm_capacity_vendorB}
    \end{figure}
\fi
%
%
%
\ifCLASSOPTIONonecolumn
    \begin{figure}
    	\centering
    	\includegraphics[width=0.7\textwidth]{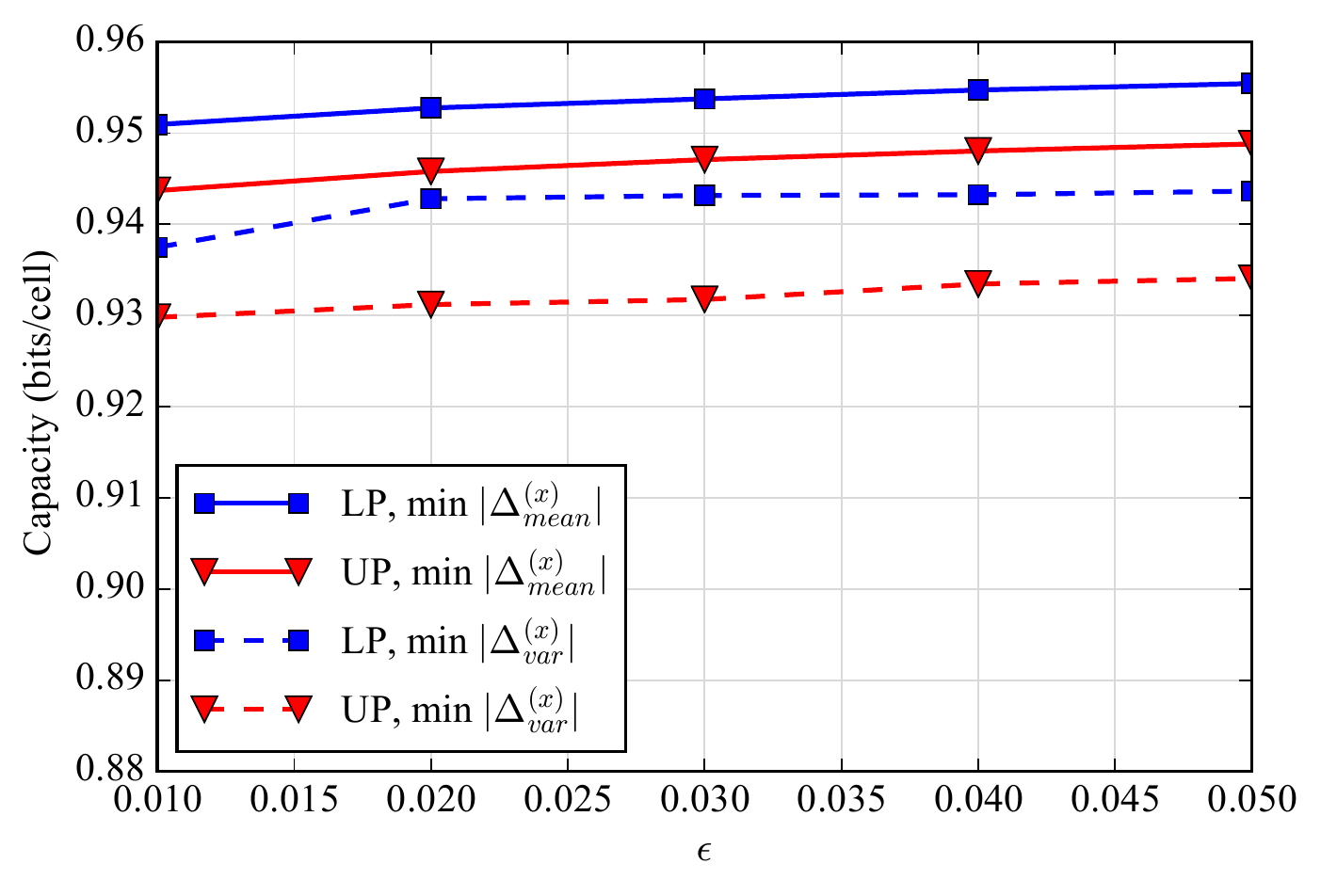}
    	\caption{Plot showing the TS-BBM channel model capacities corresponding to lower and upper pages of vendor-A flash memory chip at 8,000 P/E cycles for different values of $\epsilon$ parameter. $\mu = 10^{-8}$ for LP, $\textrm{min}~|\Delta^{(x)}_{var}|$ and $\mu = 10^{-6}$ for others.}
    	\label{fig:lp_up_tsbbm_capacity_epsilon_vendorA}
    \end{figure}
\else
    \begin{figure}
        \centering
        \includegraphics[width=0.47\textwidth]{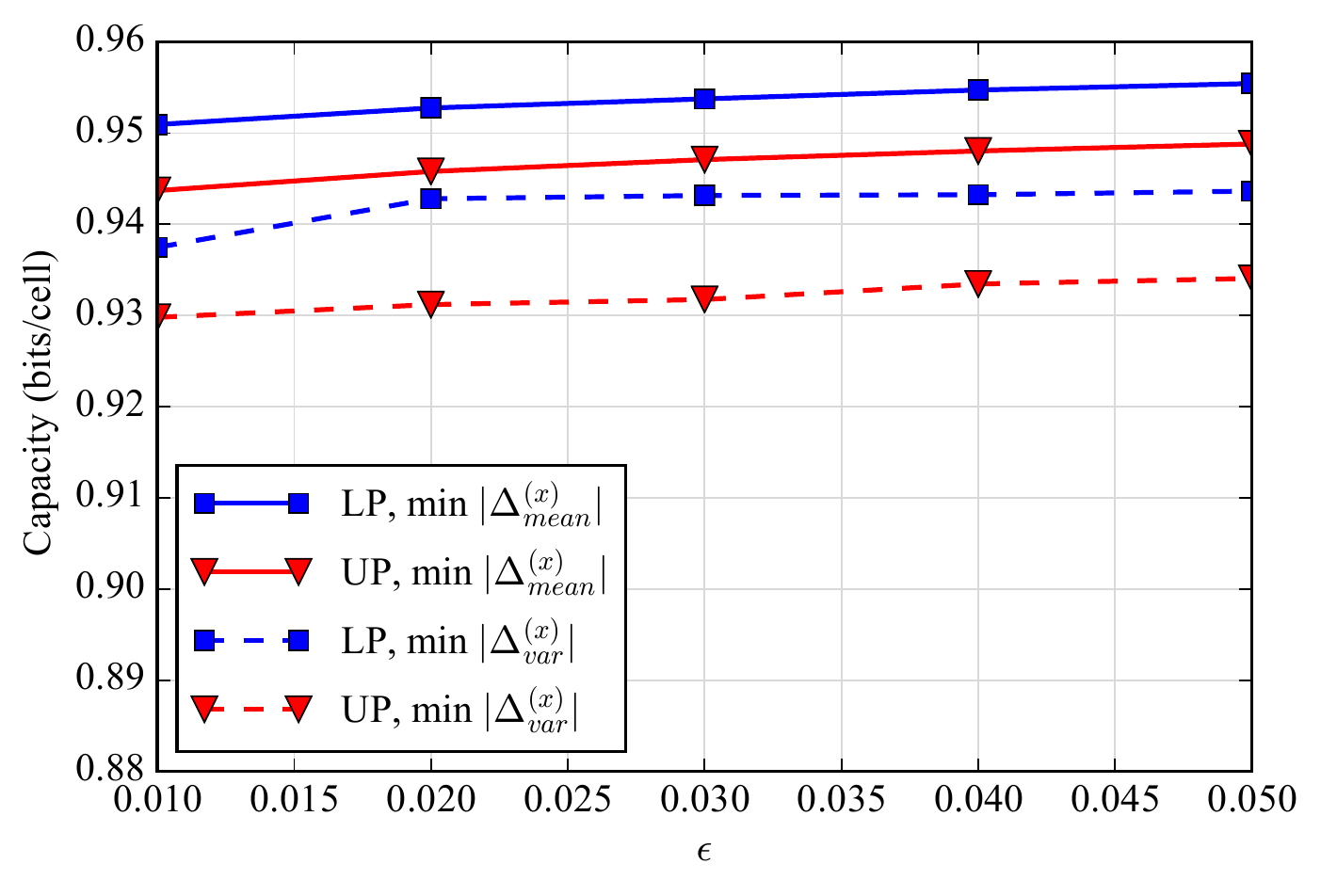}
        \caption{Plot showing the TS-BBM channel model capacities corresponding to lower and upper pages of vendor-A flash memory chip at 8,000 P/E cycles for different values of $\epsilon$ parameter. $\mu = 10^{-8}$ for LP, $\textrm{min}~|\Delta^{(x)}_{var}|$ and $\mu = 10^{-6}$ for others.}
        \label{fig:lp_up_tsbbm_capacity_epsilon_vendorA}
    \end{figure}
\fi
Figures~\ref{fig:lp_up_tsbbm_capacity_vendorA} and \ref{fig:lp_up_tsbbm_capacity_vendorB} show the  \mbox{TS-BBM} channel model capacity estimates for both the lower and upper pages of vendor-A and vendor-B flash memory chips respectively. The changing \mbox{TS-BBM} channel model capacity estimates across P/E cycles are indicative of the rate of degradation of the underlying flash memory channel, for e.g., the \mbox{TS-BBM} channel model capacity for the upper page of vendor-A chip decreases from $\ca 0.99$ at $3,000$ P/E cycles to $\ca 0.92$ at $10,000$ P/E cycles. We also note that for vendor-B chip, the upper page \mbox{TS-BBM} channel models consistently have larger capacity than the lower page \mbox{TS-BBM} channel models at all P/E cycles, whereas this is not the case for vendor-A chip. Such knowledge of the page capacities as a function of P/E cycle count in flash memories could be useful and advantageous for ECC design. For example, the estimates of the page capacities can be directly utilized for designing rate adaptive ECC schemes and for optimally distributing redundancy between the lower and upper pages at different lifetime stages of the MLC flash memory. We also observe that the TS-BBM channel models optimized for min $|\Delta^{(x)}_{mean}|$ have larger capacities than those optimized for min $|\Delta^{(x)}_{var}|$. Fig.~\ref{fig:lp_up_tsbbm_capacity_epsilon_vendorA} shows the variation of the TS-BBM channel model capacities with respect to the $\epsilon$ parameter. Choosing a larger $\epsilon$ generally leads to a larger capacity TS-BBM channel model; however a larger $\epsilon$ also results in a narrower truncation interval of the beta pdf. Hence a TS-BBM channel model corresponding to large $\epsilon$ may not be able to accurately model the variance of the number of bit errors per frame resulting in optimistic ECC FER performance estimates.
\subsection{Coding for the TS-BBM Channel Model}
Recall that the \mbox{TS-BBM} channel model is a compound channel consisting of a set of BACs with varying states $s~\in~ \mathcal{S}$ where $\mathcal{S}=\{(p, q)|p\sim\textrm{TS-Beta}(p_l, p_u; a, b), q\sim\textrm{TS-Beta}(q_l, q_u; c, d)\} $. From Theorem~\ref{theorem:tsbbm_cap}, the capacity of a \mbox{TS-BBM} channel model is equal to the capacity of the most noisy channel i.e., $\textrm{BAC}(\mathcal{X}, \mathcal{Y}, (p_u, q_u))$. Although coding techniques based on polar codes and sparse graph codes for achieving the capacity of a single BAC have been proposed in the literature~\cite{Sutter_2012, Honda_2013, Mondelli_2014}, the application of such techniques to a set of BACs as represented by the \mbox{TS-BBM} compound channel appears to be a difficult problem.

Therefore, we look at existing coding techniques that can achieve the symmetric information rate (SIR) of the \mbox{TS-BBM} channel. The SIR of a binary DMC is defined as its mutual information with a uniform input probability distribution. It is known that the difference between the capacity and the SIR of a binary DMC $(\Delta_{C, SIR})$ is at most $\ca 5.8\%$ of its capacity~\cite{Majani_1991}. Hence, we compute and show this difference between the capacity and the SIR of the \mbox{TS-BBM} channel model as a function of the P/E cycle count, in Figures~\ref{fig:lp_tsbbm_capacity_SIR} and \ref{fig:up_tsbbm_capacity_SIR}, for the lower page and upper page models, respectively. We observe that this difference between the capacity and the SIR of the \mbox{TS-BBM} channel model is extremely small for both pages. Therefore for practical applications, the loss (in capacity) is almost negligible if we use linear codes that achieve the SIR of the \mbox{TS-BBM} channel.

Under the assumption that both the encoder and decoder have no knowledge of the channel state, we require a single code that can achieve the SIR of all the component BACs of the \mbox{TS-BBM} channel. Polar codes have been shown to achieve the SIR of any binary DMC~\cite{Arikan_2009}. The SIR of a single BAC can be achieved by a polar code constructed as shown in~\cite{Arikan_2009}, with a suitable choice of values for the frozen bits. However, finding such a set of frozen bit values of a polar code for asymmetric binary DMCs such as the BAC is an open problem. Even though the component BACs in the \mbox{TS-BBM} channel model are ordered by degradation, it is not clear if the frozen bit indices of polar codes corresponding to these component BACs are aligned. We leave these problems open for future work.
\ifCLASSOPTIONonecolumn
    \begin{figure}
    	\centering
    	\includegraphics[width=0.7\textwidth]{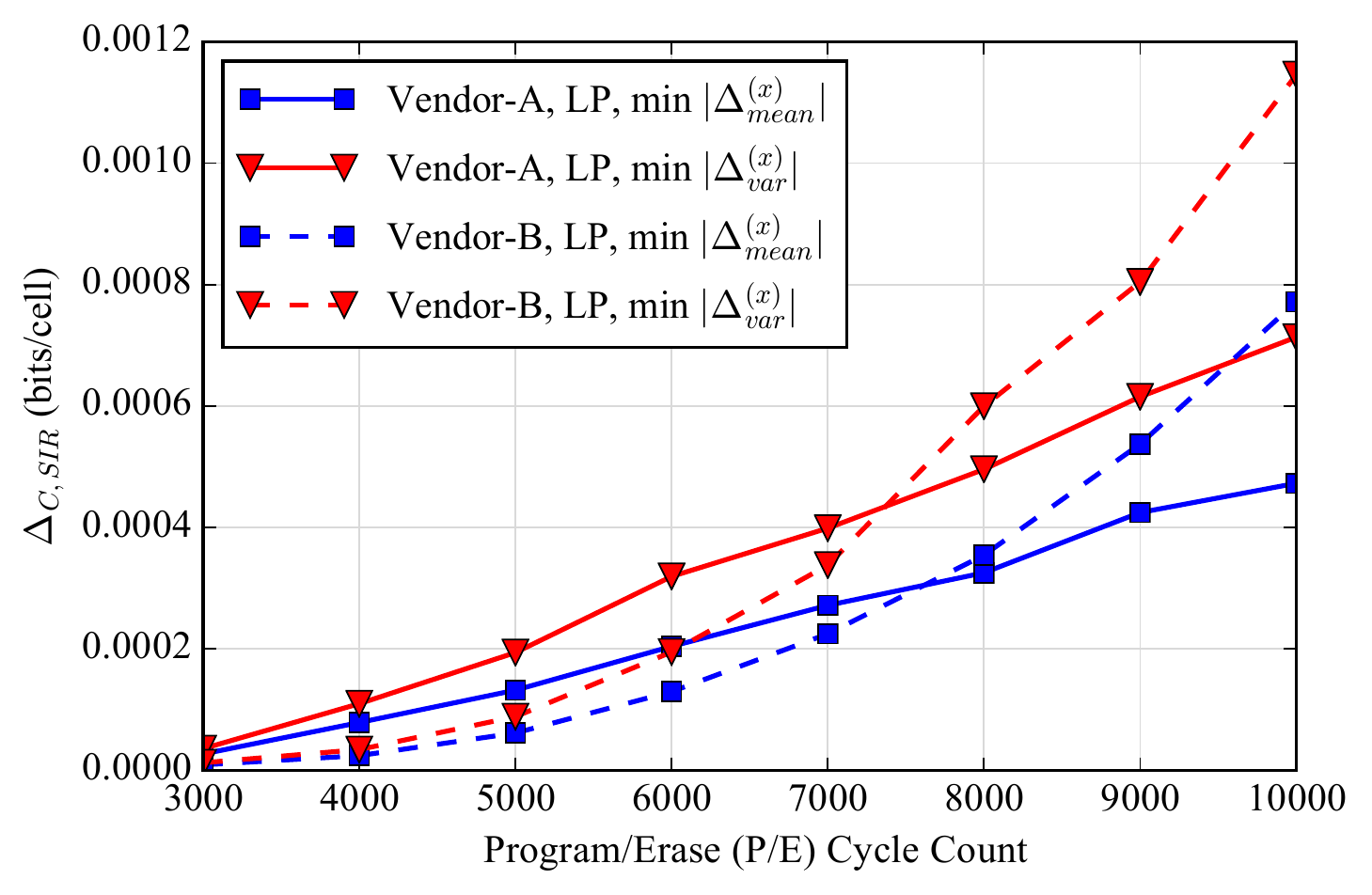}
    	\caption{Plot showing the difference between the capacity and the symmetric information rate (SIR) for the lower page TS-BBM channel models corresponding to vendor-A and vendor-B flash memory chips. $\epsilon = 0.01$, $\mu = 10^{-6}$.}
    	\label{fig:lp_tsbbm_capacity_SIR}
    \end{figure}
\else
    \begin{figure}
        \centering
        \includegraphics[width=0.47\textwidth]{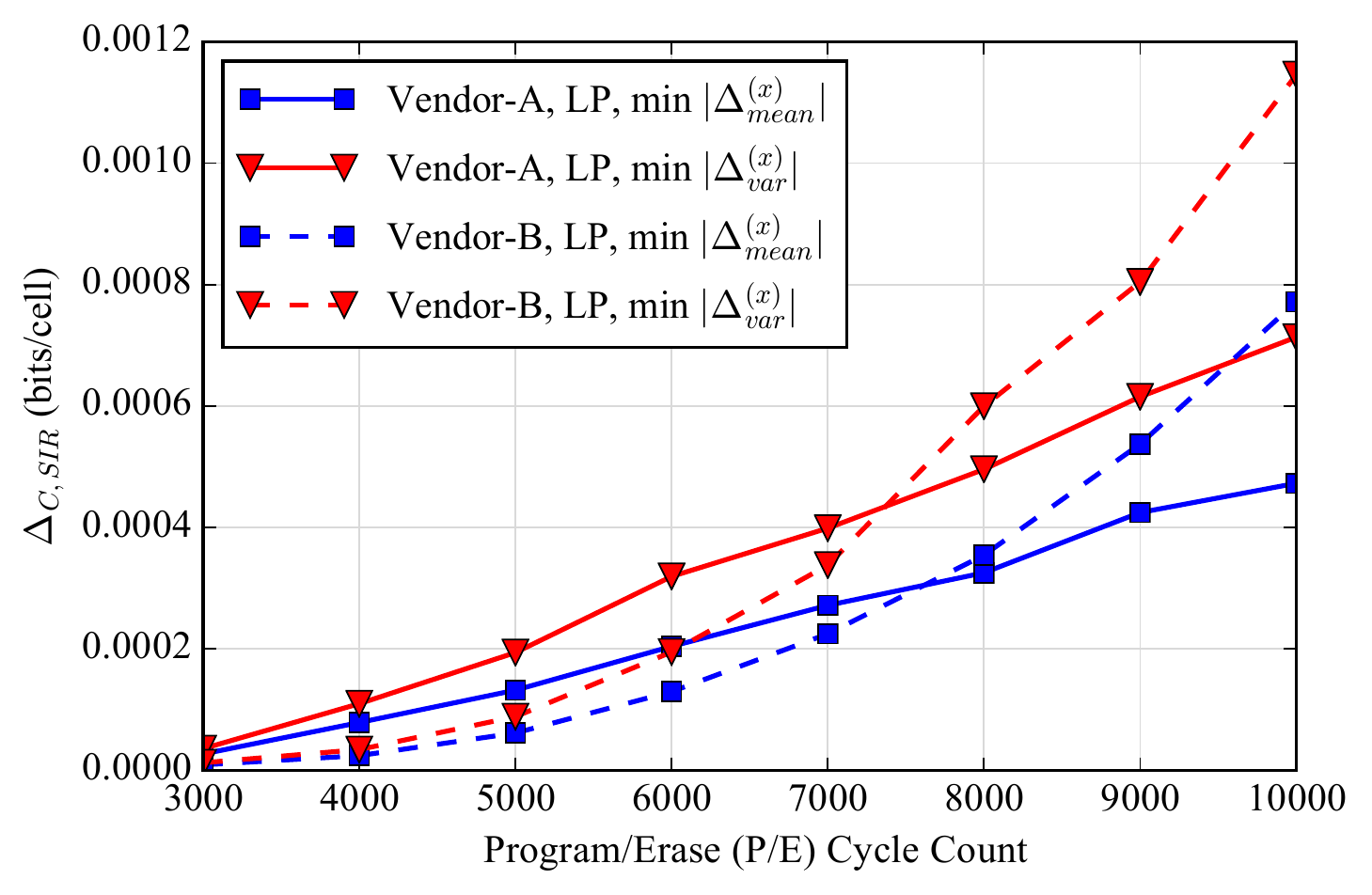}
        \caption{Plot showing the difference between the capacity and the symmetric information rate (SIR) for the lower page TS-BBM channel models corresponding to vendor-A and vendor-B flash memory chips. $\epsilon = 0.01$, $\mu = 10^{-6}$.}
        \label{fig:lp_tsbbm_capacity_SIR}
    \end{figure}
\fi
\ifCLASSOPTIONonecolumn
    \begin{figure}
    	\centering
    	\includegraphics[width=0.7\textwidth]{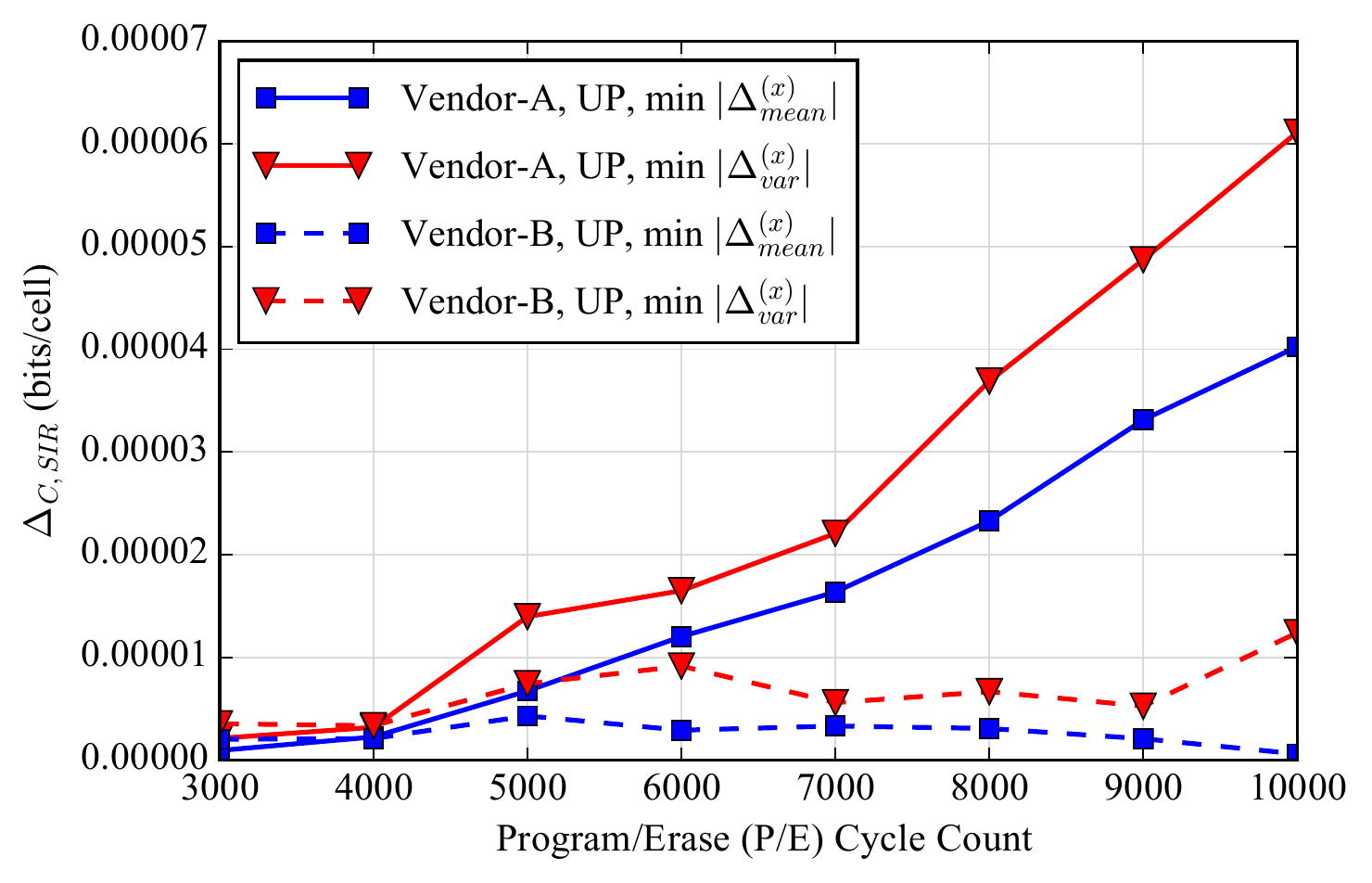}
    	\caption{Plot showing the difference between the capacity and the symmetric information rate (SIR) for the upper page TS-BBM channel models corresponding to vendor-A and vendor-B flash memory chips. $\epsilon = 0.01$, $\mu = 10^{-6}$.}
    	\label{fig:up_tsbbm_capacity_SIR}
    \end{figure}
\else
    \begin{figure}
        \centering
        \includegraphics[width=0.47\textwidth]{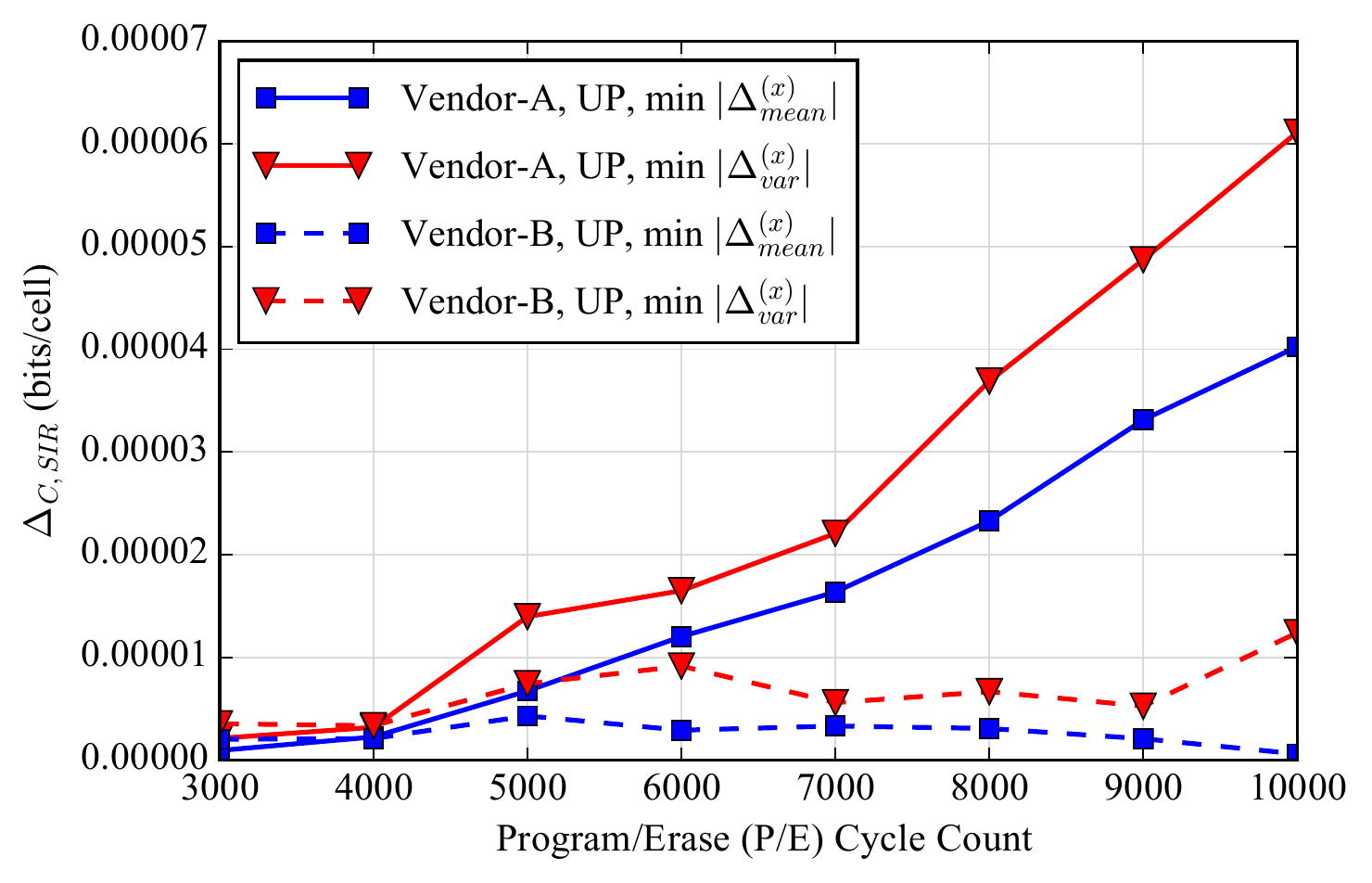}
        \caption{Plot showing the difference between the capacity and the symmetric information rate (SIR) for the upper page TS-BBM channel models corresponding to vendor-A and vendor-B flash memory chips. $\epsilon = 0.01$, $\mu = 10^{-6}$.}
        \label{fig:up_tsbbm_capacity_SIR}
    \end{figure}
\fi

%% file: sections/conclusion.tex
\section{Conclusion}
\label{sec:conclusion}
Using the compound channel model approach, we showed that the beta-binomial (\mbox{BBM}) channel model for MLC flash memories proposed in~\cite{Taranalli_2016} has zero capacity. Flash memories in practice do have non-zero positive capacities and hence the BBM channel model appears to be a very pessimistic model. As an alternative, based on empirical observations, we proposed the truncated-support beta-binomial (\mbox{TS-BBM}) channel model for MLC flash memories and derived its capacity. We used empirical error characterization data from \mbox{1X-nm} and \mbox{2Y-nm} MLC flash memories~\cite{Taranalli_2016} to obtain the \mbox{TS-BBM} channel model parameters and estimate its capacity. Using FER performance results for a regular QC-LDPC code and a polar code, we also showed that the \mbox{2-TS-BBM} channel model is almost as good as the BBM channel model for ECC FER performance estimation.  When using a single binary ECC for all pages in MLC flash memories, the \mbox{TS-BBM} channel model capacity represents an upper bound on the rate of the ECC. This is because, the ECC must be able to correct all the errors resulting from the most noisy channel in the set of channels represented by the \mbox{TS-BBM} channel model.

%% file: sections/appendix.tex
\appendices
\section{Proof of Proposition \ref{prop:ts_bbm_mean_var}}
\label{app:proof_ts_bbm_mean_var}
The mean of a \mbox{TS-BBM} random variable $Z$ is given by
\ifCLASSOPTIONonecolumn
    \begin{IEEEeqnarray}{rCl}
        \E[Z] & = & \sum_{z = 0}^{n} z \Pr(Z = z) \nonumber \\
              & = & \sum_{z = 0}^{n} z {n \choose z} \Bigg(\frac{B_{\theta_u}(\alpha+z, \beta+n-z) - B_{\theta_l}(\alpha+z, \beta+n-z)}{B_{\theta_u}(\alpha, \beta) - B_{\theta_l}(\alpha, \beta)} \Bigg). 
        \label{eqn:proof_prop_1_1}
    \end{IEEEeqnarray}
\else
    \begin{IEEEeqnarray}{rCl}
        \E[Z] & = & \sum_{z = 0}^{n} z \Pr(Z = z) \nonumber \\
              & = & \sum_{z = 0}^{n} z {n \choose z} \Bigg(\frac{B_{\theta_u}(\alpha+z, \beta+n-z)}{B_{\theta_u}(\alpha, \beta) - B_{\theta_l}(\alpha, \beta)} - \nonumber \\
              & & \frac{B_{\theta_l}(\alpha+z, \beta+n-z)}{B_{\theta_u}(\alpha, \beta) - B_{\theta_l}(\alpha, \beta)} \Bigg).
        \label{eqn:proof_prop_1_1}
    \end{IEEEeqnarray}
\fi
Let
\begin{IEEEeqnarray}{rCl}
    t_{\theta_u} & = & \sum_{z = 0}^{n} z {n \choose z} B_{\theta_u}(\alpha+z, \beta+n-z) \nonumber \\
                 & = & \sum_{z = 0}^{n} z {n \choose z} \int_{0}^{\theta_u} \lambda^{\alpha+z-1} (1 - \lambda)^{\beta+n-z-1} \textrm{d}\lambda \nonumber \\
                 & = & \int_{0}^{\theta_u} \left(\sum_{z = 0}^{n} z {n \choose z} \lambda^{z-1} (1 - \lambda)^{n-z} \right) \lambda^{\alpha-1} (1 - \lambda)^{\beta-1} \textrm{d}\lambda. \nonumber
\end{IEEEeqnarray}
The term in parentheses in the above equation is the expected value of a binomial random variable with parameters $n$ and $\lambda$, namely $n\lambda$. Substituting, we have
\begin{IEEEeqnarray}{rCl}
    t_{\theta_u} & = & n \int_{0}^{\theta_u} \lambda^{\alpha+1-1}(1 - \lambda)^{\beta-1} \textrm{d}\lambda \nonumber \\
                 & = & n B_{\theta_u}(\alpha+1, \beta).
    \label{eqn:proof_prop_1_2}
\end{IEEEeqnarray}
Similarly,
\begin{IEEEeqnarray}{rCl}
    t_{\theta_l} & = & n B_{\theta_l}(\alpha+1, \beta).
    \label{eqn:proof_prop_1_3}
\end{IEEEeqnarray}
Substituting (\ref{eqn:proof_prop_1_2}), (\ref{eqn:proof_prop_1_3}) in (\ref{eqn:proof_prop_1_1}), we find
\begin{IEEEeqnarray}{rCl}
    \E[Z] & = & n\Bigg( \frac{B_{\theta_u}(\alpha+1, \beta) - B_{\theta_l}(\alpha+1, \beta)}{B_{\theta_u}(\alpha, \beta) - B_{\theta_l}(\alpha, \beta)}  \Bigg).
    \label{eqn:proof_prop_1_4}
\end{IEEEeqnarray}
To relate $\E[Z]$ and $\E[\tilde{Z}]$ where $\tilde{Z}$ is a beta-binomial random variable, we use the recurrence relations~\cite{Gupta_2004}
\begin{IEEEeqnarray}{rCl}
    \frac{B_{x}(\alpha+1, \beta)}{B(\alpha+1, \beta)} & = & \frac{B_{x}(\alpha, \beta)}{B(\alpha, \beta)} - \frac{x^{\alpha}(1-x)^{\beta}}{\alpha B(\alpha, \beta)} \label{eqn:proof_prop_1_5} \\
    B(\alpha+1, \beta) & = & \left(\frac{\alpha}{\alpha+\beta}\right) B(\alpha, \beta)     \label{eqn:proof_prop_1_6} \\
    B(\alpha, \beta+1) & = & \left(\frac{\beta}{\alpha+\beta}\right) B(\alpha, \beta).     \label{eqn:proof_prop_1_7}
\end{IEEEeqnarray}
Using (\ref{eqn:proof_prop_1_5}), (\ref{eqn:proof_prop_1_6}), (\ref{eqn:proof_prop_1_7}) in (\ref{eqn:proof_prop_1_4}), we get
\begin{IEEEeqnarray}{rCl}
    \E[Z] & = & \frac{n\alpha}{\alpha+\beta} - \left(\frac{n}{\alpha+\beta}\right)\frac{\delta_\theta}{\eta_\theta} \nonumber \\
          & = & \E[\tilde{Z}] - \left(\frac{n}{\alpha+\beta}\right)\frac{\delta_\theta}{\eta_\theta}.
    \label{eqn:proof_prop_1_14}
\end{IEEEeqnarray}
To compute the variance, we compute the second moment of $Z$, $\E[Z^2]$, as follows,
\ifCLASSOPTIONonecolumn
    \begin{IEEEeqnarray}{rCl}
        \E[Z^2] & = & \sum_{z = 0}^{n} z^2 \Pr(Z = z) \nonumber \\
                & = & \sum_{z = 0}^{n} z^2 {n \choose z} \Bigg(\frac{B_{\theta_u}(\alpha+z, \beta+n-z) - B_{\theta_l}(\alpha+z, \beta+n-z)}{B_{\theta_u}(\alpha, \beta) - B_{\theta_l}(\alpha, \beta)} \Bigg). 
        \label{eqn:proof_prop_1_8}
    \end{IEEEeqnarray}
\else
    \begin{IEEEeqnarray}{rCl}
        \E[Z^2] & = & \sum_{z = 0}^{n} z^2 \Pr(Z = z) \nonumber \\
                & = & \sum_{z = 0}^{n} z^2 {n \choose z} \Bigg(\frac{B_{\theta_u}(\alpha+z, \beta+n-z)}{B_{\theta_u}(\alpha, \beta) - B_{\theta_l}(\alpha, \beta)} - \nonumber \\
                & & \frac{B_{\theta_l}(\alpha+z, \beta+n-z)}{B_{\theta_u}(\alpha, \beta) - B_{\theta_l}(\alpha, \beta)} \Bigg).
        \label{eqn:proof_prop_1_8}
    \end{IEEEeqnarray}
\fi
Let,
\begin{IEEEeqnarray}{rCl}
    \bar{t}_{\theta_u} & = & \sum_{z = 0}^{n} z^2 {n \choose z} B_{\theta_u}(\alpha+z, \beta+n-z) \nonumber \\
                  & = & \sum_{z = 0}^{n} z^2 {n \choose z} \int_{0}^{\theta_u} \lambda^{\alpha+z-1} (1 - \lambda)^{\beta+n-z-1} \textrm{d}\lambda \nonumber \\
                 & = & \int_{0}^{\theta_u} \left(\sum_{z = 0}^{n} z^2 {n \choose z} \lambda^{z-1} (1 - \lambda)^{n-z} \right) \lambda^{\alpha-1} (1 - \lambda)^{\beta-1} \textrm{d}\lambda . \nonumber
\end{IEEEeqnarray}
The term in parentheses in the above equation is the second moment of a binomial random variable with parameters $n$ and $\lambda$, namely $n(n-1)\lambda^2 + n\lambda$. Substituting, we have
\ifCLASSOPTIONonecolumn
    \begin{IEEEeqnarray}{rCl}
        \bar{t}_{\theta_u} & = & n(n-1) \int_{0}^{\theta_u} \lambda^{\alpha+2-1}(1 - \lambda)^{\beta-1} \textrm{d}\lambda + n \int_{0}^{\theta_u} \lambda^{\alpha+1-1}(1 - \lambda)^{\beta-1} \textrm{d}\lambda \nonumber \\
                           & = & n(n-1)B_{\theta_u}(\alpha+2, \beta) + nB_{\theta_u}(\alpha+1, \beta).
        \label{eqn:proof_prop_1_10}
    \end{IEEEeqnarray}
\else
    \begin{IEEEeqnarray}{rCl}
        \bar{t}_{\theta_u} & = & n(n-1) \int_{0}^{\theta_u} \lambda^{\alpha+2-1}(1 - \lambda)^{\beta-1} \textrm{d}\lambda \nonumber \\
        & & + n \int_{0}^{\theta_u} \lambda^{\alpha+1-1}(1 - \lambda)^{\beta-1} \textrm{d}\lambda \nonumber \\
                           & = & n(n-1)B_{\theta_u}(\alpha+2, \beta) + nB_{\theta_u}(\alpha+1, \beta).
        \label{eqn:proof_prop_1_10}
    \end{IEEEeqnarray}
\fi
Similarly,
\begin{IEEEeqnarray}{rCl}
    \bar{t}_{\theta_l} & = & n(n-1)B_{\theta_u}(\alpha+2, \beta) + nB_{\theta_u}(\alpha+1, \beta).
    \label{eqn:proof_prop_1_11}
\end{IEEEeqnarray}
Substituting (\ref{eqn:proof_prop_1_10}), (\ref{eqn:proof_prop_1_11}) in (\ref{eqn:proof_prop_1_8}), we find
\ifCLASSOPTIONonecolumn
    \begin{IEEEeqnarray}{rCl}
        \E[Z^2] & = & n(n-1) \frac{B_{\theta_u}(\alpha+2, \beta) - B_{\theta_l}(\alpha+2, \beta)}{B_{\theta_u}(\alpha, \beta) - B_{\theta_l}(\alpha, \beta)} + n  \frac{B_{\theta_u}(\alpha+1, \beta) - B_{\theta_l}(\alpha+1, \beta)}{B_{\theta_u}(\alpha, \beta) - B_{\theta_l}(\alpha, \beta)};
        \label{eqn:proof_prop_1_12} \\
        \Var[Z] & = & \E[Z^2] - (\E[Z])^2 \nonumber \\
                & = & n \Bigg( \frac{B_{\theta_u}(\alpha+1, \beta) - B_{\theta_l}(\alpha+1, \beta)}{B_{\theta_u}(\alpha, \beta) - B_{\theta_l}(\alpha, \beta)} \Bigg) \Bigg(1 - n \Bigg( \frac{B_{\theta_u}(\alpha+1, \beta) - B_{\theta_l}(\alpha+1, \beta)}{B_{\theta_u}( \alpha, \beta) - B_{\theta_l}(\alpha, \beta)}\Bigg)\Bigg) + \nonumber \\
                & & n(n-1)\Bigg( \frac{B_{\theta_u}(\alpha+2, \beta) - B_{\theta_l}(\alpha+2, \beta)}{B_{\theta_u}(\alpha, \beta) - B_{\theta_l}(\alpha, \beta)} \Bigg).
        \label{eqn:proof_prop_1_13}
    \end{IEEEeqnarray}
\else
    \begin{IEEEeqnarray}{rCl}
        \E[Z^2] & = & n(n-1)\Bigg(\frac{B_{\theta_u}(\alpha+2, \beta) - B_{\theta_l}(\alpha+2, \beta)}{B_{\theta_u}(\alpha, \beta) - B_{\theta_l}(\alpha, \beta)} \Bigg) \nonumber \\
        & & + n\Bigg( \frac{B_{\theta_u}(\alpha+1, \beta) - B_{\theta_l}(\alpha+1, \beta)}{B_{\theta_u}(\alpha, \beta) - B_{\theta_l}(\alpha, \beta)}  \Bigg);
        \label{eqn:proof_prop_1_12} \\
        \Var[Z] & = & \E[Z^2] - (\E[Z])^2 \nonumber \\
                & = & n \Bigg( \frac{B_{\theta_u}(\alpha+1, \beta) - B_{\theta_l}(\alpha+1, \beta)}{B_{\theta_u}(\alpha, \beta) - B_{\theta_l}(\alpha, \beta)} \Bigg) \nonumber \\
                & & \Bigg(1 - n \Bigg( \frac{B_{\theta_u}(\alpha+1, \beta) - B_{\theta_l}(\alpha+1, \beta)}{B_{\theta_u}( \alpha, \beta) - B_{\theta_l}(\alpha, \beta)}\Bigg)\Bigg) + \nonumber \\
                & & n(n-1)\Bigg( \frac{B_{\theta_u}(\alpha+2, \beta) - B_{\theta_l}(\alpha+2, \beta)}{B_{\theta_u}(\alpha, \beta) - B_{\theta_l}(\alpha, \beta)} \Bigg).
        \label{eqn:proof_prop_1_13}
    \end{IEEEeqnarray}
\fi
Using the recurrence relations (\ref{eqn:proof_prop_1_5}), (\ref{eqn:proof_prop_1_6}), (\ref{eqn:proof_prop_1_7}) in (\ref{eqn:proof_prop_1_13}), we get
\ifCLASSOPTIONonecolumn
    \begin{IEEEeqnarray}{rCl}
        \Var[Z] & = & \Var[\tilde{Z}] - \left(\frac{n\beta(\alpha + \beta + n)}{(\alpha + \beta)^2(\alpha + \beta + 1)}\right)\frac{\delta_\theta}{\eta_\theta} - \frac{n^2}{(\alpha+\beta)^2}\left(\frac{\delta_\theta}{\eta_\theta}\right)^2 - \left(\frac{n(n-1)}{\alpha+\beta+1}\right) \frac{\phi_\theta}{\eta_\theta}
            \label{eqn:proof_prop_1_15}
    \end{IEEEeqnarray}
\else
    \begin{IEEEeqnarray}{rCl}
        \Var[Z] & = & \Var[\tilde{Z}] - \left(\frac{n\beta(\alpha + \beta + n)}{(\alpha + \beta)^2(\alpha + \beta + 1)}\right)\frac{\delta_\theta}{\eta_\theta} \nonumber \\
        & & -  \frac{n^2}{(\alpha+\beta)^2}\left(\frac{\delta_\theta}{\eta_\theta}\right)^2 - \left(\frac{n(n-1)}{\alpha+\beta+1}\right) \frac{\phi_\theta}{\eta_\theta}
            \label{eqn:proof_prop_1_15}
    \end{IEEEeqnarray}
\fi
where $\Var[\tilde{Z}]$ is the variance of a BBM random variable and $\eta_\theta$, $\delta_{\theta}$, $\phi_\theta$ are defined in equations (\ref{eqn:etatheta_def}), (\ref{eqn:deltatheta_def}), (\ref{eqn:phitheta_def}), respectively.~\hfill\IEEEQED
%
\section{Proof of Proposition \ref{prop:mean_var_k0_k1_ts_bbm_model}}
\label{app:proof_mean_var_k0_k1_ts_bbm_model}
The probability distribution of $K^{(0)}$ in terms of the probability distribution of $\kcount{m}{0}$ is given by~\cite{Taranalli_2016}
\ifCLASSOPTIONonecolumn
    \begin{IEEEeqnarray}{rCl}
        \Pr(K^{(0)} = k) & = & \sum_{m=k}^{N} \frac{{N \choose m}}{2^N} \Pr(\kcount{m}{0} = k) \nonumber \\
                         & = & \sum_{m=k}^{N} \frac{{N \choose m}}{2^N} {m \choose k} \Bigg(\frac{B_{p_u}(a+k, b+m-k) - B_{p_l}(a+k, b+m-k)}{B_{p_u}(a, b) - B_{p_l}(a, b)} \Bigg).
    \end{IEEEeqnarray}
\else
    \begin{IEEEeqnarray}{rCl}
        \Pr(K^{(0)} = k) & = & \sum_{m=k}^{N} \frac{{N \choose m}}{2^N} \Pr(\kcount{m}{0} = k) \nonumber \\
                         & = & \sum_{m=k}^{N} \frac{{N \choose m}}{2^N} {m \choose k} \Bigg(\frac{B_{p_u}(a+k, b+m-k)}{B_{p_u}(a, b) - B_{p_l}(a, b)} - \nonumber \\
                         & & \frac{B_{p_l}(a+k, b+m-k)}{B_{p_u}(a, b) - B_{p_l}(a, b)} \Bigg).
    \end{IEEEeqnarray}
\fi
The mean of $K^{(0)}$ is given by
\begin{IEEEeqnarray}{rCl}
    \E[K^{(0)}] & = & \sum_{k=0}^{N} k \Pr(K^{(0)} = k) \nonumber \\
                & = & \frac{1}{2^N} \sum_{m=0}^{N} {N \choose m} \E[\kcount{m}{0}] \nonumber \\
                & = & \frac{1}{2^N} \sum_{m=0}^{N} {N \choose m} m \Bigg( \frac{B_{p_u}(a+1, b) - B_{p_l}(a+1, b)}{B_{p_u}(a, b) - B_{p_l}(a, b)}  \Bigg) \nonumber \\
                & = & \frac{N}{2} \Bigg(\frac{U_{p_u}^{(1)} - U_{p_l}^{(1)}}{U_{p_u}^{(0)} - U_{p_l}^{(0)}}\Bigg).
    \label{eqn:proof_prop_2_5}
\end{IEEEeqnarray}
The second moment of $K^{(0)}$ is given by
\begin{IEEEeqnarray}{rCl}
    \E[(K^{(0)})^2] & = & \sum_{k=0}^{N} k^2 \Pr(K^{(0)} = k) \nonumber \\
                & = & \frac{1}{2^N} \sum_{m=0}^{N} {N \choose m} \E[(\kcount{m}{0})^2].
    \label{eqn:proof_prop_2_3}
\end{IEEEeqnarray}
Using (\ref{eqn:proof_prop_1_12}),
\ifCLASSOPTIONonecolumn
    \begin{IEEEeqnarray}{rCl}
        \E[(\kcount{m}{0})^2] & = & m(m-1)\Bigg(\frac{B_{p_u}(a+2, b) - B_{p_l}(a+2, b)}{B_{p_u}(a, b) - B_{p_l}(a, b)} \Bigg) \nonumber \\
        & & + m\Bigg( \frac{B_{p_u}(a+1, b) - B_{p_l}(a+1, b)}{B_{p_u}(a, b) - B_{p_l}(a, b)}  \Bigg).
        \label{eqn:proof_prop_2_2}
    \end{IEEEeqnarray}
\else
    \begin{IEEEeqnarray}{rCl}
        \E[(\kcount{m}{0})^2] & = & m(m-1)\Bigg(\frac{B_{p_u}(a+2, b) - B_{p_l}(a+2, b)}{B_{p_u}(a, b) - B_{p_l}(a, b)} \Bigg) \nonumber \\
        & & + m\Bigg( \frac{B_{p_u}(a+1, b) - B_{p_l}(a+1, b)}{B_{p_u}(a, b) - B_{p_l}(a, b)}  \Bigg).
        \label{eqn:proof_prop_2_2}
    \end{IEEEeqnarray}
\fi
Substituting (\ref{eqn:proof_prop_2_2}) in (\ref{eqn:proof_prop_2_3}) and simplifying, we get
\ifCLASSOPTIONonecolumn
    \begin{IEEEeqnarray}{rCl}
        \E[(K^{(0)})^2] & = & \frac{N(N-1)}{4} \Bigg( \frac{B_{p_u}(a+2, b) - B_{p_l}(a+2, b)}{B_{p_u}(a, b) - B_{p_l}(a, b)}  \Bigg) + \frac{N}{2}\Bigg( \frac{B_{p_u}(a+1, b) - B_{p_l}(a+1, b)}{B_{p_u}(a, b) - B_{p_l}(a, b)}  \Bigg) \nonumber \\
        & = & \frac{N(N-1)}{4} \Bigg(\frac{U_{p_u}^{(2)} - U_{p_l}^{(2)}}{U_{p_u}^{(0)} - U_{p_l}^{(0)}}\Bigg) + \frac{N}{2} \Bigg(\frac{U_{p_u}^{(1)} - U_{p_l}^{(1)}}{U_{p_u}^{(0)} - U_{p_l}^{(0)}}\Bigg).
        \label{eqn:proof_prop_2_4}
    \end{IEEEeqnarray}
\else
    \begin{IEEEeqnarray}{rCl}
        \E[(K^{(0)})^2] & = & \frac{N(N-1)}{4} \Bigg( \frac{B_{p_u}(a+2, b) - B_{p_l}(a+2, b)}{B_{p_u}(a, b) - B_{p_l}(a, b)}  \Bigg) \nonumber \\ & & + \frac{N}{2}\Bigg( \frac{B_{p_u}(a+1, b) - B_{p_l}(a+1, b)}{B_{p_u}(a, b) - B_{p_l}(a, b)}  \Bigg) \nonumber \\
        & = & \frac{N(N-1)}{4} \Bigg(\frac{U_{p_u}^{(2)} - U_{p_l}^{(2)}}{U_{p_u}^{(0)} - U_{p_l}^{(0)}}\Bigg) \nonumber \\ & & + \frac{N}{2} \Bigg(\frac{U_{p_u}^{(1)} - U_{p_l}^{(1)}}{U_{p_u}^{(0)} - U_{p_l}^{(0)}}\Bigg).
        \label{eqn:proof_prop_2_4}
    \end{IEEEeqnarray}
\fi
Therefore, using (\ref{eqn:proof_prop_2_5}) and (\ref{eqn:proof_prop_2_4}), we get
\ifCLASSOPTIONonecolumn
    \begin{IEEEeqnarray}{rCl}
        \Var[K^{(0)}] & = & \E[(K^{(0)})^2] - (\E[K^{(0)}])^2 \nonumber \\
                      & = & \frac{N}{2}\Bigg(\frac{U_{p_u}^{(1)} - U_{p_l}^{(1)}}{U_{p_u}^{(0)} - U_{p_l}^{(0)}}\Bigg) \Bigg(1 - \frac{N}{2}\Bigg(\frac{U_{p_u}^{(1)} - U_{p_l}^{(1)}}{U_{p_u}^{(0)} - U_{p_l}^{(0)}}\Bigg)\Bigg) + \frac{N(N-1)}{4} \Bigg(\frac{U_{p_u}^{(2)} - U_{p_l}^{(2)}}{U_{p_u}^{(0)} - U_{p_l}^{(0)}}\Bigg).
        \label{eqn:proof_prop_2_1}
    \end{IEEEeqnarray}
\else
    \begin{IEEEeqnarray}{rCl}
        \Var[K^{(0)}] & = & \E[(K^{(0)})^2] - (\E[K^{(0)}])^2 \nonumber \\
                      & = & \frac{N}{2}\Bigg(\frac{U_{p_u}^{(1)} - U_{p_l}^{(1)}}{U_{p_u}^{(0)} - U_{p_l}^{(0)}}\Bigg) \Bigg(1 - \frac{N}{2}\Bigg(\frac{U_{p_u}^{(1)} - U_{p_l}^{(1)}}{U_{p_u}^{(0)} - U_{p_l}^{(0)}}\Bigg)\Bigg) \nonumber \\ & & + \frac{N(N-1)}{4} \Bigg(\frac{U_{p_u}^{(2)} - U_{p_l}^{(2)}}{U_{p_u}^{(0)} - U_{p_l}^{(0)}}\Bigg).
        \label{eqn:proof_prop_2_1}
    \end{IEEEeqnarray}
\fi
Note that we have used the following combinatorial identities to simplify the summations in this proof:
\begin{IEEEeqnarray}{L}
   \sum_{m=0}^{N}{N \choose m}~m = N 2^{N-1} \label{eqn:comb_id_1} \\
   \sum_{m=0}^{N}{N \choose m}~m^2 = (N + N^2) 2^{N-2}. \label{eqn:comb_id_2}
\end{IEEEeqnarray}
The expressions for $\E[K^{(1)}]$ and $\Var[K^{(1)}]$ can be derived similarly.~\hfill\IEEEQED
%

\section{Proof of Proposition \ref{prop:mean_var_ts_bbm_model}}
\label{app:proof_mean_var_ts_bbm_model}
By definition, we have
\begin{IEEEeqnarray}{rCl}
    K & = & K^{(0)} + K^{(1)} \\
    \E[K] & = & \E[K^{(0)}] + \E[K^{(1)}].
\end{IEEEeqnarray}
Using the results of Proposition~\ref{prop:mean_var_k0_k1_ts_bbm_model}, we have
\begin{IEEEeqnarray}{rCl}
    \E[K] & = & \frac{N}{2}\Bigg(\frac{U_{p_u}^{(1)} - U_{p_l}^{(1)}}{U_{p_u}^{(0)} - U_{p_l}^{(0)}} + \frac{V_{q_u}^{(1)} - V_{q_l}^{(1)}}{V_{q_u}^{(0)} - V_{q_l}^{(0)}} \Bigg). \\
\end{IEEEeqnarray}
To compute $\Var[K]$, we first need $\E[K^2]$, derived as follows:
\begin{IEEEeqnarray}{rCl}
    \E[K^2] & = & \frac{1}{2^N} \sum_{m = 0}^{N} {N \choose m} \E[K^{2}_m] \label{eqn:proof_prop_4_1} \\
    \textrm{where}~~~\E[K^{2}_m] & = & \Var[K_m] + (\E[K_m])^2. \label{eqn:proof_prop_4_2}\\
    \E[K_m] & = & \E[K_m^{(0)}] + \E[K_{N-m}^{(1)}] \\
    \Var[K_m] & = & \Var[K_m^{(0)}] + \Var[K_{N-m}^{(1)}].
\end{IEEEeqnarray}
Using the results of Proposition~\ref{prop:ts_bbm_mean_var}, we obtain
\ifCLASSOPTIONonecolumn
    \begin{IEEEeqnarray}{rCl}
        \E[K_m] & = &  m \frac{B_{p_u}(a+1, b) - B_{p_l}(a+1, b)}{B_{p_u}(a, b) - B_{p_l}(a, b)} + (N-m) \frac{B_{q_u}(c+1, d) - B_{q_l}(c+1, d)}{B_{q_u}(c, d) - B_{q_l}(c, d)}
        \label{eqn:proof_prop_4_3}
    \end{IEEEeqnarray}
\else
    \begin{IEEEeqnarray}{rCl}
        \E[K_m] & = &  m\Bigg( \frac{B_{p_u}(a+1, b) - B_{p_l}(a+1, b)}{B_{p_u}(a, b) - B_{p_l}(a, b)}  \Bigg) + \nonumber \\ & &(N-m)\Bigg( \frac{B_{q_u}(c+1, d) - B_{q_l}(c+1, d)}{B_{q_u}(c, d) - B_{q_l}(c, d)}  \Bigg)
        \label{eqn:proof_prop_4_3}
    \end{IEEEeqnarray}
\fi
\ifCLASSOPTIONonecolumn
\begin{IEEEeqnarray}{rCl}
    \Var[K_m] & = & m \Bigg( \frac{B_{p_u}(a+1, b) - B_{p_l}(a+1, b)}{B_{p_u}(a, b) - B_{p_l}(a, b)} \Bigg) - m^2 \Bigg( \frac{B_{p_u}(a+1, b) - B_{p_l}(a+1, b)}{B_{p_u}(a, b) - B_{p_l}(a, b)}\Bigg)^2 + \nonumber \\
    & & m(m-1)\Bigg( \frac{B_{p_u}(a+2, b) - B_{p_l}(a+2, b)}{B_{p_u}(a, b) - B_{p_l}(a, b)} \Bigg) +
    \nonumber \\
    & & (N-m)\Bigg( \frac{B_{q_u}(c+1, d) - B_{q_l}(c+1, d)}{B_{q_u}(c, d) - B_{q_l}(c, d)}  \Bigg) - \nonumber \\
    & & (N-m)^2 \Bigg( \frac{B_{q_u}(c+1, d) -  B_{q_l}(c+1, d)}{B_{q_u}(c, d) - B_{q_l}(c, d)}\Bigg)^2 + \nonumber \\ & & (N-m)(N-m-1) \Bigg( \frac{B_{q_u}(c+2, d) - B_{q_l}(c+2, d)}{B_{q_u}(c, d) - B_{q_l}(c, d)} \Bigg).
    \label{eqn:proof_prop_4_4}
\end{IEEEeqnarray}
\else
\begin{IEEEeqnarray}{rCl}
    \Var[K_m] & = & m \Bigg( \frac{B_{p_u}(a+1, b) - B_{p_l}(a+1, b)}{B_{p_u}(a, b) - B_{p_l}(a, b)} \Bigg) - \nonumber \\
    & & m^2 \Bigg( \frac{B_{p_u}(a+1, b) - B_{p_l}(a+1, b)}{B_{p_u}(a, b) - B_{p_l}(a, b)}\Bigg)^2 + \nonumber \\
    & & m(m-1)\Bigg( \frac{B_{p_u}(a+2, b) - B_{p_l}(a+2, b)}{B_{p_u}(a, b) - B_{p_l}(a, b)} \Bigg) + \nonumber \\
    & & (N-m)\Bigg( \frac{B_{q_u}(c+1, d) - B_{q_l}(c+1, d)}{B_{q_u}(c, d) - B_{q_l}(c, d)}  \Bigg) - \nonumber \\
    & & (N-m)^2 \Bigg( \frac{B_{q_u}(c+1, d) -  B_{q_l}(c+1, d)}{B_{q_u}(c, d) - B_{q_l}(c, d)}\Bigg)^2 + \nonumber \\
    & & (N-m)(N-m-1) \nonumber \\ & & \Bigg( \frac{B_{q_u}(c+2, d) - B_{q_l}(c+2, d)}{B_{q_u}(c, d) - B_{q_l}(c, d)} \Bigg).
    \label{eqn:proof_prop_4_4}
\end{IEEEeqnarray}
\fi
$\E[K_m^{2}]$ can be computed using equations (\ref{eqn:proof_prop_4_2}), (\ref{eqn:proof_prop_4_3}) and (\ref{eqn:proof_prop_4_4}). Using $\E[K_m^{2}]$ in equation (\ref{eqn:proof_prop_4_1}), $\E[K^2]$ can be computed and thus $\Var[K]$ can be written as
\ifCLASSOPTIONonecolumn
    \begin{IEEEeqnarray}{rCl}
        \Var[K] & = & \E[K^2] - (\E[K])^2 \nonumber \\
        & = & \frac{N}{2}\Bigg(\frac{U_{p_u}^{(1)} - U_{p_l}^{(1)}}{U_{p_u}^{(0)} - U_{p_l}^{(0)}}\Bigg) \Bigg(1 - \frac{N}{2}\Bigg(\frac{U_{p_u}^{(1)} - U_{p_l}^{(1)}}{U_{p_u}^{(0)} - U_{p_l}^{(0)}}\Bigg)\Bigg) \nonumber \\
        & & + \frac{N}{2}\Bigg(\frac{V_{q_u}^{(1)} - V_{q_l}^{(1)}}{V_{q_u}^{(0)} - V_{q_l}^{(0)}}\Bigg) \Bigg(1 - \frac{N}{2}\Bigg(\frac{V_{q_u}^{(1)} - V_{q_l}^{(1)}}{V_{q_u}^{(0)} - V_{q_l}^{(0)}}\Bigg)\Bigg) \nonumber \\
        & & + \frac{N(N-1)}{4} \Bigg(\frac{U_{p_u}^{(2)} - U_{p_l}^{(2)}}{U_{p_u}^{(0)} - U_{p_l}^{(0)}} + \frac{V_{q_u}^{(2)} - V_{q_l}^{(2)}}{V_{q_u}^{(0)} - V_{q_l}^{(0)}} \Bigg) \nonumber \\
        & & - \frac{N}{2} \Bigg(\frac{U_{p_u}^{(1)} - U_{p_l}^{(1)}}{U_{p_u}^{(0)} - U_{p_l}^{(0)}}\Bigg)\Bigg(\frac{V_{q_u}^{(1)} - V_{q_l}^{(1)}}{V_{q_u}^{(0)} - V_{q_l}^{(0)}} \Bigg).
    \end{IEEEeqnarray}
\else
    \begin{IEEEeqnarray}{rCl}
        \Var[K] & = & \E[K^2] - (\E[K])^2 \nonumber \\
        & = & \frac{N}{2}\Bigg(\frac{U_{p_u}^{(1)} - U_{p_l}^{(1)}}{U_{p_u}^{(0)} - U_{p_l}^{(0)}}\Bigg) \Bigg(1 - \frac{N}{2}\Bigg(\frac{U_{p_u}^{(1)} - U_{p_l}^{(1)}}{U_{p_u}^{(0)} - U_{p_l}^{(0)}}\Bigg)\Bigg) \nonumber \\
        & & + \frac{N}{2}\Bigg(\frac{V_{q_u}^{(1)} - V_{q_l}^{(1)}}{V_{q_u}^{(0)} - V_{q_l}^{(0)}}\Bigg) \Bigg(1 - \frac{N}{2}\Bigg(\frac{V_{q_u}^{(1)} - V_{q_l}^{(1)}}{V_{q_u}^{(0)} - V_{q_l}^{(0)}}\Bigg)\Bigg) \nonumber \\
        & & + \frac{N(N-1)}{4} \Bigg(\frac{U_{p_u}^{(2)} - U_{p_l}^{(2)}}{U_{p_u}^{(0)} - U_{p_l}^{(0)}} + \frac{V_{q_u}^{(2)} - V_{q_l}^{(2)}}{V_{q_u}^{(0)} - V_{q_l}^{(0)}} \Bigg) \nonumber \\
        & & - \frac{N}{2} \Bigg(\frac{U_{p_u}^{(1)} - U_{p_l}^{(1)}}{U_{p_u}^{(0)} - U_{p_l}^{(0)}}\Bigg)\Bigg(\frac{V_{q_u}^{(1)} - V_{q_l}^{(1)}}{V_{q_u}^{(0)} - V_{q_l}^{(0)}} \Bigg).
    \end{IEEEeqnarray}
\fi
Note that we have used the following combinatorial identities to simplify the summations in this proof:
\begin{IEEEeqnarray}{L}
   \sum_{m=0}^{N}{N \choose m}~m = N 2^{N-1} \label{eqn:comb_id_1} \\
   \sum_{m=0}^{N}{N \choose m}~m^2 = (N + N^2) 2^{N-2}. \label{eqn:comb_id_2}
\end{IEEEeqnarray}
This completes the proof.~\hfill\IEEEQED
%
\section{Proof of Proposition \ref{prop:mean_var_k0_k1_k_ts_bbm_model_bbm_model}}
\label{app:proof_mean_var_k0_k1_k_ts_bbm_model_bbm_model}
Using (\ref{eqn:etatheta_def}), (\ref{eqn:deltatheta_def})-(\ref{eqn:phitheta_def}) and the recurrence relations in (\ref{eqn:proof_prop_1_5})-(\ref{eqn:proof_prop_1_7}), we have
\ifCLASSOPTIONonecolumn
\begin{IEEEeqnarray}{rCl}
    \frac{U_{p_u}^{(1)} - U_{p_l}^{(1)}}{U_{p_u}^{(0)} - U_{p_l}^{(0)}} & = & \frac{a}{a+b} - \frac{1}{(a+b)} \frac{\delta_p}{\eta_p} \label{eqn:proof_prop_3_1} \\
    \frac{U_{p_u}^{(2)} - U_{p_l}^{(2)}}{U_{p_u}^{(0)} - U_{p_l}^{(0)}} & = & \frac{a(a+1)}{(a+b)(a+b+1)} - \frac{a+1}{(a+b)(a+b+1)} \frac{\delta_p}{\eta_p} - \frac{1}{(a+b+1)} \frac{\phi_p}{\eta_p} \label{eqn:proof_prop_3_2} \\
    \frac{V_{q_u}^{(1)} - V_{q_l}^{(1)}}{V_{q_u}^{(0)} - V_{q_l}^{(0)}} & = & \frac{c}{c+d} - \frac{1}{(c+d)} \frac{\delta_q}{\eta_q} \label{eqn:proof_prop_3_3} \\
    \frac{V_{q_u}^{(2)} - V_{q_l}^{(2)}}{V_{q_u}^{(0)} - V_{q_l}^{(0)}} & = & \frac{c(c+1)}{(c+d)(c+d+1)} - \frac{c+1}{(c+d)(c+d+1)} \frac{\delta_q}{\eta_q} - \frac{1}{(c+d+1)} \frac{\phi_q}{\eta_q}. \label{eqn:proof_prop_3_4}
\end{IEEEeqnarray}
\else
\begin{IEEEeqnarray}{rCl}
    \frac{U_{p_u}^{(1)} - U_{p_l}^{(1)}}{U_{p_u}^{(0)} - U_{p_l}^{(0)}} & = & \frac{a}{a+b} - \frac{1}{(a+b)} \frac{\delta_p}{\eta_p} \label{eqn:proof_prop_3_1} \\
    \frac{U_{p_u}^{(2)} - U_{p_l}^{(2)}}{U_{p_u}^{(0)} - U_{p_l}^{(0)}} & = & \frac{a(a+1)}{(a+b)(a+b+1)} -  \frac{a+1}{(a+b)(a+b+1)} \frac{\delta_p}{\eta_p} \nonumber \\ & & - \frac{1}{(a+b+1)} \frac{\phi_p}{\eta_p} \label{eqn:proof_prop_3_2} \\
    \frac{V_{q_u}^{(1)} - V_{q_l}^{(1)}}{V_{q_u}^{(0)} - V_{q_l}^{(0)}} & = & \frac{c}{c+d} - \frac{1}{(c+d)} \frac{\delta_q}{\eta_q} \label{eqn:proof_prop_3_3} \\
    \frac{V_{q_u}^{(2)} - V_{q_l}^{(2)}}{V_{q_u}^{(0)} - V_{q_l}^{(0)}} & = & \frac{c(c+1)}{(c+d)(c+d+1)} -  \frac{c+1}{(c+d)(c+d+1)} \frac{\delta_q}{\eta_q} \nonumber \\ & & - \frac{1}{(c+d+1)} \frac{\phi_q}{\eta_q}. \label{eqn:proof_prop_3_4}
\end{IEEEeqnarray}
\fi
Substituting (\ref{eqn:proof_prop_3_1}) and (\ref{eqn:proof_prop_3_2}) in the expressions for $\E[K^{(0)}]$ and $\Var[K^{(0)}]$ given by Proposition~\ref{prop:mean_var_k0_k1_ts_bbm_model} and simplifying, we derive the relationships between $\E[K^{(0)}]$, $\E[\tilde{K}^{(0)}]$ and between $\Var[K^{(0)}]$, $\Var[\tilde{K}^{(0)}]$ as
\begin{IEEEeqnarray}{rCl}
    \E[K^{(0)}] & = & \E[\tilde{K}^{(0)}] - \Delta^{(0)}_{mean} \\
    \Var[K^{(0)}] & = & \Var[\tilde{K}^{(0)}] - \Delta^{(0)}_{var}
\end{IEEEeqnarray}
The relationships among $\E[K^{(1)}]$, $\E[\tilde{K}^{(1)}]$ and $\Var[K^{(1)}]$, $\Var[\tilde{K}^{(1)}]$ can be derived similarly. Substituting (\ref{eqn:proof_prop_3_1})-(\ref{eqn:proof_prop_3_4}) in the expressions for $\E[K]$ and $\Var[K]$ given by Proposition~\ref{prop:mean_var_ts_bbm_model} and simplifying, we derive the relationships between $\E[K]$, $\E[\tilde{K}]$ and between $\Var[K]$, $\Var[\tilde{K}]$ as
\begin{IEEEeqnarray}{rCl+x*}
    \E[K] & = & \E[\tilde{K}] - \Delta_{mean} \\
    \Var[K] & = & \Var[\tilde{K}] - \Delta_{var}. \\* &&& \IEEEQED \nonumber
\end{IEEEeqnarray}

%% file: capacity_analysis_bbm_channel_model.bbl
\begin{thebibliography}{10}
\providecommand{\url}[1]{#1}
\csname url@samestyle\endcsname
\providecommand{\newblock}{\relax}
\providecommand{\bibinfo}[2]{#2}
\providecommand{\BIBentrySTDinterwordspacing}{\spaceskip=0pt\relax}
\providecommand{\BIBentryALTinterwordstretchfactor}{4}
\providecommand{\BIBentryALTinterwordspacing}{\spaceskip=\fontdimen2\font plus
\BIBentryALTinterwordstretchfactor\fontdimen3\font minus
  \fontdimen4\font\relax}
\providecommand{\BIBforeignlanguage}[2]{{%
\expandafter\ifx\csname l@#1\endcsname\relax
\typeout{** WARNING: IEEEtran.bst: No hyphenation pattern has been}%
\typeout{** loaded for the language `#1'. Using the pattern for}%
\typeout{** the default language instead.}%
\else
\language=\csname l@#1\endcsname
\fi
#2}}
\providecommand{\BIBdecl}{\relax}
\BIBdecl

\bibitem{Taranalli_2016}
V.~Taranalli, H.~Uchikawa, and P.~H. Siegel, ``Channel models for multi-level
  cell flash memories based on empirical error analysis,'' \emph{{IEEE} Trans.
  Commun.}, vol.~64, no.~8, pp. 3169--3181, Aug. 2016.

\bibitem{Alajaji_1994}
F.~Alajaji and T.~Fuja, ``A communication channel modeled on contagion,''
  \emph{{IEEE} Trans. Inf. Theory}, vol.~40, no.~6, pp. 2035--2041, Nov. 1994.

\bibitem{Tse_2005}
D.~Tse and P.~Viswanath, \emph{Fundamentals of Wireless Communication}.\hskip
  1em plus 0.5em minus 0.4em\relax Cambridge University Press, 2005.

\bibitem{Dong_2012}
G.~Dong, Y.~Pan, N.~Xie, C.~Varanasi, and T.~Zhang, ``Estimating
  information-theoretical {NAND} flash memory storage capacity and its
  implication to memory system design space exploration,'' \emph{{IEEE} Trans.
  {VLSI} Syst.}, vol.~20, no.~9, pp. 1705--1714, Sep. 2012.

\bibitem{Huang_2013}
X.~Huang, A.~Kavcic, X.~Ma, G.~Dong, and T.~Zhang, ``Multilevel flash memories:
  Channel modeling, capacities and optimal coding rates,'' \emph{Int. J. Adv.
  Syst. Meas.}, vol.~6, no. 3--4, pp. 364--373, 2013.

\bibitem{Li_2016}
\BIBentryALTinterwordspacing
Y.~Li, A.~Kavcic, and G.~Han, ``On the capacity of multilevel {NAND} flash
  memory channels,'' \emph{CoRR}, vol. abs/1601.05677, 2016. [Online].
  Available: \url{http://arxiv.org/abs/1601.05677}
\BIBentrySTDinterwordspacing

\bibitem{Wolfowitz_1959}
J.~Wolfowitz, ``Simultaneous channels,'' \emph{Arch. Rational Mech. Anal.},
  vol.~4, no.~1, pp. 371--386, Jan. 1959.

\bibitem{Blackwell_1959}
D.~Blackwell, L.~Breiman, and A.~J. Thomasian, ``The capacity of a class of
  channels,'' \emph{Ann. Math. Stat.}, vol.~30, no.~4, pp. 1229--1241, Dec.
  1959.

\bibitem{Taranalli_2015}
V.~Taranalli, H.~Uchikawa, and P.~H. Siegel, ``Error analysis and inter-cell
  interference mitigation in multi-level cell flash memories,'' in \emph{Proc.
  IEEE Int. Conf. Commun. (ICC)}, London, U.K., Jun. 2015, pp. 271--276.

\bibitem{Skellam_1948}
J.~G. Skellam, ``A probability distribution derived from the binomial
  distribution by regarding the probability of success as variable between the
  sets of trials,'' \emph{J. Roy. Statist. Soc. B (Methodol.)}, vol.~10, no.~2,
  pp. 257--261, 1948.

\bibitem{Boyd_2004}
S.~Boyd and L.~Vandenberghe, \emph{Convex Optimization}.\hskip 1em plus 0.5em
  minus 0.4em\relax Cambridge University Press, 2004.

\bibitem{Gupta_2004}
A.~K. Gupta and S.~Nadarajah, \emph{Handbook of Beta Distribution and Its
  Applications}.\hskip 1em plus 0.5em minus 0.4em\relax CRC Press, 2004.

\bibitem{SciPy_2016}
\BIBentryALTinterwordspacing
E.~Jones, T.~Oliphant, P.~Peterson \emph{et~al.}, ``{SciPy}: Open source
  scientific tools for {Python},'' 2001--. [Online]. Available:
  \url{http://www.scipy.org/}
\BIBentrySTDinterwordspacing

\bibitem{Tal_Vardy_2013}
I.~Tal and A.~Vardy, ``{H}ow to construct polar codes,'' \emph{{IEEE} Trans.
  Inf. Theory}, vol.~59, no.~10, pp. 6562--6582, Oct. 2013.

\bibitem{Tal_Vardy_2015}
I.~Tal and A.~Vardy, ``{L}ist decoding of polar codes,'' \emph{{IEEE} Trans.
  Inf. Theory}, vol.~61, no.~5, pp. 2213--2226, May 2015.

\bibitem{Cover_2006}
T.~M. Cover and J.~A. Thomas, \emph{Elements of Information Theory},
  2nd~ed.\hskip 1em plus 0.5em minus 0.4em\relax New York: John Wiley, 2006.

\bibitem{Sutter_2012}
D.~Sutter, J.~M. Renes, F.~Dupuis, and R.~Renner, ``Achieving the capacity of
  any {DMC} using only polar codes,'' in \emph{Proc. IEEE Inf. Theory Workshop
  (ITW)}, Sep. 2012, pp. 114--118.

\bibitem{Honda_2013}
J.~Honda and H.~Yamamoto, ``Polar coding without alphabet extension for
  asymmetric models,'' \emph{{IEEE} Trans. Inf. Theory}, vol.~59, no.~12, pp.
  7829--7838, Dec. 2013.

\bibitem{Mondelli_2014}
\BIBentryALTinterwordspacing
M.~Mondelli, S.~H. Hassani, and R.~L. Urbanke, ``How to achieve the capacity of
  asymmetric channels,'' \emph{CoRR}, vol. abs/1406.7373, 2014. [Online].
  Available: \url{http://arxiv.org/abs/1406.7373}
\BIBentrySTDinterwordspacing

\bibitem{Majani_1991}
E.~E. Majani and H.~Rumsey, Jr., ``Two results on binary-input discrete
  memoryless channels,'' in \emph{Proc. IEEE Symp. Inf. Theory (ISIT)}, Jun.
  1991, p. 104.

\bibitem{Arikan_2009}
E.~Arikan, ``{C}hannel {P}olarization: {A} method for constructing
  capacity-achieving codes for symmetric binary-input memoryless channels,''
  \emph{{IEEE} Trans. Inf. Theory}, vol.~55, no.~7, pp. 3051--3073, Jul. 2009.

\end{thebibliography}
